\documentstyle[preprint,aps]{revtex}

\preprint{\bf PREPRINT}

\newcommand{\be}{\begin{equation}}
\newcommand{\ee}{\end{equation}}
\newcommand{\bea}{\begin{eqnarray}}
\newcommand{\eea}{\end{eqnarray}}
\newcommand{\eq}[1]{eq.~(\ref{#1})}
\newcommand{\eqs}[1]{eqs.~(\ref{#1})}
\newcommand{\Eq}[1]{Eq.~(\ref{#1})}

\input{psfig}

\begin{document}

\columnsep0.1truecm
%\flushbottom
\draft
%\preprint{ }
\title{Hysteresis, Avalanches, and Disorder Induced Critical Scaling:
A Renormalization Group Approach}
\author{Karin Dahmen and James P. Sethna}
\address{Laboratory of Atomic and Solid State Physics,\\
Cornell University, Ithaca, NY, 14853-2501}
\maketitle

\begin{abstract}
Hysteresis loops are often seen in experiments at first order phase
transformations, when the system goes out of equilibrium.  They may
have a macroscopic jump (roughly as in the supercooling of liquids)
or they may be smoothly varying (as seen in most magnets).  We have
studied the nonequilibrium zero-temperature random-field
Ising-model as a model for hysteretic behavior at first order phase
transformations. As disorder is added, one finds a transition where
the jump in the magnetization (corresponding to an infinite
avalanche) decreases to zero. At this transition we find a diverging
length scale, power law distributions of noise (avalanches)
and universal behavior.
We expand the critical exponents about mean-field theory in
$6-\epsilon$ dimensions. Using a mapping to the pure Ising model, we
Borel sum the $6-\epsilon$ expansion to $O(\epsilon^5)$ for the
correlation length exponent.  We have developed a new method for
directly calculating avalanche distribution exponents,
which we perform to $O(\epsilon)$.  Numerical exponents in three,
four, and
five dimensions are in good agreement with the analytical
predictions.
Some suggestions for further analyses and experiments
are also discussed.

\end{abstract}


\narrowtext

%%%%%%%%%%%%%%%%%%%%%%%%%%%%%%%%%%%%%%%%%%%%%%%%%%%%%%%%%%%%%%%%%%%%%%%
%% This is a REVTEX file, which hopefully will run on your system too.%
%%%%%%%%%%%%%%%%%%%%%%%%%%%%%%%%%%%%%%%%%%%%%%%%%%%%%%%%%%%%%%%%%%%%%%%

\section{Introduction}
\label{chap:introduction}
The modern field of disordered systems has its roots in dirt.
An important effect of disorder is the slow relaxation to
equilibrium seen in many experimental systems \cite{glasses}.
This paper is an attempt to unearth {\it universal, nonequilibrium}
collective behavior buried in the muddy details of real materials
and inherently due to their tendency to remain far from equilibrium
on experimental time scales. In particular, we focus on two distinctly
nonequilibrium effects: (a) the avalanche response to an external
driving force and (b) the internal history dependence of the system
(hysteresis).

Systems far from equilibrium often show interesting memory effects
not present in equilibrium systems.
Far from equilibrium, the system will usually occupy some
metastable state that has been selected according to the history of
the system. Jumps over large free energy barriers to reach a more
favorable state are unlikely.
The system will move through the most easily accessible
local minima in the free energy landscape
as an external driving field is ramped, because it
cannot sample other, probably lower lying minima, from which its
current state is separated by large (free energy) barriers.
The complexity of the free energy landscape is usually greatly
enhanced by the presence of disorder. It is well
known \cite{glasses,RFIM,dynamics-RFIM,dyn-scal-RFIM,Fisher},
that disorder can lead to diverging
barriers to relaxation and consequent nonequilibrium behavior and
glassiness.

{\bf (a) Avalanches:} In some systems,
collective behavior in the form of avalanches is found when the
system is pushed by the driving field into a region
of descending slope in the free energy surface.
In experiments avalanches are often associated with
crackling noises as in acoustic emission and Barkhausen
noise \cite{Barkhausen-review,Jiles,Martensites,non-SOC}.
There are other nonequilibrium systems where no such collective
behavior is seen. Bending a copper bar for example causes a
sluggish, creeping
response due to the entanglement of dislocation lines. In contrast, wood
snaps and crackles under stress due to ``avalanches'' of fiber
breakings \cite{fiber-breaking}.

Although avalanches are collective events of processes happening on
microscales, in many systems they can become monstrously large
so that we - in spite of being large, slow creatures - can actually
perceive them directly without technical devices.
This reminds one of the behavior observed near continuous phase
transitions, where critical fluctuations
do attain human length and time scales
if a tunable parameter is close enough to its critical value.
Correspondingly one might expect to find universal features
when the sizes and times of the avalanches get large compared to
microscopic scales.

In fact, interesting questions
concerning the {\it distribution} of avalanche
sizes arise. Many experiments show power law distributions over several
decades.
For example, experiments measuring Barkhausen-pulses in an amorphous
alloy, iron and alumel revealed several decades of power law scaling
for the distribution of pulse areas, pulse durations and
pulse-energies \cite{Cote}.
Similarly, Field, Witt, and Nori recorded superconductor vortex
avalanches in $Nb_{47\%}Ti_{53\%}$ in the Bean-state as the system
was driven to the threshold of instability by the slow ramping of the
external magnetic field. The avalanche sizes ranged from
50 to $10^7$ vortices. The corresponding distribution of avalanche
sizes revealed about three decades of power law scaling.
Numerous other systems show similar power law scaling
behavior
\cite{Cote,Stuart-Field,Ortin-martensite-aval,Gutenberg-Richter,Hallock}.

Why should there be avalanches of many sizes? Power laws suggest
a scaling relationship between different length scales with universal
exponents.

There has been much recent progress studying avalanches near
(continuous) {\it depinning transitions}.
In these systems a single, preexisting interface or ``rubber sheet''
is pushed through a disordered medium
by an external driving force.
When the randomness is in some sense weak, the interface
distorts elastically without breaking over a wide range of
length scales \cite{Coppersmith}.
The distortions occur in the form of avalanches on increasing
size as the external driving force is raised to a critical
threshold field at which the interface
starts to slide (``depins'').
Examples studied include charge density waves
\cite{FisherCDW,CDW-review,NarayanI,NarayanII,Myers,Middleton},
weakly pinned Abrikosov flux lattices \cite{flux-lattice},
single vortex-lines \cite{flux-line,flux-line-theory,Ertas2},
preferentially wetting fluids invading porous media
\cite{fluid-invasion,Robbins},
a single advancing domain wall in weakly disordered
magnets \cite{Robbins,NarayanIII,Nattermann} and
fluids advancing across dirty surfaces \cite{Ertas}.
Their analytical description turns out to be rather involved,
demanding functional renormalization groups (see appendix
\ref{ap:related}).
If the disorder is in some sense strong, the elastic interface
can tear. It then responds much more inhomogeneously and
like a plastic
or fluid, to the external driving force. This is the case
for strongly pinned vortex lines in the mixed state
of superconducting films \cite{flux-lattice-torn,flux-line-theory},
the invasion of nonwetting fluids into porous media
\cite{nonwetting,Robbins} and nonlinear fluid flow across dirty surfaces
\cite{NarayanIV}, and others.

Many hysteretic systems exhibit a wide distribution of avalanche
sizes {\it without} an underlying depinning transition.
They usually have {\it many} interacting advancing
interfaces and in some cases also new interfaces
created spontaneously in the bulk.
We propose that the large range of observed avalanche sizes
in these systems might be a manifestation of a nearby critical
point with both disorder and external magnetic field
as tunable parameters.
In the class of models, which we have been studying
near the critical point, we find not only
universal scaling behavior in the avalanche size distribution,
but also in the shape of the associated hysteresis loops.

{\bf (b) Hysteresis (response lags the force):}
Hysteresis is often observed
at first order transitions, when the
system goes out of equilibrium.
It probably has been best studied in magnetic systems. The origin of
magnetic hysteresis lies in the spontaneous magnetization of the
microscopic Weiss-domains \cite{Chikazumi}.
In the demagnetized state, the Weiss
domains are irregularly oriented in different directions. The
orientation of the magnetization of each domain is a function of
the energy of the magnetic field and the elastic energy in the
crystal (magnetostriction) and it is chosen such that the free
energy is at a (local) minimum.
Since there are many minima available, the specific choice depends
on the history of the system. Thermal vibrations of the lattice are
usually not sufficient to rotate the magnetization of a Weiss-domain
into another preferred direction, since in most cases these directions
are separated from each other by large energy barriers.
As a weak magnetic field is
applied, first those Weiss-domains which are most closely aligned
with the field will grow at the expense of the others.
At higher fields, up to saturation, entire domains will rotate in the
direction of the field. The ultimate cause for hysteresis are the
irreversible domain wall motion and domain rotation,
which happen suddenly, as ``avalanches'' without
further field increase, when the corresponding threshold fields are
exceeded.
The resulting jumps in the magnetization are called Barkhausen
jumps. Under a magnifying glass the magnetization curve looks like a
staircase:
the slope of the flat parts is due to the reversible part of the
susceptibility, the step height is given by the irreversible
avalanche-like changes of the magnetization.
If the magnetization curve has the shape of
a rectangle, the change of the
magnetization happens in enormous, system sweeping avalanches,
the so called large Barkhausen effect.
Experimentally the Barkhausen jumps can be observed by magnetic
induction or through the associated acoustic emission.

Analogous effects are found in ferroelectric materials,
where avalanches of flipping ferroelectric domains can be observed
in response to a changing external electric
field \cite{ferroelectrica,Rudyak}.
Hysteresis curves with step-like noise
are also found in elastic
transformations, for example in athermal shape-memory alloys
ramping temperature or stress. The noise
is due to avalanches of regions transforming from martensite to
austenite or vice versa \cite{Ortin-martensite-aval}.
Similar behavior has been observed for vortices moving in
avalanches in type II superconductors as the external magnetic
field is increased \cite{Stuart-Field},
for liquid helium leaving
Nuclepore in avalanches as the chemical potential is
reduced \cite{Hallock}, and for some earthquake models
\cite{Nagel,earthquakes,Carlson,Gutenberg-Richter}.
In section \ref{sec:experiments} we discuss recent experiments performed
in some of these systems.

We have modeled the long wavelength, low frequency behavior of these
systems using the nonequilibrium zero temperature random field Ising
model (RFIM).
Some of our results have been published
previously \cite{hysterI,hysterII}. In contrast to some other
hysteresis models, like the Preisach model \cite{Preisach}
and the Stoner-Wohlfarth model \cite{Jiles},
where interactions between the
individual hysteretic units (grains) are not included and collective
behavior is not an issue, in the RFIM the
intergrain coupling is the essential ingredient and cause for
hysteresis and avalanche effects.
Tuning the amount of disorder in the system we find a second order
critical point with an associated diverging length scale,
measuring the spatial extent of the avalanches of spin
flips.

{\bf (c) Avalanches in the RFIM:}
A power law distribution with avalanches of {\it all} sizes
is seen only at the critical value of the disorder.
However, our numerical simulations indicate that the critical region
is remarkably large: almost three decades of power law scaling in the
avalanche
size distribution remain when measured 40\% away from the critical
point. At 2\% away, we extrapolate seven decades of scaling.
One reason for this large critical range is trivial:
avalanche sizes are expressed in
terms of volumes rather than lengths, so one decade of length scales
translates to at least
three decades of size (or more if the avalanches are not
compact,
{\it i.e.} if the Hausdorff dimension is less than three).
Some experiments that revealed three decades of power law
scaling have been interpreted as being
spontaneously self-similar
(``self-organized critical'')
\cite{Stuart-Field,Cote,OBrien-Weissman}.\footnote{The name ``self-organized
critical'' is in fact an oxymoron: "Critical" means that you have to
be just at the right place,
"self-organized" indicates that
the system does not need to be tuned anywhere special.
While it is true that many ``self organized critical'' systems are
regular critical points where the boundary conditions stabilize the
system at the transition, the same is true of ice in a glass of
water. The tools for studying the critical point are the RG methods
in development over many years. We discuss here our doubts that the
experiments are self-organized even in this limited sense.}
Our model suggests that many of the samples might just have disorders
within 40\% of the critical value. Tuning the amount of disorder in
these systems might reveal a plain old critical point rather than
self organized criticality.

{\bf (d) Hysteresis in the RFIM:} At the critical disorder we also find a
transition in the shape of the associated hysteresis loops:
Systems with low disorder relative to the coupling strength,
have rectangle-shaped hysteresis loops
and a big (Barkhausen) discontinuity,
while systems with large disorder relative to the coupling
show smooth hysteresis loops without macroscopic jumps.
At the critical disorder $R_c$ separating these two regimes,
the size of the jump seen in the low disorder
hysteresis loops shrinks to a point at a critical magnetic field
$H_c(R_c)$, where the magnetization curve $M(H)$ has infinite slope.
The power law with which it approaches this point
is universal.

{\bf (e) Results:} We have extracted the universal exponents near
this transition point from a history dependent renormalization group
(RG) description for the nonequilibrium zero temperature
random field Ising model. The calculation turns out to be much
simpler than for related depinning transitions
\cite{NarayanI,NarayanII,NarayanIII,Nattermann,Ertas,Ertas2} (see
also appendix \ref{ap:related}).
Above 6 dimensions the exponents are
described by mean field theory.
We expand the critical exponents around mean-field theory in
$6-\epsilon$ dimensions and discover a mapping to the perturbation
expansion for the critical exponents in the pure
equilibrium Ising model in two
lower dimensions.
The mapping does not, however, apply to the exponents governing the
avalanche size distribution, which
to our knowledge,
have not yet been calculated directly in the depinning transitions.
The simplicity of the RG calculation allowed us to
develop a new method to calculate these avalanche exponents directly
in the $\epsilon$-expansion, involving replicas of the
system in a very physical way. We have used it to calculate the
avalanche exponents to first order in $\epsilon$.
We report under separate cover \cite{Perkovic} extensive numerical
simulations used to extract exponents in 3, 4, and 5 dimensions (see
section \ref{sec:numerics-and-experiment}).
We find good agreement between the two approaches.

This paper is organized as follows:
In section \ref{sec:experiments}
we discuss several experiments in magnetic systems,
shape memory alloys, porous media and superconductors that have
close connections to the model studied here.
The model is introduced in section \ref{sec:the-model} and
a summary of our results is given in
section \ref{sec:results}.
In section \ref{sec:motives}
we pause for a moment and reflect upon our
real motives. In section \ref{sec:analytic-description}
the RG description is set up using
the Martin-Siggia-Rose formalism, and a description of the
perturbative expansion and the results for the
exponents to $O(\epsilon)$ is given in
section \ref{sec:perturbative-expansion}.
section \ref{sec:mapping} contains
a discussion of the mapping of the
expansion to the
expansion for the thermal RFIM.
We extract
corrections to $O(\epsilon^5)$ for most of the exponents and show a
comparison between the Borel resummation of the $\epsilon$-expansion
and numerical results. In section \ref{sec:epsilon-expansion}
a new method to
calculate avalanche exponents directly in an $\epsilon$-expansion
is described and performed to $O(\epsilon)$.
Finally, in section \ref{sec:numerics-and-experiment} we
compare the results to
% XXXX
our numerical simulation \cite{Perkovic}.

Some of the details of the mean-field calculation are given in
appendix \ref{ap:mean-field}.
The expected tilting of the scaling axes in finite
dimensions is discussed in appendix \ref{ap:turning-of-axes}.
Details on the implementation
of the history in the RG calculation are given in appendix
\ref{ap:RG-details}. Appendix \ref{ap:Borel} contains a description of
the Borel summation of the results for $\eta$ and $1/\nu$ to
$O(\epsilon^5)$ (which is relevant also for the pure Ising model).
The behavior near the infinite avalanche line in systems with less
than critical randomness is discussed in appendix
\ref{ap:inf-aval-line}.
Appendix \ref{ap:RG-details-avalanches} renders
details on the calculation of the avalanche exponents
by the use of replicas.
Related problems are finally discussed in appendix \ref{ap:related}.

\section{Experiments}
\label{sec:experiments}

In this section we will discuss several experiments
that reveal scaling behavior which might be related to the
critical point studied in this paper.
The critical exponents found in real experiments do not
necessarily have to be the same as in our model, since long-range
interactions, different conserved quantities and other changes
are likely to alter the universality class in some cases.
We do propose however that the qualitative features, in particular
the existence of an underlying plain old critical point
with disorder and driving field as tunable parameters are likely to
be the same as in our model (see also appendix \ref{ap:related}).
A more detailled discussion of these and other
Barkhausen experiments
%% FOLLOWING LINE CANNOT BE BROKEN BEFORE 80 CHAR
\cite{Stierstadt,Bertotti94,Bittel,Lieneweg,Lieneweg-and-Grosse-Nobis,Bertotti90,Montalenti,Urbach},
and related experiments in
non-magnetic avalanching systems (in shape memory alloys
\cite{Ortin-martensite-aval,Olson}, superconductors
\cite{Stuart-Field,Adams}, liquid helium in Nuclepore \cite{Hallock},
and others), and a quantitative comparison with our theory
will be given in a forthcoming publication
\cite{hysterIII,eps-big,thesis}.

\subsection{Magnetic hysteresis loops for different annealing
temperatures}
\label{subsub:magnetic-films-hysteresis}
%
%\begin{figure}
%\vspace{18cm}
%\caption
%[Experiment: Magnetic hysteresis loops in a thin Gd film
%after annealing the sample at different temperatures]
%{{\bf Experiment: Magnetic hysteresis loops of a 60 nm thick Gd film for
%various annealing temperatures} (as indicated next to each loop)
%and constant annealing time (3 minutes each).
%All measurements are performed at $200 \pm 5 K$. The sweeping
%frequency of the external magnetic field is $0.5$ Hz (from A.
%Berger, unpublished).
%\label{fig:Berger}}
%\end{figure}
%
A beautiful, qualitative illustration of the crossover
from smooth hysteresis loops at large disorder to hysteresis
loops with macroscopic jumps at low disorder is shown in figure
\ref{fig:Berger}.
The hysteresis loops were measured by Berger
\cite{Berger,Berger-setup} for a 60 nm thick Gd film which had been grown
onto a tungsten single crystal with a (110) surface orientation.
The substrate as well as the film are highly purified
(contaminants are less than 1/20 of a monolayer).
Gd films prepared in this way exhibit a hcp structure with the
(0001) direction perpendicular to the surface. The substrate
temperature during deposition was $T=350 K$, which
results in smooth films with large atomically flat terraces,
but also produces films with locally varying
strain and therefore with locally varying anisotropy.
Subsequent annealing at higher temperatures improves the
crystallographic order, which is accompanied by a strain relaxation.
Thus, by varying the annealing temperature the authors are able to
change the variation of the anisotropy defect density, which is
somewhat analogous to the disorder parameter $R$ in our model.
Higher annealing temperatures correspond to lower values of $R$.
If there is a second order critical point of the kind described
in the introduction
underlying the crossover from hysteresis loops with a jump to
smooth hysteresis loops, it should be possible to extract a scaling
form for the magnetization curves similar to the one given in
\eq{magn-scaling}. (The annealing temperature minus some critical
value would play the role of the tunable reduced disorder parameter
$r$.) Under appropriate rescaling of the axes near the critical
point the magnetization curves should all collapse one onto another.
The necessary amount of rescaling as a function of distance from the
critical point determines the (presumably universal)
exponents \cite{Perkovic}.
Further measurements near the crossover to extract potential scaling
behavior are currently being performed by Berger \cite{Berger}.

\subsection{Barkhausen noise for different annealing temperatures}
\label{subsub:magn-annealing-Barkhausen}
Scaling behavior has also been recorded
in the Barkhausen pulse duration and pulse area distribution
in a related experiment \cite{Barkhausen-review}.
The pulse area gives the total change in the magnetization due to
the corresponding Barkhausen pulse. It is analogous to the
avalanche size
given by the number of spins participating in an avalanche in our
model.
It was found that the
distribution of pulse areas integrated over the hysteresis loop
of an $81 \%$ Ni-Fe wire (50 cm long, 1mm diameter) was well
described by a power law up to a certain cutoff size
(see figure \ref{fig:Barkhausen-annealing}).
The cutoff appeared to be smaller at higher annealing
temperatures. It would be interesting to see whether the cutoff
takes a system-size dependent maximum value at a critical
annealing temperature $T_c^{ann}$ and decreases again at higher
and lower
annealing temperatures. This would be expected if varying the
annealing temperature would correspond to tuning the system through a
critical region with a diverging length scale at $T_c^{ann}$.
Near $T_c^{ann}$ the Barkhausen pulse area distributions should then
be described by a scaling form that would allow a scaling collapse
of all distributions onto one single curve for appropriate stretching
of the axes. Again, potentially universal critical exponents could
be extracted from such a collapse. They would be predicted by our
avalanche critical exponents if our model is in the same
universality class.

%\begin{figure}
%\vspace{18cm}
%\caption
%[Experiment: Distribution of Barkhausen pulse areas in a $81 \%$
%Ni-Fe wire
%after annealing the sample at different temperatures]
%{{\bf Experiment: Distribution of pulse areas ($p$), integrated over
%the hysteresis
%loop for $81 \%$ Ni-Fe wires after various heat treatments.}
%The originally hard drawn wires have been subjected to a one-hour
%heat treatment in high vacuum at temperatures of $240^\circ C$ or
%$460^\circ C$ and cooled down in the furnace.
%(From U. Lieneweg and W. Grosse-Nobis, {\it Int. J. Magn} {\bf 3},
%11 (1972).)
%\label{fig:Barkhausen-annealing}}
%\end{figure}

\subsection{Barkhausen pulse size distributions at fixed disorder}
\label{subsub:Barkh-fixed-R}
There are other experiments which revealed power law decays for
Barkhausen pulse size distributions in various samples, as we have
mentioned earlier \cite{Cote,OBrien-Weissman}.
To our knowledge
the amount of disorder (or another parameter)
was not varied in these experiments.
The power law scaling over several decades found in these systems
has in some cases been interpreted as a manifestation of
self-organized criticality \cite{Cote}.
According to our simulations however,
three decades of scaling occur when the disorder is as far
as $40\%$ away from the critical value. Tuning the
amount of disorder
in the system (for example by annealing the sample as in
the previous two experiments, or by introducing random strain fields)
might lead to a larger (or smaller) cutoff in the power law pulse
size distribution. It seems rather plausible
that the observed scaling behavior would be due
to a plain old critical point rather than self-organized criticality.

\subsection{Remarks}
The first experiment that showed the crossover in the shape of the
magnetic hysteresis loops was performed in an effectively
two dimensional system, while the experiments on Barkhausen noise
used effectively three dimensional systems.
Interestingly, two might be
the lower critical dimension of the transition which we are
studying in this paper.\footnote{In one dimension there will
still be a crossover from
hysteresis loops with a macroscopic jump to smooth hysteresis loops
for a {\it bounded}
distribution of random fields \cite{Chayes-private-com},
however the potential scaling behavior found near the transition will
not be universal, but rather depend on the exact shape of the
tails of the distribution
of random fields.} These conjectures are currently being tested with
our numerical simulations \cite{Perkovic}.

Real experiments may involve long-range fields which may in principle
alter the universality class. Different kinds of
disorder, such as correlated disorder rather than point disorder and
random anisotropies or random bonds rather than
random fields, may also be present.
Furthermore, the symmetries can be changed if there are
more than two available (``spin'') states at each site in the
lattice. In appendix \ref{ap:related} we discuss related models,
and which of these
changes are expected to change the universality class relative to
our model.

What is the moral of this story? Power laws with cutoffs are anything but
sufficient evidence for self-organized criticality. In the systems
discussed here, they are more likely due to a nearby {\it plain old
critical} point
with disorder as a tunable parameter
and a large critical region,
than spontaneous
self-organization towards a self-similar state.
For related future experiments and analysis one would recommend the
search for tunable parameters other than system size that allow to
change the cutoff in the power law distributions. For the analysis
of our simulation results,
scaling collapses and other techniques from equilibrium
critical phenomena turned out to be
very useful for extracting critical exponents \cite{Perkovic}.

\section{The model}
\label{sec:the-model}

As we have explained in the introduction, the goal is to describe
the long-wavelength behavior of hysteretic
systems with noise due to microscopic avalanches triggered by the
external driving field. We will focus in particular on the scaling
regime,
where collective behavior is observed on many length scales.
Conventional hysteresis models like the Preisach model
\cite{Preisach},
which do not take into account interaction between the smallest
hysteretic units (grains), would not be suitable for this purpose.
The Preisach model could only be used to fit a certain measured
distribution of avalanche sizes  --- the power law scaling
would be the {\it input} determining free parameters of the
model rather than the {\it output} with universal predictive power.

The key ingredient is interaction.\footnote{For
systems exhibiting return-point-memory
(also called ``subloop-closure'' or ``wiping-out'' property)
\cite{hysterI,RPME,Preisach} there is in fact a well established method
to verify whether
interactions play an important role, and whether the
Preisach model is applicable at all.
It involves testing for subloop-congruency.
For further
details we refer the reader to the
literature \cite{RPME,Preisach,Hallock}.
This test has been conducted for the experiment on
liquid He in Nuclepore \cite{Hallock} revealing the importance
of interactions between the pores, and the failure of the
conventional Preisach model to describe the hysteresis curve.}
As is well known from equilibrium phenomena,
behavior on long length scales can often be well described by simple
microscopic models that only need describe a few basic properties
correctly, such as symmetries, interaction range and effective
dimensions. This notion has been successfully
applied in particular to equilibrium magnetic systems:
the scaling behavior found in
some pure anisotropic ferromagnets near the
Curie-temperature is mimicked reliably by the regular Ising
model \cite{Goldenfeld,Yeomans}.
At each site $i$ in a simple cubic lattice there is a variable $s_i$,
in this context called a spin, which can take two different values,
$s_i=+1$ or $s_i=-1$ \cite{quantum}.
(This corresponds to a real magnet where
a crystal anisotropy prefers the magnetic moments
(spins) to point along a certain easy axis.)
Each spin interacts with its nearest neighbors
on the lattice through an exchange interaction, $J_{ij}=J/z$, which favors
parallel alignment. z is the coordination number of the
lattice and $J$ is a positive constant. (For the behavior on long
length scales the exact range of the microscopic interaction
is irrelevant, so long as it is finite.)
One can write the Hamiltonian as
\be
\label{pure-Ising}
{\cal H} = - \sum_{ij} J_{ij} s_i s_j - H \sum_{i} s_i \, ,
\ee
where it is understood that
the sum runs over nearest neighbor pairs of spins on sites $i$ and $j$.
$H$ is a homogeneous external magnetic field.
In two and higher dimensions this model exhibits an equilibrium
ferromagnetic state at
temperatures $T<T_c$, where $T_c$ is the Curie temperature.
Figure \ref{fig:equil-M(H)} shows the corresponding equilibrium
magnetization
curve at zero temperature: All spins are pointing up at positive external
magnetic fields, and all spins are pointing down at negative external
magnetic fields. At $H=0$ the curve is discontinuous.

%XXXXXXXXXXXXXXXX figure fig:equil-M(H) XXXXXXXXXXXXXXX
%\begin{figure}[p]
%%figure.tex - produces the figure of two BW tilings.
%%\normalfigure{equilibrium-magnetization-vs-H}
%%\psfig{figure=equilibrium-magnetization-vs-H.ps}
%\epsffile{equilibrium-magnetization-vs-H.ps}
%\epsfsize=\hsize
%\epsfbox{equilibrium-magnetization-vs-H.ps}
%\psfig{figure=equilibrium-magnetization-vs-H.ps, width=figwidth}
%\hbox{
%\vbox to 3in{
% \vfil
% \special{psfile=equilibrium-magnetization-vs-H.ps
%          hoffset=0
%          voffset=270
%          hscale=50
%          vscale=50}
%}}
%\caption
%[Equilibrium magnetization curve $M(H)$
%for the pure Ising model at zero temperature.]
%{{\bf Equilibrium magnetization curve $M(H)$}
%for the pure Ising model at zero temperature.
%\label{fig:equil-M(H)}}
%\end{figure}

What would the magnetization curve look like for the same model,
but far from equilibrium, as is the case for most real magnets?
The answer is shown in figure \ref{fig:rectangle}.
We have imposed a certain local dynamics, assuming that
each spin $s_i$ will flip only when the
total effective field at its site, given by
\be
h_i^{\it eff} = - \sum_{j} J_{ij} s_j -H \, ,
\ee
changes sign.
We find that the resulting magnetization curve becomes history
dependent. The system will typically be in some metastable state
rather than the ground state.
The upper branch of the hysteresis curve in figure \ref{fig:rectangle}
corresponds to the case where we have monotonically and adiabatically
lowered the
external magnetic field, starting from $H=+\infty$, where all spins
were pointing up. At the coercive field $H_c^l=-2d J_{ij} \equiv -J$
all spins flip in
a single system spanning event or ``avalanche''. Similarly, for
increasing external magnetic field, they all flip at $H_c^u= 2d
J_{ij} \equiv +J$.
It becomes clear that the underlying cause for hysteresis in this model is the
interaction between the spins.

%XXXXXXXXXXXXXXXX figure fig:rectangle XXXXXXXXXXXXXXX
%\begin{figure}[p]
%\vspace{1cm}
%\hbox{
%\vbox to 3in{
% \vfil
% \special{psfile=rectangle.ps
%          hoffset=0
%          voffset=270
%          hscale=50
%          vscale=50}
%}}
%\caption
%[Nonequilibrium magnetization curve $M(H)$ for the pure Ising model
%at zero temperature with a local dynamics.]
%{{\bf nonequilibrium magnetization curve $M(H)$}
%in the pure Ising model at zero temperature for the dynamics
%defined in the text.
%\label{fig:rectangle}}
%\end{figure}

So far, however, an essential feature
of real materials is missing: there is no account for {\it dirt}.
Usually there will be
inhomogeneities and disorder in the form of defects,
grain boundaries, impurities, leading to random crystal anisotropies,
and varying interaction strengths in the system. Consequently not
all spins will flip at the same value of the external magnetic field.
Instead, they will flip in avalanches of various sizes that can be
broken up or stopped by strongly ``pinned'' spins or clusters of
previously flipped spins.

If the disorder in the system is small, the picture will not deviate
dramatically from the pure case.
One would expect only a few small precursors to the
macroscopic avalanche of figure \ref{fig:rectangle}.
If however, the disorder is large compared to the coupling strength
in the system, one might expect no system sweeping avalanche at all,
but only small clusters of spins flipping over a broad range of the
external magnetic field.

A simple way to implement a certain kind of
uncorrelated, quenched disorder is by
introducing uncorrelated random fields into the model.
(Other kinds of disorder are discussed in appendix \ref{ap:related}.)
The energy function is replaced by
\be
\label{model1}
{\cal H}= -\sum_{ij} J_{ij} s_i s_j - \sum_i (H+f_i) s_i  \, .
\ee
The local dynamics remains unchanged, except for a modification
in the expression for the total effective field at site $i$, which
now also has to take into account the random field $f_i$:
\be
h_i^{\it eff} = - \sum_{j} J_{ij} s_j -H - f_i   \, .
\ee
We assume a Gaussian
distribution $\rho(f_i)$ of standard deviation $R$ for the fields
$f_i$, which is
centered at $f_i=0$:
\be
\rho(f_i) ={ 1\over{\sqrt{2 \pi} R}} \exp{(-{f_i^2\over{2 R^2}})} \, .
\ee
As we will show, the critical exponents do not depend
on the exact shape of the distribution of random fields. To pick a
Gaussian is a standard choice, which (due to the central limit
theorem \cite{vanKampen}) is also more likely
to be found in some real experiments than, for example, rectangular
distributions \cite{Olami}.

In magnets the random fields might model frozen-in magnetic clusters
with net magnetic moments that remain fixed even if the surrounding
spins change their orientation. In contrast to random anisotropies they
break time reversal invariance by coupling to the order parameter
(rather than its square).
In shape memory alloys, ramping temperature, the random fields
can be thought of as concentration fluctuations that prefer
martensite over
the austenite phase \cite{Kartha}.
In the martensitic phase, ramping stress, they model strain fields
that prefer one martensitic variant over another \cite{Kartha}.

In appendix \ref{ap:related} we discuss related systems with
different kinds of disorder, symmetries, interactions and dynamics,
and the possible effects of such changes on the associated long-wavelength
behavior and critical properties at transitions analogous to the one
studied here.

\section{Results}
\label{sec:results}

It is relatively easy to solve the nonequilibrium model
of \eq{model1} in the mean-field
approximation
where every spin interacts equally strongly
with every other spin in the system.
The coupling is of size $J_{ij}=J/N$, where $N$ is the total number
\label{page:footJ} of
spins\footnote{We warn the reader that in all plots of
{\it numerical simulation} results in finite dimensions $J$ will
denote the strength of the nearest neighbor coupling $J_{ij}$ in the
simulated crystal, while in the analytic calculation and in mean-field
theory it denotes $\sum_j J_{ij}$, {\it i.e.} the two definitions
differ by the coordination number of the lattice. \label{foot:J}}
({\it i.e.} all spins act as nearest neighbors).
The Hamiltonian then takes the form
\be
\label{Hamiltonian}
{\cal H} = -\sum_i (J M + H + f_i) s_i \, ,
\ee
Just as in the Curie-Weiss mean field theory for the
Ising model, the interaction of a spin with its neighbors
is replaced by its interaction with the magnetization of the
system.

It turns out that the mean-field theory already reflects most of the
essential features of the long-length scale behavior of the system in
finite dimensions:
Sweeping the external field through zero, the model exhibits
hysteresis.
As disorder is
added, one finds a continuous transition where the jump in the magnetization
(corresponding to an infinite avalanche) decreases to zero. At this
transition power law distributions of noise
(avalanches) and universal behavior are observed.

As we will show later in an RG description of the model, the
critical exponents describing the scaling behavior near the critical
point are correctly given by mean-field theory for systems in 6 and
higher spatial dimensions.
The RG allows us to calculate their values in $(6-\epsilon)$
dimensions in a power series expansion in $\epsilon > 0$
around their mean-field values at $\epsilon=0$.
In the following we briefly present the results from mean-field
theory, from the $\epsilon$-expansion and from numerical simulations
in 3 dimensions. More details will be given in later sections.

\subsection{Results on the magnetization curve}

Figure \ref{fig:MFT-magnetization} shows the hysteresis curve
in mean field theory at various
values of the disorder $R<R_c=\sqrt{(2/\pi)} J$, $R=R_c$, and $R>R_c$.
For $R<R_c$, where the coupling is important relative to the
amount of disorder in the system, the hysteresis curve displays a
jump due to an infinite avalanche of spin flips, which spans the
system. Close to $R_c$ the size of the jump scales as
$\Delta M \sim r^\beta$, with $r=(R_c-R)/R$, and $\beta=1/2$ in mean
field theory. Using a mapping to the pure Ising model, we
find in $6-\epsilon$ dimensions \cite{Kleinert}
\bea
\nonumber
\beta &=& 1/2 -\epsilon/6 + 0.00617685\epsilon^2 - 0.035198 \epsilon^3
+ 0.0795387 \epsilon^4  \\
& & - 0.246111 \epsilon^5 + O(\epsilon^6)\, .
\eea

At $R=R_c$ the magnetization curve scales as
$M-M(H_c(R_c)) \sim h^{(1/\delta)}$, where $h=H-H_c(R_c)$ and
$H_c(R_c)$ is the (nonuniversal) magnetic field value
at which the magnetization
curve has infinite slope. In this mean field theory $H_c(R_c)=0$, and
$M(H_c(R_c))=0$, and
$\beta \delta =3/2$. In $6-\epsilon$ dimensions \cite{Kleinert}
\be
\beta\delta = 3/2 + 0.0833454 \epsilon^2 - 0.0841566 \epsilon^3
+ 0.223194 \epsilon^4 - 0.69259 \epsilon^5 + O(\epsilon^6)\, .
\ee
Numerical simulations in 3 dimensions yield $\beta=0.036\pm0.036$
and $\beta \delta= 1.81\pm 0.36$ \cite{Perkovic}.

For $R>R_c$ the disorder can be considered more important than the
coupling. Consequently there are no system spanning avalanches
(for infinite system size) and the magnetization curve is smooth.

Note that the hard spin mean-field theory does not show any hysteresis
for $R\geq R_c$. This is only an artifact of its particularly simple
structure and not a universal feature.
For example, the analogous soft spin model, which is introduced for the
RG description in section \ref{sec:analytic-description},
has the same exponents in mean-field
theory, but shows hysteresis at {\it all} disorders $R$, even for
$R>R_c$ as seen in figure \ref{fig:MFT-magn-soft-spin}.

Close to $R_c$ and $H_c(R_c)$
the magnetization curve is described by a scaling form:
\be
\label{magn-scaling}
M-M(H_c(R_c)) \equiv m(r,h) \sim r^\beta {\cal
M}_\pm(h/r^{\beta\delta})
\ee
where ${\cal M}_\pm$ is a universal scaling function
($\pm$ refers to the sign of $r$).
It is computed in mean-field theory in appendix \ref{ap:mean-field}.
Corrections\footnote{The scaling form in finite dimensions may
depend not on
$r$ and $h$, but on rotated variables $r'$ and $h'$,
which are
linear combinations of $r$ and
$h$. This applies to all scaling relations derived in the infinite
range model.
The amount by which the
scaling axes $r'=0$ and $h'=0$ are turned relative to $r=0$
and $h=0$ is a nonuniversal quantity and has no effect on the
critical exponents (see appendix \ref{ap:turning-of-axes}).}
to the mean-field equation of state in
$6-\epsilon$ dimensions are calculated to $O(\epsilon)$ in section
\ref{subsub:equation-of-state}. Results to $O(\epsilon^2)$
are quoted in section \ref{sec:mapping}.

\subsection{Results on the mean-field phase diagram}
\label{sub:results-MFT-phase-diagram}

Figure \ref{fig:MFT-phase-diagram} shows the phase diagram for
the lower branch
of the hysteresis curve as obtained from the simple hard spin mean-field
theory, defined through \eq{Hamiltonian}.
The bold line with the critical endpoint $(R_c,H_c(R_c))$
indicates the function $H_c^u(R)$ for the onset of the infinite
avalanche for the history of an {\it increasing} external magnetic
field. The dashed line describes $H_c^l(R)$ for a
{\it decreasing} external
magnetic field. The three dotted vertical lines marked (a), (b), and
(c) describe the paths in parameter space which lead to the
corresponding hysteresis loops shown in figure
\ref{fig:MFT-magnetization}.
Figure \ref{fig:MFT-soft-spin-phase-diagram} shows the
corresponding phase diagram
for the soft-spin model. As before, the three dotted vertical lines
marked (a), (b) and (c) indicate the paths through parameter space
associated with the three hysteresis loops shown in figure
\ref{fig:MFT-magn-soft-spin}.
Note that in the soft-spin model as well as in
simulations in finite dimensions the two infinite avalanche lines
$H_c^u(R)$ and $H_c^l(R)$ do not touch at $R_c$ --- this is another
way of saying that in these cases there is hysteresis at $R=R_c$,
and, because of continuity also at $R>R_c$.

%XXXXXXXXXXX figure fig:MFT-phase-diagram XXXXXXXXXXXXXX

%XXXXXXXXXXX figure fig:MFT-soft-spin-phase-diagram XXXX

\subsection{Results on scaling near the onset $H_c(R)$ of the
infinite avalanche line $(R < R_c)$ }

The mean-field magnetization curve scales near the onset of the
infinite avalanche as
\be
(M-M_c(H_c(R))) \sim (H-H_c(R))^\zeta
\ee
with $\zeta = 1/2$.
(In the following
$H_c(R)$ always stands for $H_c^u(R)$ for the history
of an increasing external magnetic field, and for $H_c^l(R)$ for
a decreasing external magnetic field.)

Curiously we do not observe this scaling behavior in
numerical simulations with short range interactions
in 2, 3, 4 and 5 dimensions.
Indeed,
the RG description suggests that the
onset of the infinite avalanche would be an
abrupt (``first order'' type) transition
for all dimensions $d<8$ (see appendix \ref{ap:inf-aval-line}),
and a continuous transition for $d>8$.
We \cite{Perkovic} have
performed initial numerical simulations in 7 and 9 dimensions
for system sizes $7^7$ and $5^9$ at less than critical
disorders. The simulation results do in fact seem to confirm
the RG prediction \cite{Perkovic}.
In the following we will mostly focus on the critical endpoint
at $(R_c,H_c(R_c))$, where the mean-field scaling behavior is
expected to persist in finite dimensions with slightly changed critical
exponents.

\subsection{Results on avalanches}
\label{sec:results-on-avalanches}

Figure \ref{fig:MFT-simul-m(h)} shows the magnetization
curves from simulations of two 3 dimensional systems
with only $5^3$ spins.
The curves are not smooth. They display steps of various sizes.
Each step in the magnetization curve corresponds to an
avalanche of spin flips during which the external
magnetic field is kept constant.
Figure \ref{fig:MFT-D(s)} shows histograms $D(S,r)$
of all avalanche sizes $S$ observed in mean-field
systems at various disorders
$r$ when sweeping through the entire hysteresis loop.
For small $r$ the distribution roughly follows a power law
$D(S, r) \sim S^{-(\tau+\sigma\beta\delta)}$ up to a certain
cutoff size $S_{\it max} \sim |r|^{-1/\sigma}$ which scales to
infinity as $r$ is taken to zero.

%XXXX figure fig:MFT-simul-m(h) XXXXX

%XXXX figure fig:MFT-D(s) XXXXXXXXXXX

In appendix \ref{ap:mean-field} we derive
a scaling form for the avalanche size
distribution for systems near the critical point:
Let $D(S,r,h)$ denote the probability to find an avalanche of size
$S$ in a system with disorder $r$ at magnetic field $h$ upon an
infinitesimal increase of the external magnetic field.
For large $S$ one finds
\be
\label{D-S-r-h}
D(S,r,h) \sim 1/S^{\tau} {\cal D}_\pm(S r^{1/\sigma}, h/r^{\beta
\delta}) \, .
\ee
The scaling form for $D(S,r)$ of the histograms in
figure \ref{fig:MFT-D(s)}
is obtained by integrating $D(S,r,h)$ over the external magnetic field.

In mean field theory we find $\sigma = 1/2$ and $\tau=3/2$.
In $6-\epsilon$ dimensions we obtain
from the RG calculation
$\sigma = 1/2 - \epsilon/12 + O(\epsilon^2)$
and $\tau = 3/2 + O(\epsilon^2)$.
Numerical simulations in 3 dimensions \cite{Perkovic}
render $\sigma= 0.238\pm 0.017$
and $\tau = 1.60\pm 0.08$ \cite{Perkovic}.

\subsection{Results on correlations near $(R_c,H_c(R_c))$}

With the mean field approximation we have lost all information about
length scales in the system. The RG description, which involves a
coarse graining transformation to longer and longer length scales,
provides a natural means to extract scaling forms for various
correlation functions and the correlation length of the system.

\subsubsection{Avalanche correlations}
\label{subsub:results-avalanche-correlations}

The avalanche correlation
function $G(x,r,h)$ measures the probability for the configuration
of random fields in the system to be such that a flipping spin will
trigger another at relative distance $x$ through an avalanche of
spin flips.
Close to the critical point and for
large $x$ the function $G(x,r,h)$ scales as
\be
\label{scaling-avalanche-correlations}
G(x,r,h) \sim 1/x^{d-2+\eta} {\cal G}_{\pm}(x /\xi(r,h)) \, ,
\ee
where $\eta$ is called ``anomalous dimension'' and
${\cal G}_{\pm}$ is a universal scaling function.
The correlation length $\xi(r,h)$ is the important (macroscopic)
length scale of the system. At the critical point, where it diverges,
the correlation function $G(x,0,0)$ decays algebraically --- there will be
avalanches on all length scales. Close to the critical point the
correlation length scales as
\be
\label{xi-scaling}
\xi(r,h) \sim r^{-\nu} {\cal Y}_\pm(h/r^{\beta \delta}) \, ,
\ee
where ${\cal Y}_\pm$ is the corresponding scaling function.
{}From the $\epsilon$-expansion one obtains \cite{Kleinert}
\be
\label{nu-result}
1/\nu = 2-\epsilon/3 - 0.1173 \epsilon^2 + 0.1245 \epsilon^3 -0.307
\epsilon^4 + 0.951\epsilon^5 + O(\epsilon^6)\, ,
\ee
and
\be
\label{eta-result}
\eta= 0.0185185 \epsilon^2 + 0.01869 \epsilon^3 - 0.00832876
\epsilon^4 +  0.02566 \epsilon^5 + O(\epsilon^6)\, .
\ee
The numerical values in 3 dimensions are
$1/\nu = 0.704\pm 0.085$ and $\eta = 0.79\pm 0.29$.

\subsubsection{Spin-spin (``cluster'') correlations }

There is another correlation function which measures correlations in
the fluctuations of the spin orientation at different sites.
It is related to the probability that two spins $s_i$ and $s_j$
at two different sites $i$ and $j$, that are distanced by $x$,
have the same value \cite{Goldenfeld}.
It is defined as
\be
\label{spin-spin-cor-def}
C(x,r,h) = \langle (s_i-\langle s_i\rangle_f) (s_j-\langle
s_j\rangle_f) \rangle_f \, ,
\ee
where $\langle \rangle_f$ indicates the average over the random
fields.
{}From the RG description we find that
for large $x$ it has the scaling form
\be
C(x,r,h) \sim x^{-(d-4+\bar\eta)} {\cal C}_\pm(x/\xi(r,h)) \, ,
\ee
where
$\xi(r,h)$ scales as given in \eq{xi-scaling} and
${\cal C}_\pm$ is a universal scaling function. At the critical
point $C(x,0,0)$ decays algebraically --- there will be clusters
of equally oriented spins on all length scales. The
$\epsilon$-expansion renders \cite{Kleinert}
\be
\label{etabar-result}
\bar\eta= 0.0185185 \epsilon^2 + 0.01869 \epsilon^3 - 0.00832876
\epsilon^4 +  0.02566 \epsilon^5 + O(\epsilon^6)\, .
\ee
which is in fact the same perturbation expansion as for $\eta$
to all orders in $\epsilon$.
(The two exponents do not have to be equal beyond perturbation
theory, see also \cite{thesis,eps-big2}).

\subsection{Results on avalanche durations}

Avalanches take a certain amount of time to spread, because the
spins are flipping sequentially. The further the
avalanche spreads, the longer it takes till its completion.
The RG treatment suggests that there is a scaling relation between
the duration $T$ of an avalanche and its linear extent $l$
\be
T(l) \sim l^z
\ee
with $z=2 +2 \eta$ to $O(\epsilon^3)$ \cite{Krey}, {\it i.e.}
\be
\label{z-result}
z= 2 + 0.037037 \epsilon^2 + 0.03738 \epsilon^3 +
O(\epsilon^4) \, .
\ee
Our numerical result in 3 dimensions is $z= 1.7 \pm
0.3$.\footnote{While
we expect the $6-\epsilon$ results for the static exponents
$\beta$, $\delta$, $\nu$, $\eta$, $\bar\eta$, $\tau$, $\sigma$, etc.
to agree with our hard-spin simulation results close to 6 dimensions,
this is not necessarily so for the dynamical exponent $z$. There
are precedences for the dynamics being sensitive to the exact shape
of the potential, sometimes only in mean-field theory
\protect{\cite{FisherCDW,NarayanI,NarayanII}}
(charge density waves
in a smooth-potential versus a linear cusp potential),
and sometimes even in the $\epsilon$-expansion
\protect{\cite{NarayanII,ParisiCDW}} (charge density waves
in a sawtooth potential).}

The fractal dimension for the biggest avalanches
$S_{\it max} \sim r^{-1/\sigma} \sim \xi^{1/(\sigma\nu)}$ is
\be
d_{\it fractal} = 1/(\sigma\nu) \, ,
\ee
so that the time for the biggest finite avalanches scales as
$T(S_{\it max}) \sim S_{\it max}^{\sigma\nu z}$.

\subsection{Results on the area of the hysteresis loop}
\label{sec:area-of-hysteresis}
In some analogy to the free energy density in equilibrium
systems, one can extract
the scaling of the area of the hysteresis
loop for this system near the critical endpoint. (This is the energy
dissipated in the loop per unit volume.)
{}From the fact that the singular part of the
magnetization curve scales as $m(h,r) \sim r^\beta {\cal
M}_\pm(h/r^{\beta\delta})$ (see \eq{magn-scaling})
we conjecture that the
singular part of the area would scale as
$A_{sing} \sim \int m(h,r) dh \sim r^{2-\alpha}$ with
$2-\alpha=\beta+\beta\delta$.
(The scaling form for the total area $A_{tot}$
will also have an analytical piece:
$A_{tot}= c_0 + c_1 r^n + c_2 r^{(n+1)} + ... + A_{sing}$;
for any data analysis one needs to keep all terms with
$n \leq 2-\alpha$.)
In mean field theory $\alpha=0$.
Numerical and analytical results can be derived from the results
for $\beta$ and $\beta \delta$ quoted earlier.

\subsection{Results on the number of system-spanning avalanches at the critical
disorder $R=R_c$}

In percolation in any dimensions less than 6,
there is at most {\it
one} infinite cluster present at any value of the concentration
parameter $p$, in particular also at its critical value $p_c$
\cite{Stauffer}.
In contrast, in our system
at the critical point $R=R_c$ the number $N_{\infty}$ of ``infinite
avalanches'' found during one sweep through the hysteresis
loop,
diverges with system size as $N_{\infty} \sim L^\theta$ in all
dimensions $d >2$.

The $\epsilon$-expansion for our system yields
\be
\theta \nu = 1/2 - \epsilon/6 + O(\epsilon^2)
\ee
Numerical simulations \cite{Perkovic} show clearly that
$\theta >0 $ in 4 and 5
dimensions. In three dimensions one finds $\theta \nu = 0.021\pm
0.021$ and $\theta = 0.015 \pm 0.015$.
(For more details on $\theta$ see \cite{thesis,eps-big2}.)

\subsection{List of exponent relations}
\label{sec:exponent-equalities}
In the following sections we list various exponent relations, for
which we give detailled arguments in references
\cite{thesis,eps-big2}.

\subsubsection{Exponent equalities}

The exponents introduced above are
related by the following exponent equalities:

\be
\label{exp-eq1}
\beta -\beta\delta = (\tau-2)/\sigma \qquad {\it if} \quad \tau <
2 \, ,
\ee

\be
\label{fluct-diss}
(2-\eta) \nu = \beta \delta -\beta \, ,
\ee

\be
\beta = {\textstyle {\nu \over 2}}(d-4+\bar\eta)\,,
\ee

and
\be
\delta = (d-2 \eta + \bar\eta)/(d-4+\bar\eta)\,.
\ee
(The latter three equations are not independent and are
also valid in the equilibrium random field Ising model
\cite{Nattermann-review,Rieger-Young,BWR-Newman}).\footnote{Also,
using these relations one finds that
the inequality $\nu/\beta\delta \geq 2/d$
(which applies in our model \protect{\cite{eps-big2,thesis,Chayes}})
goes over into
the Schwartz-Soffer inequality $\bar\eta \leq 2 \eta$ that has been
derived for the corresponding equilibrium model
\protect{\cite{SchwartzSoffer}}.}

\subsubsection{Incorrect exponent equalities}
\paragraph{Breakdown of hyperscaling}
\label{Breakdown-of-hyperscaling}
In our system there are two different violations of hyperscaling.

1. In references \cite{thesis,eps-big2}, we show that
the connectivity hyperscaling relation
$1/\sigma = d \nu - \beta$ from percolation is violated in our system.
There is a new exponent $\theta$ defined by $1/\sigma=(d-\theta)\nu
- \beta$ with
$\theta \nu = 1/2 -\epsilon/6 + O(\epsilon^2)$  and $\theta \nu =
0.021 \pm 0.021$ in three dimensions \cite{Perkovic}.
$\theta$ is related to the number of system spanning avalanches
observed during a sweep through the hysteresis loop.

2. As we will discuss in
section \ref{sec:mapping}
there is a mapping of the perturbation theory for our problem
to that of the equilibrium random field Ising model to all orders in
$\epsilon$. From that mapping we deduce the breakdown of an
infamous (``energy'')-hyperscaling relation, which has caused much
controversy in the case of the {\it equilibrium} random-field Ising model
\cite{Nattermann-review}
\be
\beta+\beta\delta = (d-\tilde\theta) \nu \, ,
\ee
with a new exponent $\tilde\theta$.
In \cite{thesis,eps-big2} we discuss
the relation of the exponent $\tilde\theta$ to the energy output of the
avalanches. The $\epsilon$-expansion yields $\tilde \theta =2$ to
all orders in $\epsilon$.
Non-perturbative corrections are expected to lead to
deviations of $\tilde\theta$ from 2 as the dimension is lowered.
The same is true as is the case in the equilibrium RFIM
\cite{thesis,eps-big2}. The numerical result in three
dimensions is $\tilde \theta = 1.5 \pm 0.5$ \cite{Perkovic}. (In the
three-dimensional Ising model it is ${\tilde \theta}_{\it eq} = 1.5
\pm 0.4$ \cite{Nattermann-review,BWR-thesis}.)

\paragraph{Breakdown of perturbative exponent equalities}

There is another
strictly perturbative exponent equality, which is also
obtained from the perturbative mapping to the random-field
Ising-model \cite{thesis,eps-big2},
\be
\bar\eta = \eta \, .
\ee
It, too, is expected to be violated by non-perturbative corrections
below $6$ dimensions.

\subsubsection{Exponent inequalities}
In references \cite{thesis,eps-big2} we
give arguments for
the following two exponent-inequalities\footnote{From the
normalization of the avalanche size distribution $D(s,r,h)$
(see \protect{\eq{D-S-r-h}}) follows that $\tau >1$.}:
% 	and that
%	XXX $\tau+\sigma\beta\delta >2$ by integration over the
%	hysteresis loop}
\be
\label{ineq-strong}
\nu/\beta\delta \geq 2/d \, ,
\ee
which is formally equivalent to the ``Schwartz-Soffer''
inequality, $\bar\eta \leq 2 \eta$, first derived
for the equilibrium random field Ising model \cite{SchwartzSoffer},
and
\be
\label{ineq-weak}
\nu \geq 2/d \, ,
\ee
which is a weaker bound than \eq{ineq-strong} so long as
$\beta\delta \geq 1$, as
appears to be the case both theoretically and numerically at least
for $d \geq 3$.

\subsection{Results on the upper critical dimension of the critical
endpoint $(R_c,H_c(R_c))$}

The consistency of the mean-field theory exponents for $d\geq 6$
can be shown by a Harris criterion type of argument \cite{Goldenfeld},
which also leads to \eq{ineq-strong} \cite{eps-big,thesis}.
Approaching the critical point along the $r'=0$ line, one finds
a well defined transition point only if the fluctuations $\delta h'$ in the
critical field $H_c$ due to fluctuations in the random fields are
always small compared to to the distance $h'$ from the critical
point, i.e.
$\delta h'/h' << 1$ as $ h' \rightarrow 0$.
With $\delta h' \sim \xi^{-d/2}$ and $\xi\sim (r')^{-\nu}
f_\pm(h'/(r')^{\beta\delta}) \sim (h')^{-\nu/(\beta \delta)}$
at $r'=0$,
one obtains $\delta h'/h' \sim \xi^{-d/2}/\xi^{-{\beta\delta/\nu}} << 1$, or
$\nu/{\beta\delta} \geq 2/d$.
For $\nu=1/2$ and $\beta\delta=3/2$ this is only fulfilled if $d\geq
6$, {\it i.e.} $d=6$ is the upper critical dimension.

\section{Why an $\epsilon$-expansion?}
\label{sec:motives}

Perturbation theory has proven an invaluable tool for practical
calculations in many branches of physics. The asymptotic expansion
in powers of the electronic charge in electrodynamics is an example
where perturbation theory not only gives extremely accurate results
but also provides a qualitative insight in the underlying physical
processes such as absorption and emission of photons.
Critical phenomena in general are among the cases where perturbation
theory cannot be applied (at least not below a certain ``upper critical
dimension'' $d_c$). The fluctuations in the order parameter near the
critical point become too large to extract useful information from a
perturbation expansion in some physical coupling constant.

The RG has been developed specifically for such cases.
It is a means to extend the derivation of critical scaling forms
from mean-field theory to finite dimensions, and to obtain
information about the effective behavior of the system on long
length scales. The Wilson-Fisher momentum shell renormalization group
is an iterative coarse graining transformation. In each step the
shortest wavelength degrees of freedom are integrated out, leading
to an effective action for the modes on longer length scales. If a
theory is renormalizable, the coarse grained action will be in the
same form as the original action, but with rescaled parameters.

Under coarse graining many of these parameters will flow
to a certain fixed point, which is independent of their original
microscopic value. The properties of the system on long length scales
only depend on the fixed point.
This is the key to universality: different microscopic systems
flow to the same fixed point and are therefore
described by the same effective action on long length scales.
They will consequently show the same critical behavior.

Second order critical points are fixed points of the coarse graining
transformation. They are characterized by a diverging correlation
length and consequent scale invariance. There are fluctuations (such as
clusters, avalanches, etc.) on all length scales. In the same sense
power laws are scale free functions.\footnote{When a function
$f(x)\sim x^{|\alpha|}$ is measured over three pairs of octaves, say
over the intervals $[1,4]$, $[10,40]$, and $[100,400]$, the ratio of the
largest to the smallest value is always $4^{|\alpha|}$, so the three
graphs of $f(x)$ can be superimposed by simple change of scale.
In this sense, power laws are scale invariant \cite{Binney}.}
It is therefore not surprising that systems near their critical
point are described by power laws. One of the triumphs of the RG is
to show from first principles that near the critical point the
interesting long-wavelength properties are given by {\it homogeneous
functions}\footnote{A function $f(x_1,\cdot \cdot \cdot, x_n)$ is
	homogeneous of degree $D$ in the variables $x_1,\cdot \cdot \cdot,
	x_n$, if on multiplying each $x_i$ by an arbitrary factor $b$ the
	value of $f$ is multiplied by $b^D$, {\it i.e.}
	$f(b x_1,\cdot \cdot \cdot, b x_n) = b^D f(x_1,\cdot \cdot \cdot,
	x_n)$ \cite{Binney}.Here we use the definition
	$f(b^{\lambda_1} x_1, \cdot \cdot \cdot, b^{\lambda_n} x_n) =
	b^y f(x_1, \cdot \cdot \cdot, x_n)$ for
	$\lambda_1, \cdot \cdot \cdot, \lambda_n$ real constants.}
with respect to a change of length scale in the
system. This observation leads to Widom scaling forms for the
various macroscopic quantities, some of which we already obtained
from simple expansions in mean field theory.

The RG thus provides a formal justification of the scaling ansatz
used in the data analysis and an explanation for the
universality of the critical exponents. It is the ultimate
justification for the attempt to extract useful predictions about
real complex materials from extremely simple
caricatures of the microscopic physics.

In particular
it has been used to derive an expansion for the critical exponents
around their mean-field values in powers
of the dimensional parameter $\epsilon=d_c-d$, where $d$ is the
dimension of the system. Note that the $\epsilon$-expansion is an
asymptotic expansion in terms of a quite unphysical parameter.
Nevertheless it has proven very successful for mathematical
extrapolations.
In this paper we shall apply its basic ideas to our problem
and refer the reader for further details to excellent reviews in the
existing literature
%% FOLLOWING LINE CANNOT BE BROKEN BEFORE 80 CHAR
\cite{Ma,Fisher-SA-notes,Goldenfeld,Binney,Zinn-Justin,Amit,Halperin-review,real-space}.

The calculation turns out to be interesting in its own right.
In contrast to RG treatments of equilibrium critical phenomena,
a calculation for our hysteresis problem has to take into account
the entire history of the system.
It reveals formal similarities to related {\it single interface}
depinning transitions
\cite{NarayanI,NarayanII,NarayanIII,Nattermann,Ertas,Ertas2}.
Although our problem deals with the seemingly more complex case
of {\it many} advancing interfaces or domain walls, the calculation
turns out
to be rather simple, much simpler in fact than in the single interface
depinning problem. More details are given in appendix
\ref{ap:related}.
The techniques employed here are likely to be applicable to other
nonequilibrium systems as well.

\section{Analytical Description}
\label{sec:analytic-description}

In equilibrium systems one defines a partition function as the sum
over the thermal weights or probabilities of all possible states.
{}From this partition function all ensemble averaged correlation
functions can be obtained. It is also usually the quantity used to
calculate the critical exponents in an RG treatment.
Our system is at zero temperature and far from equilibrium.
For a given configuration of random fields, the system will follow
a deterministic path through the space of spin microstates as the
external magnetic field is raised adiabatically.
Systems with different configurations of random fields will
follow different paths.
%The magnetization curves corresponding to
%the paths in two different configurations of random fields are
%shown in figure \ref{fig:M(H)-for-diff-paths}.
If we assign a $\delta$-function weight
to the correct path for each configuration and then average over
the distribution of random fields, we obtain a probability
distribution for the possible {\it paths} of the system, which is the
analogue of the probability distributions for the possible states
in equilibrium systems. The analogue of ensemble averaging for
equilibrium systems, is random field averaging in our system.
The sum over all possible paths weighted by their corresponding
probability will play the role of a partition function for our
nonequilibrium system when we set up a Wilson-Fisher momentum shell
renormalization group transformation to calculate the
critical exponents.

%XXXXXXXX figure fig:M(H)-for-diff-paths (removed) XXXXX

How can one formally describe the path that a system takes for a
specific configuration of random fields as the external magnetic
field is increased from $-\infty$ to $+\infty$?
A convenient way is to introduce a time $t$ into the otherwise
adiabatic problem via $H(t) = H_0 + \Omega t$. $H_0$ is the magnetic
field at time $t=0$. $\Omega >0$ is the sweeping rate for a
monotonically increasing external magnetic field.
The idea is to write down an equation of motion for each spin
such that the resulting set of coupled differential equations has a
unique solution which corresponds to the correct path the system
takes for a given history. By taking $\Omega$ to zero in the end one
obtains the adiabatic or ``static'' limit, in which we are
interested. For convenience we introduce soft spins that can
take values ranging from $-\infty$ to $+\infty$.
Later on this will allow us to replace traces over all possible
spin configurations by path integrals over the range of definition
of the spins. We assume that each spin is moving
in a
double well potential $V(s_i)$ with minima at the ``discrete'' spin
values $s_i= \pm 1$:
\vspace{.3cm}
\be
\label{V-def}
V(s_i) = \left\{ \begin{array}{ll}
                            k/2~(s_i+1)^2 & \mbox{for $s<0$} \\
                            k/2~(s_i-1)^2 & \mbox{for $s>0$}
                            \end{array}
                   \right.
\ee
To guarantee that the system takes a finite magnetization at any
magnetic field, one needs $k>0$ and $k/J>1$.
The Hamiltonian of the soft spin model is then given by:
\begin{equation}
\label{model}
{\cal H}=-\sum_{ij}J_{ij}s_i s_j-\sum_i ( f_i s_i+H s_i-V(s_i)) \, .
\end{equation}
All terms in the Hamiltonian are as before, except for the additional
$V(s_i)$ term.
A spin flip in this model corresponds to a spin moving from the
lower to the upper potential well.
\footnote{We believe
the calculation could just as well have been performed for discrete
spins, maybe by using the ideas of Lubensky {\it et
al.} \cite{Lubensky}. At the time it seemed easier to use the
established formalism for continuous spins. However, in mean-field
theory the formulas
look very similar to the hard-spin model.
The new parameter $k$ from the double well
potential never seems to come into play for any of the universal
properties of the system.}

We impose purely relaxational
dynamics, given by
\be
\label{dynamics}
(1/\Gamma_0) \partial_t s_i(t) = - \delta {\cal H}/\delta s_i(t) \, .
\ee
$\Gamma_0$ is a ``friction constant''.

This model shows qualitatively similar behavior to real magnets: As
the external magnetic field is ramped, we observe spin flips, which
correspond to irreversible domain wall motions. The linear relaxation
between the spin flips corresponds to the reversible domain wall
motion, which we have described in the introduction.

The soft spin mean-field theory, where every spin interacts equally
with every other spin
yields the same static critical exponents
as we have obtained earlier for the
hard spin model. We have also checked that replacing
the linear cusp potential by
the more common, smooth $s^4$  double well potential
does not change
the static mean field exponents.\footnote{The form of the potential
does change the (dynamical) mean-field exponents in the CDW
depinning transition at zero temperature \cite{NarayanI}.
In finite dimensions it changes some properties that are associated
with the thermal rounding of the CDW transition
\cite{Middleton-thermal}.}

\subsection{Formalism}
\label{sub:formalism}

We use the formalism introduced by Martin Siggia and Rose \cite{MSR}
to treat
dynamical critical phenomena, which is similar to the
Bausch--Janssen--Wagner
method \cite{BJW}.
One defines the generating functional $Z$ for the dynamical
problem as an integral over a product of $\delta$--functions
(one for each spin), each of which imposes the equation of motion
at all times on its particular spin \cite{NarayanI}:
\be
\label{Z}
1 \equiv Z= \int [ds] {\cal J}[s] \prod_{i} \delta
(\partial_t s_i/\Gamma_0 + \delta {\cal H}/\delta s_i) \, .
\ee
$[ds]$ symbolizes the path integral over all spins in the
lattice at all times, and ${\cal J}[s]$ is the necessary Jacobian,
which
fixes the measure of the integrations over the $s_i$ such that the
integral over each delta--function yields $1$ \cite{NarayanI}.
One can show that ${\cal J}[s]$ merely cancels the equal time response
functions \cite{dominicis,NarayanI}.\footnote{To that end, one chooses the
following regularization (when discretizing in time):
Let $t=n \epsilon$ with $\epsilon$ a small number taken to zero
later when $n$ is taken to infinity, such that their product remains
fixed.
Then $\partial_t s_i(t)$ becomes $( s_i(n \epsilon) - s_i((n-1)
\epsilon) )/ \epsilon $ and we integrate over $ds_i(n \epsilon)$.
Because of the analyticity we are free to take the rest of the
argument of the $\delta$--function at the lower value
$s_i((n-1)\epsilon)$, or the average value
$( s_i(n \epsilon) + s_i((n-1)\epsilon) )/ 2 $, or
the upper value $s_i(n \epsilon)$.
If we choose the first possibility, one finds that the Jacobian
will be only a constant, since each argument only depends on
$s_i(n \epsilon)$, and is independent of any other
$s_j(n \epsilon)$ with $i \neq
j$. This corresponds to allowing a force at time $(n-1)\epsilon$
to have an effect only {\it after} some time $\epsilon$,
i.e. equal times response functions are manifestly zero.}

In order to write $Z$ in an exponential form in analogy to the
partition function in equilibrium problems,
we express the $\delta$-functions in their Fourier-representation,
introducing an unphysical auxiliary field $\hat s_j(t)$:
\be
\label{MSR-trick}
\delta( \partial_t/\Gamma_0
s_i(t) + \delta {\cal H}/\delta s_i(t)) \sim
1/2 \pi \int d{\hat s} \exp ( i
\sum_j \hat{s}_j(t)(\partial_t
s_j(t)/\Gamma_0 +\delta {\cal H}/\delta s_j(t)) \, .
\ee
Absorbing any constants into ${\cal J}[s]$, this yields for the
(not yet random--field averaged) generating
functional (in continuous time):
\be
\label{Z-before-averaging}
1 \equiv Z = \int \int [ds] [d{\hat s}] {\cal J}[s] \exp ( i \sum_j
\int dt
\hat s_j(t)(\partial_t s_j(t)/\Gamma_0+\delta {\cal H}/\delta
s_j(t))) \, ,
\ee
or
\be
1=Z= \int \int [ds] [d{\hat s}] {\cal J}[s] \exp(W) \, ,
\ee
with the action
\bea
\label{action}
W &=& i \sum_j
\int dt
\hat s_j(t)(\partial_t s_j(t)/\Gamma_0+\delta {\cal H}/\delta
s_j(t)) \\
\nonumber
&=& i \sum_j
\int dt
\hat s_j(t)(\partial_t s_j(t)/\Gamma_0- \sum_l J_{jl} s_l -H -f_j +
\delta V/\delta s_j) \, .
\eea

We can express correlation and response functions of $s_j(t)$
as path integrals in
terms of $W$, because solely the unique deterministic path
of the system for the given configuration of random fields makes a
nonzero contribution to the path integral over $[ds]$ in
\eq{Z-before-averaging}.
For example the value of spin $s_j$ at time $t'$ is given by
\be
\label{*1}
s_j(t') = Z^{-1} \int \int [ds'] [d{\hat s}'] {\cal J}[s'] s'_j(t')\exp(W)
\, .
\ee
Similarly, correlation functions are given by
\be
\label{*2}
s_j(t') s_k(t'') = Z^{-1} \int \int [ds'] [d{\hat s}'] {\cal J}[s'] s'_j(t')
s'_k(t'') \exp(W)
\, .
\ee
To calculate the response of $s_j$ at time $t'$ to a perturbative field
$J \epsilon_k(t',t'')$ switched on at site $k$ at time $t''$, we
add the perturbation to the magnetic field at site $k$, such that
the action becomes
\bea
\label{action-with-perturbation}
W_{\epsilon} &=& i \sum_{j \neq k}
\int dt
\hat s_j(t)(\partial_t s_j(t)/\Gamma_0- \sum_l J_{jl} s_l -H -f_j +
\delta V/\delta s_j) \\
\nonumber
 & & + i \int dt
\hat s_k(t)(\partial_t s_k(t)/\Gamma_0- \sum_l J_{kl} s_l -H -f_k +
\delta V/\delta s_k -J \epsilon_k(t,t''))
 \, .
\eea
Taking the derivative with respect to $J \epsilon_k$
and the limit $\epsilon_k \rightarrow 0$ afterwards\footnote{
An exact definition of the functional derivative which is consistent
with the history of a monotonically increasing magnetic field is
given in appendix \ref{ap:RG-details}.}
one obtains
\be
\label{*3}
\delta s_j(t')/\delta \epsilon_k(t'')
 = (-i) Z^{-1} \int \int [ds] [d{\hat s}] {\cal J}[s] s_j(t')
{\hat s}_k(t'') \exp(W)
\, ,
\ee
so $\hat s$ acts as a ``response field''.
Henceforth we shall suppress ${\cal J}[s]$, keeping in mind that its
only effect is to cancel equal time response functions (see previous
footnote).

Since $Z=1$ independent of the random fields, we could have
left out the $Z^{-1}$ factors in \eqs{*1}, (\ref{*2}), and (\ref{*3}).
This greatly facilitates averaging over the random fields:
The average response and correlation functions are generated by
averaging $Z$ directly over the random fields. Unlike in equilibrium
problems with quenched randomness it is not necessary to calculate
the (more complicated) average of $\ln Z$.
One obtains for the random field averaged correlation functions
\be
\langle s_j(t') s_k(t'') \rangle_f = \int \int [ds] [d{\hat s}]
s_j(t')
s_k(t'') \langle \exp(W) \rangle_f \, ,
\ee
and similarly
\be
\delta s_j(t')/\delta \epsilon_k(t'') =
\langle s_j(t') {\hat s}_k(t'') \rangle_f = \int \int [ds] [d{\hat s}]
s_j(t')
{\hat s}_k(t'') \langle \exp(W) \rangle_f \, .
\ee

It is not obvious how to calculate $\langle \exp(W) \rangle_f$
directly, since $W$ involves terms like
$J_{ij} s_i {\hat s}_j$ which couple different sites.
Following Sompolinsky and Zippelius \cite{Zippelius}, and
Narayan and Fisher \cite{NarayanI}, we can circumvent this problem by
performing a change of variables from the spins $s_j$ to local
fields $J {\tilde \eta}_i= \sum_j J_{ij} s_j$.
(We introduce the coefficient $J$ on the left hand side to keep the
dimensions right.)
At the saddle point of the
associated action the new variables ${\tilde \eta}_j$ (for all j) are given
by the mean-field magnetization and the different sites become
decoupled. A saddle point expansion becomes possible, because the
coefficients in the expansion can be calculated in mean-field
theory --- they are also the same for all sites $j$.

Here is how it works:
we insert into $Z$ the expression
\be
\label{transformation}
1=1/2 \pi \int \int [d{\hat {\tilde \eta}}] [d{\tilde \eta}] {\cal
J}[{\tilde \eta}] \exp (i
\sum_j \int dt
{\hat {\tilde \eta}}_j(t) (s_j(t) - \sum_i J_{ij}^{-1} J {\tilde \eta}_i(t)))
\ee
where ${\cal J}[{\tilde \eta}]$ stands for the suitable Jacobian,
which
is simply a constant and will be suppressed henceforth.
Integrating out the auxiliary fields ${\hat {\tilde \eta}}_j$,
one recovers that
the expression in \eq{transformation} is the integral over a
product of $\delta$-functions which impose the definitions
$J {\tilde \eta}_i(t)= \sum_j J_{ij} s_j(t)$ at all times for all $i$.

After some reshuffling of terms and introducing some redefinitions
that are motivated by the attempt to separate the nonlocal from the
local terms, one obtains
\be
\label{Z-as-a-function-of-eta}
 Z= \int \int [d{\hat {\tilde \eta}}] [d{\tilde \eta}]
\prod_j {\bar Z}_j[{\tilde \eta}_j,
{\hat {\tilde \eta}}_j] \exp [-\int dt {\hat {\tilde \eta}}_j(t)(\sum_l
J_{jl}^{-1}
J {\tilde \eta}_l(t))]
\ee
where ${\bar Z}_j[{\tilde \eta}_j,{\hat {\tilde \eta}}_j]$ is a {\it local}
functional
\bea
\label{Z-bar-j}
\lefteqn{{\bar Z}_j[{\tilde \eta}_j, {\hat {\tilde \eta}}_j] =
 \int \int [ds_j] [d{\hat s}_j]
\langle \exp \{J^{-1} \int dt [ J {\hat {\tilde \eta}}_j(t)
s_j(t) + } \\
\nonumber
& &
i {\hat s}_j(t) (\partial_t s_j(t)/\Gamma_0 -J {\tilde \eta}_j(t) -
H -f_j +
\delta V/\delta s_j )]
\}  \rangle_f \,
\eea
(we have absorbed a factor $i$ in the definition of ${\hat
{\tilde \eta}}_j$).
In short this can also be written as
\be
\label{Z-and-S-eff}
Z \equiv \int [d{\tilde \eta}] [d{\hat {\tilde \eta}}]
\exp(\tilde S_{\it eff} )
\ee
with the effective action $\tilde S_{\it eff}$, now expressed in terms of the
``local field'' variables ${\tilde \eta}$ and ${\hat {\tilde \eta}}$

\begin{eqnarray}
\label{S-bar-eff}
{\tilde S}_{\it eff} &=& - \int dt \sum_j {\hat {\tilde \eta}}_j(t) \sum_l
J_{jl}^{-1} J {\tilde \eta}_l(t) + \sum_j \ln({\bar Z}_j[{\tilde \eta}_j,
{\hat
{\tilde \eta}}_j]) \, .
\end{eqnarray}
Physically we can interpret the functional
\be
\Phi[\eta]\equiv \int [d{\hat {\tilde \eta}}]
\exp(\tilde S_{\it eff} )
\ee
as the random field averaged probability distribution for the
possible paths the system can take through the spin configuration
space as the external magnetic field is slowly increased.
Each path is specified by a set of $N$ effective field functions
$J \eta_i(t)$ ($i$ runs over the lattice with $N \rightarrow
\infty$ spins). $Z$ is the integral of this (normalized) probability
distribution over all possible paths of the system and is therefore
equal to $1$.

The stationary point $[{\tilde \eta}_j^0, {\hat {\tilde \eta}}_j^0]$
of the effective action
is given by
\be
\label{stationary-point1}
[\delta {\tilde S}_{\it eff} /\delta {\tilde \eta}_j ]_{
{\tilde \eta}_j^0,{\hat {\tilde \eta}}_j^0} =
0 \, ,
\ee
and
\be
\label{stationary-point2}
[\delta {\tilde S}_{\it eff} /
\delta {\hat {\tilde \eta}}_j ]_{{\tilde \eta}_j^0,{\hat
{\tilde \eta}}_j^0} = 0 \, .
\ee
With \eqs{Z-bar-j} and (\ref{S-bar-eff})
we find the
saddle--point equations:
\be
\label{saddle-point1}
(-i) \langle {\hat s}_i \rangle_{l,{\hat {\tilde \eta}}^0,{\tilde \eta}^0}
- \sum_j J J_{ij}^{-1} {\hat {\tilde \eta}}_j^0 = 0 \, ,
\ee
and
\be
\label{saddle-point2}
\langle s_i \rangle_{l,{\hat {\tilde \eta}}^0,{\tilde \eta}^0} -
\sum_j J J_{ij}^{-1}
{\tilde \eta}_j^0 = 0 \, .
\ee
The notation $\langle \rangle_{l,{\hat {\tilde \eta}}^0,{\tilde \eta}^0}$
here denotes
a {\it local} average, obtained from the {\it local} partition
function ${\bar Z}_i$, after having fixed ${\hat {\tilde \eta}}_i$
and ${\tilde \eta}_i$
to their stationary--point solutions ${\hat {\tilde \eta}}_i^0$ and
${\tilde \eta}_i^0$.

For example (from \eqs{Z-bar-j}, (\ref{S-bar-eff}),
(\ref{stationary-point2}), and
(\ref{saddle-point2}))
\bea
\label{s-average-local-magn}
\nonumber
\lefteqn{\langle s_i(t) \rangle_{l,{\hat {\tilde \eta}}^0,{\tilde
\eta}^0} =
[{{\delta}\over \delta {\hat {\tilde \eta}}_i} \sum_j
\ln ({\bar
Z}_j[{\tilde \eta}_j,{\hat{\tilde \eta}}_j]) ]_{{\hat {\tilde
\eta}}^0,{\tilde \eta}^0}
= {1 \over {\bar Z}_i} \int \int [ds] [d{\hat s}] s_i(t)} \\
\nonumber
&&\langle \exp \{ J^{-1} \int dt [ \sum_j J {\hat {\tilde \eta}}^0_j(t)
s_j(t) \\
&& +i {\hat s}_j(t) (\partial_t s_j(t)/\Gamma_0
-J {\tilde \eta}^0_j(t) - H -f_j +
\delta V/\delta s_j )] \} \rangle_f \, .
\eea

\Eq{saddle-point1} and \eq{saddle-point2} have the self-consistent
solution
\be
\label{self-consistent1}
\hat{\tilde \eta}^0_i(t) = 0 \, ,
\ee
and
\be
\label{self-consistent2}
{\tilde \eta}^0_i(t) = M(t) = \langle s_i(t) \rangle_{l,
{\hat {\tilde \eta}}^0,{\tilde \eta}^0}
\, ,
\ee
where $M(t)$ is the random field average of the solution of the mean
field equation of motion
\be
\label{mean-field-equ-of-motion}
\partial_t s_j(t)/\Gamma_0 = J {\tilde \eta}^0_j(t) + H  + f_j -
\delta V/\delta s_j \, .
\ee
(This can be seen by setting $\hat {\tilde \eta}^0=0$ in
\eq{s-average-local-magn}. Integrating out the $\hat s$ fields we
see that ${\bar Z}_j$ is the random field average over a product of
$\delta$-functions which impose \eq{mean-field-equ-of-motion} by their
argument. \Eq{self-consistent2} is the self-consistency condition
for this mean-field equation of motion.)

We can now expand the effective action ${\tilde S}_{\it eff}$ in the
variables ${\hat \eta}_j\equiv ({\hat {\tilde \eta}}_j -
{\hat {\tilde \eta}}_j^0)$ and
$\eta_j \equiv({\tilde \eta}_j-{\tilde \eta}_j^0)$,
which corresponds to an expansion around mean--field--theory:
\be
\label{Z-bar}
Z = \int \int [d\eta] [d{\hat \eta}] \exp( S_{\it eff})
\ee
with an effective action (expressed in the new variables $\eta$ and
$\hat \eta$):
\begin{eqnarray}
\label{effective-action-final}
\nonumber
\lefteqn{S_{\it eff} =
- \sum_{j,l} \int dt J_{jl}^{-1} J\hat{\eta}_j(t) \eta_l(t) +
\sum_j \sum_{m,n=0}^{\infty} \frac{1}{m! n!} \int dt_1\cdot \cdot
\cdot dt_{m+n}} \\
& &u_{mn}(t_1,...,t_{m+n})
\hat{\eta}_j(t_1)\cdot \cdot \cdot \hat{\eta }_j(t_m)
\eta_j(t_{m+1}) \cdot \cdot \cdot \eta_j(t_{m+n})
%\nonumber
\end{eqnarray}
Here, as seen by inspection from \eq{S-bar-eff} and \eq{Z-bar-j}
\bea
\label{u-mn}
u_{m,n} &=&
\frac{\partial}{\partial \eta_j(t_{m+1})} \cdot \cdot
\cdot
\frac{\partial}{\partial \eta_j(t_{m+n})}
\left[\frac{\delta^m [\ln {\bar Z}_j - \hat{\eta}_j(t) \eta^0_j(t)]}
{\delta \hat{\eta}_j(t_1) \cdot \cdot \cdot \delta
\hat{\eta}_j(t_m)}\right]_{\hat{\eta}=0, \eta=0}
\\
\nonumber
 &=& \frac{\partial}{\partial \epsilon(t_{m+1})} \cdot \cdot
\cdot
\frac{\partial}{\partial \epsilon(t_{m+n})} \langle
(s(t_1)-\eta^0(t_1)) \cdot \cdot \cdot
(s(t_m)-\eta^0(t_m)) \rangle_{l,\hat{\eta}^0, \eta^0} \, ,
\eea
{\it i.e.}, the coefficients $u_{mn}$ are equal to the local (l),
connected responses and correlations in mean field theory.

Again, {\it local} ($l$) means \cite{NarayanI} that we do not vary
the local field
$\eta^0_j$ in the mean--field equation
\begin{equation}
\label{response-in-local-mft}
\frac{1}{\Gamma_0} \partial_t s_j(t) = J \eta^0_j(t)+H+f_j-
\frac{\delta V}{\delta s_j(t)}+ J \epsilon (t)
\end{equation}
when we perturb with the infinitesimal force $J \epsilon(t)$.

\subsection{Source terms}
\label{source-terms}
Correlations of $s$ and $\hat s$ can be related to correlations
of $\eta$ and $\hat \eta$ \cite{NarayanI}.
If we introduce the source terms
\be
\int dt (s_j(t) {\hat l}_j(t) - i {\hat s}_j(t) l_j(t))
\ee
into the action, we can write the
correlations of $s$ and $\hat s$ as functional derivatives with
respect to ${\hat l}$ and $l$ at $l=\hat l=0$.
A shift in the variables $\eta$ and $\hat \eta$ by $l$ and $\hat l$
respectively leads to a source term of the kind
\be
J J_{ij}^{-1} ({\hat \eta}_i(t) - {\hat l}_i(t)) ( \eta_j(t) -
l_j(t))
\ee
so that derivatives with respect to $l$ and $\hat l$ give
correlation functions of $\hat \eta$ and $\eta$. For low momentum
behavior the factor $J J_{ij}^{-1}$
can be replaced by one since $\sum_i J_{ij}^{-1} = J^{-1}$.

\subsection{Implementing the history}
\label{sub:Implementing-the-history}

Up to here the effective action $S_{\it eff}$ manifestly involves
the entire magnetic field range $-\infty < H < +\infty$.
As we discuss in appendix \ref{ap:RG-details}
it turns out, however, that in the adiabatic limit a separation of
time scales emerges. The relaxation rate
$k \Gamma_0$ in response to a perturbation is
fast compared to the driving rate $\Omega/k$ of the external
magnetic field.
The static critical exponents
can then be extracted self-consistently
from a RG calculation performed at a single, fixed
value $H$ of the external magnetic field.
The analysis is much simpler than one might have expected.
Instead of dealing with the entire effective action which involves
{\it all} field values $H$,
it suffices in the adiabatic limit
to calculate all coefficients $u_{mn}$ in
\eq{effective-action-final} at one single fixed magnetic field $H$,
and then to coarse grain the resulting action
$S_{\it eff}(H) \equiv S_H$. There are no
corrections from earlier
values of the external magnetic field.

Physically this corresponds to the statement that increasing the
magnetic field within an infinite ranged model (mean field theory)
and then tuning the elastic coupling to a short ranged form (RG)
would be equivalent to the physical relevant critical behavior,
which actually corresponds to {\it first} tuning the elastic
coupling to a short ranged form and {\it then} increasing the force
within a short ranged model \cite{NarayanII}.
In their related calculation for CDWs below the depinning threshold
\cite{NarayanII},
Narayan and Middleton give an argument that this approach is
self-consistent for their problem.
In the appendix \ref{ap:RG-details} we first show that their
argument applies to our system as well, and then discuss the
consistency of the magnetic field decoupling
within the RG treatment of the entire history for separated
time scales.

Where did the history dependence go?
Note that the values of the coefficients $u_{mn}$ at field $H$ are
still history dependent (in the way the mean-field solution is).
Also, causality must be observed by the coarse graining
transformation, so that even in the adiabatic limit the intrinsic
history dependence of the problem does not get lost.

\subsection{Calculating some of the $u_{mn}$ coefficients at field
$H$}
\label{sub:calculating-u-mn}

In appendix \ref{ap:RG-details} we show that
$u_{mn}$
basically assume their static values in the adiabatic limit.
In this section we will briefly outline their derivation
and quote the relevant results.

We have to be consistent with the history of an
increasing external magnetic field,
when expanding around the ``mean--field--path'' $\eta^0(t)$.
This implies that for calculating
responses from \eq{u-mn} we must only allow a perturbing force
$J \epsilon(t)$
that {\it increases} with time in \eq{response-in-local-mft}.
For example, for $u_{1,1}$ we add a force
$J\epsilon(t)\equiv J\epsilon\Theta(t-t')$ in
\eq{response-in-local-mft},
where $\Theta(t-t')$ is the step function,
and solve for
$\langle s(t)|_{H+J \epsilon(t)} \rangle_f$.
The local response function is then
given by the derivative of
$lim_{\epsilon \rightarrow 0}[(\langle (s(t)|_{H+J
\epsilon(t)}\rangle_f-\langle
s(t)|_H\rangle_f)/\epsilon]$ with
respect to $(-t')$.
The higher response functions are calculated correspondingly.
(The most important ones are calculated in appendix
\ref{ap:RG-details}.)
One obtains in the low frequency approximation
for the first few terms
of the effective action of \eq{effective-action-final} at field $H$
\begin{eqnarray}
\label{S-H}
\nonumber
S_H &=& -\sum_{j,l} \int dt J_{jl}^{-1}J \hat\eta_j(t)
\eta_l(t)
-\sum_j \int dt\, \hat{\eta}_j(t)
        [- a \partial_t/\Gamma_0 - u_{11}^{stat}]\, \eta_j(t)\,
\\
\nonumber
& & + \sum_j \int dt\, {\textstyle {1 \over 6}} u \hat{\eta}_j(t)
(\eta_j(t))^3 \\
& & + \sum_j \int dt_1 \int dt_2\, {\textstyle {1 \over 2}} u_{2,0}
\hat{\eta}_j(t_1) \hat{\eta}_j(t_2)  \, ,
\label{effective-action-relevant}
\end{eqnarray}
with
\be
\label{a}
a=(J/k + 4 \rho(-J\eta^0 -H +k))/k \, ,
\ee
\be
\label{u11-stat}
u_{11}^{stat} = 2 J \rho(-J\eta^0 -H +k) + J/k \, ,
\ee
\be
\label{w}
w=-2J^2 \rho'(-J\eta^0 -H +k) \, ,
\ee
\be
\label{u}
u=2J^3\rho''(-J\eta^0-H+k) \, ,
\ee
and
\begin{eqnarray}
\label{u-20-short}
u_{2,0} &=& R^2/k^2 +
4 (\int_{-\infty}^{-H-\eta^0+k} \rho(h) dh )
 - 4 (\int_{-\infty}^{-H-\eta^0+k} \rho(h) dh )^2
\nonumber \\
 & & - 4 (\int_{-\infty}^{-H-\eta^0+k} (h/k) \rho(h) dh) \, .
\label{u-20}
\end{eqnarray}

\Eq{u-20-short} implies that
$u_{2,0} \geq 0$
for any normalized distribution $\rho(f)$.

\section{Perturbative expansion}
\label{sec:perturbative-expansion}

\subsection{The Gaussian theory for $d>d_c$: response and
correlation functions}
\label{sub:Gaussian-response}

One can show \cite{Goldenfeld,Fisher-SA-notes} that for
systems with dimension $d$
above the upper critical dimension $d_c$, the nonquadratic terms in
the action become less and less important on longer and longer
length (and time) scales. Near the critical point, where the behavior is
dominated by fluctuations on long length scales, the system is
then well described by the quadratic parts of the action, and
the calculation of correlation and response functions amounts to the
relatively simple task of solving Gaussian integrals.
It should come as no surprise that the mean-field exponents are
recovered, since the quadratic parts of the action represent the
lowest order terms in the saddle-point
expansion around mean-field theory.

In our problem the action $S_H$ of \eq{S-H} has the quadratic part
\bea
\label{quadratic}
Q(\eta,\hat \eta) &=& -\sum_{j,l} \int dt J_{jl}^{-1}J \hat\eta_j(t)
\eta_l(t)
-\sum_j \int dt\, \hat{\eta}_j(t)
        [-a \partial_t/\Gamma_0 - u_{11}^{stat}]\, \eta_j(t)\,
\nonumber \\
& & + (1/2) \sum_j \int dt_1 \int dt_2\, \hat{\eta}_j(t_1)
\hat{\eta}_j(t_2)
u_{2,0} \, .
\eea

In the long--wavelength limit we
can write $J^{-1}(q) = 1/J + J_2 q^2$ \cite{Binney}.

Rescaling
$\hat\eta$, $\omega$ and $q$ we can replace the constants
$J_2 J$ and $a$ by $1$.
The low frequency part of the $\hat \eta \eta$-term in
$Q(\eta,\hat \eta)$
is then given by
\be
\label{propagator}
-\int d^dq \int dt \hat\eta(-q,t) (-\partial_t/\Gamma_0 + q^2 -
\chi^{-1}/J) \eta(q,t)
\ee
where
\be
\label{chi}
\chi^{-1} = J(u_{11}^{stat} - 1) = 2J^2 \rho(-JM -H +k) -J(k-J)/k
\ee
is the negative static response to a monotonically increasing
external magnetic field, calculated in
mean--field--theory.\footnote{Note that $\chi^{-1} = J(k-J)/k t$,
where $t$ is a parameter used in the mean-field scaling functions
in appendix \protect{\ref{ap:mean-field}}. It is defined for the
soft spin model in \eq{t-def-soft-spin-mft}. In particular, this
implies that near the critical point $\chi^{-1}$ scales with
the original parameters $r$ and $h$ in the same way as $t$, see
\eq{aval_near_crit_endpoint} in the same appendix.}

In frequency space, it can be
written as
\be
\label{frequency-propagator}
-\int d^dq \int dt (-i\omega /\Gamma_0 + q^2 -
\chi^{-1}/J) \hat\eta(-q,-\omega) \eta(q,\omega) \, .
\ee

The $\hat \eta \hat \eta$ term in $Q(\eta,\hat \eta)$, given by
\be
\int d^dq \int dt_1 \int dt_2 1/2 u_{2,0} \hat\eta(-q,t_1)
\hat\eta(q,t_2) \,
\ee
can also be expressed in frequency space
\be
\label{frequency-etahat-etahat}
\int d^dq \int d\omega (1/2) u_{2,0} \delta (\omega)
\hat\eta(-q,-\omega)
\hat\eta(q,\omega) \, .
\ee
The expressions from \eq{frequency-propagator} and
\eq{frequency-etahat-etahat} can be written together as

%\be
%Q(\eta, \hat \eta) = -\int dw \int d^dq
%\begin{array}{cc}
%(\hat\eta(-q, -\omega) & \eta(-q, -\omega) ) \\
%    &
%\end{array}
%\left
%\begin{array}{ccc}
%-1/2u_{2,0} \delta(\omega) & (-i\omega/\Gamma_0+q^2 - \chi^{-1}/J)
%\\
%(i\omega/\Gamma_0+q^2 - \chi^{-1}/J & 0
%\end{array} \right
%\left
%\begin{array}{c}
%\hat\eta(q, \omega) \\
%\eta(q, \omega)
%\end{array} \right \, .
%\ee

\bea
\lefteqn{Q(\eta, \hat \eta) = -\int dw \int d^dq } \\
\nonumber
&&
% Reihenvektor:
        \lower1.1ex \hbox{ $
        {\stackrel{({\scriptstyle  \hat\eta(-q, -\omega), \;
\eta(-q, -\omega)})}{
        \scriptstyle \vphantom{|^2} }} $ } \! \!
% Matrix:
        \left( \lower.8ex \hbox{ $
        {\stackrel{\scriptstyle -1/2u_{2,0} \delta(\omega)}
{\scriptstyle \vphantom{|^2} (-i\omega/\Gamma_0+q^2- \chi^{-1}/J)}}
        {\stackrel{\scriptstyle \phantom{\;
\;-}(i\omega/\Gamma_0+q^2 - \chi^{-1}/J)}{\scriptstyle
        \vphantom{|^2} \; \; 0}}
        $ } \right)
% Spaltenvektor:
        \left( \lower.8ex \hbox{ $
        {\stackrel{\scriptstyle \hat\eta(q, \omega)}
{\scriptstyle \vphantom{|^2} \eta(q, \omega)}}
        $ } \right)
\eea

This can be used to determine the response and correlation functions
at field $H$ at low frequencies to lowest order in the expansion
around mean field theory \cite{Ramond}.
Inverting the matrix one
obtains \cite{NarayanI}
\be
\label{correlation-as-fct-of-omega}
G_{\hat \eta \eta}(q,\omega) = \langle \hat \eta(-q,-\omega)
\eta(q,\omega) \rangle \approx 1/(-i\omega/\Gamma_0 + q^2 -\chi^{-1}/J)
\ee
and
\be
\label{G-hat-hat}
G_{\hat \eta \hat \eta}(q,\omega) = \langle \eta(-q,-\omega)
\eta(q,\omega) \rangle \approx u_{2,0} \delta(\omega) /
|-i\omega/\Gamma_0 + q^2 -\chi^{-1}/J|^2 \, .
\ee
The $\delta(\omega)$ in \eq{G-hat-hat} is a consequence of the
underlying separation of time scales. It will lead to an essentially
static character of the RG analysis of the problem. This might have
been expected, since the critical phenomena we set out to describe
are essentially static in nature.
At the critical point $\chi^{-1}=0$ we have
$G_{\hat\eta \eta}(q,\omega=0) \sim q^{2-\eta}$ with $\eta=0$
and $G_{\eta \eta}(q,\omega=0) \sim q^{4-\bar\eta}$
with $\bar \eta=0$ at lowest order in perturbation theory.
One can Fourier transform the correlation functions back to time
\bea
\label{response-greens-function}
         G_{{\hat\eta}\eta}(q,t,t') &\equiv& \langle \hat\eta(q,t)
        \eta(-q,t') \rangle \\
\nonumber
&=&         \left\{ \begin{array}{ll}
                 \Gamma_0 \exp \{-\Gamma_0(q^2 -\chi^{-1}/J)(t'-t)\} &
        \mbox{for $t' > t$} \\
                 0 & \mbox{for $t'\le t$}
                 \end{array}
         \right.
\eea
and
\bea
\label{correlation-greens-function}
\nonumber
G_{\eta\eta}(q,t,t') &=& \int dt_1 \int dt_2
G_{{\hat\eta}\eta}(q,-(t'-t)+t_1+t_2) u_{2,0}
G_{{\hat\eta}\eta}(q,t_2) \\
&=& u_{2,0}/(q^2-\chi^{-1}/J)^2 \, .
\eea

\subsection{The RG analysis}
\label{sub:RG-analysis}

In dimension $d<d_c$ the nonquadratic parts of the action are no
longer negligible near the critical point. One obtains
corrections to the mean-field behavior.

The Wilson-Fisher coarse graining procedure is an iterative
transformation to calculate the effective action for the long
wavelength and low frequency degrees of freedom of the system.
In each coarse graining step  \cite{Wilson,Ma,NarayanI,Fisher-SA-notes}
one integrates out high momentum modes of {\it all}
frequencies $\hat\eta(q,\omega)$  and $\eta(q,\omega)$,
with $q$ in a momentum shell $[\Lambda/b,\Lambda]$, $b > 1$,
and afterwards rescales according to
$q=b^{-1}q'$, $\omega=b^{-1}\omega'$, $\hat\eta(q,\omega)=b^{{\hat
c}_p} \hat\eta'(q,\omega)$, and $\eta(q,\omega)=b^{c_p}
\eta'(q,\omega)$.
As usual, the field rescalings ${\hat c}_p$ and $c_p$
are chosen such that the quadratic parts of the action at the
critical point ($\chi^{-1}=0$) remain unchanged, so that the
rescaling of the response and the cluster correlation function under
coarse graining immediately gives their respective power law
dependence on momentum ({\it i.e.} this is an appropriate choice of
the scaling units.)
Without loop corrections this implies that $z=2$,
$\hat\eta(x,t) = b^{-d/2-z} \hat\eta'(x,t)$ and
$\eta(x,t) = b^{-d/2+2} \eta'(x,t)$.

Performing one coarse graining step for
the expansion for $S_H$ or \eq{S-H} yields a coarse
grained action which can be written in the original form, with
``renormalized'' vertices $u_{m,n}'$.
Without loop corrections, the vertices of the coarse-grained action
are simply rescalings of the original vertices, which can be easily
read off using the rescalings of $q$, $\omega$, $\eta$,
and $\hat\eta$.
Taking into account that each $\delta/\delta\epsilon(t)$ involves a
derivative with respect to time, and therefore another factor
$b^{-z}$ under rescaling, we arrive at
\be
\label{u-mn-rescaling}
u_{m,n}' = b^{[-(m+n)+2]d/2+2n} u_{m,n} \, .
\ee

This shows that above $8$ dimensions all vertices that are
coefficients of terms of higher than quadratic order in the fields,
shrink to zero under coarse--graining and are therefore
``irrelevant'' for the critical behavior on long length scales and at
low frequencies.

The ``mass'' term $\chi^{-1}$ in the action actually grows under
rescaling in any dimension:
\be
\label{chi-rescaling}
(\chi^{-1})' = b^2 (\chi^{-1})
\ee
(without loop corrections). At $\chi^{-1}=0$ and if all irrelevant
coefficients are set to zero,
the action does not change under coarse graining.
This case is obviously a fixed point of the coarse graining
transformation, where
the systems looks the same on all length scales. There is no
finite (correlation) length determining the long wavelength
behavior, which is just what one would expect at a critical point.

The critical exponents can be extracted from a linearization
of the transformation around the associated fixed point.
In the RG sense ``relevant'' eigenvectors of the linearized
transformation ({\it i.e.} coefficients
like $\chi^{-1}$ that grow under coarse graining)
render the corresponding scaling fields or tunable
parameters.

Below eight dimensions the vertex $u_{1,2} \equiv w= 2 J
\rho'(-J\eta^0 - H +k)$ is the first coefficient of a nonquadratic
term to become relevant. An action with the original parameters
$\chi^{-1}=0$ and $w \neq 0$ corresponds to a system with less than
critical randomness $R<R_c$ at the onset field of the infinite
avalanche.
In appendix \ref{ap:inf-aval-line}
we show how to extract the mean-field exponents
for the infinite avalanche line from the scaling above eight
dimensions and that the RG treatment suggests a first order
transition for the same systems in less than 8 dimensions.

In systems where the bare value of $w$ is zero at the critical fixed
point with $\chi^{-1}=0$, all nonquadratic terms are irrelevant
above {\it six} dimensions.
As can be seen from appendix \ref{ap:mean-field}, this case constitutes
the interesting ``critical endpoint'' at
$R=R_c$ and $H=H_c(R_c)$, which we have discussed in the
introduction.
The corresponding ``upper
critical dimension'' is $d_c=6$. For any dimension higher than $6$,
the mean--field exponents should describe the behavior on long
length scales correctly  \cite{Goldenfeld}.
For $d<6$, the vertex $u_{1,3}=u$ becomes relevant, while all higher
vertices remain irrelevant. We are left with the
effective action which includes all vertices relevant for an
expansion around 6 dimensions:
\begin{eqnarray}
\label{relevant_action}
\tilde S &=& -\int d^dq \int dt\, \hat{\eta}(-q,t)
        [-\partial_t/\Gamma_0 + q^2 - \chi^{-1}/J]\, \eta(q,t)\,
 \\
\nonumber
& & + (1/6) \sum_j \int dt\, \hat{\eta}_j(t) (\eta_j(t))^3 u
+ (1/2) \sum_j\int dt_1 \int dt_2\, \hat{\eta}_j(t_1)
\hat{\eta}_j(t_2)
u_{2,0}
\end{eqnarray}

Below $6$ dimensions, where $u$ does not scale to zero, we will
perform the coarse graining transformation in perturbation theory
in $u$.
At the fixed point, in $6-\epsilon$ dimensions, $u$ will be of
$O(\epsilon)$.
The perturbation series for the paramenters in the action and thus
also for the critical exponents, becomes a perturbation series in
powers of $\epsilon$.
{}From the form of the action one can derive Feynman rules
(see appendix \ref{ap:RG-details}), which
enable us to write down the perturbative corrections in a systematic
scheme. Examples of their derivation for the $\phi^4$ model are
given elsewhere \cite{Fisher-SA-notes,Goldenfeld,Binney}.

\subsubsection{Loop corrections}
\label{subsub:loop-corrections}

In the remaining parts of this section
we perform a coarse graining transformation to first order in
$\epsilon$.\footnote{In principle this is not necessary, since
we can read off the corrections to $O(\epsilon^5)$ from
the mapping to the $\epsilon$-expansion of the pure Ising model in
two lower dimensions, which we discuss in the next section.
However we will need the techniques introduced here later for the
calculation of the avalanche exponents.}
{}From the integration over the short wavelength degrees
of freedom (of all frequencies) one obtains loop corrections to
various vertices. Figure \ref{fig:feynman1} (a) and (b)
shows the corrections to $\chi^{-1}$
and $u$ which are important for an $O(\epsilon)$ calculation. The
dots correspond to the vertices with the indicated names.
An outgoing arrow corresponds to an $\hat \eta$ operator, an
incoming arrow corresponds to an $\eta$ operator.

We consider the $\hat \eta \eta $ term in the action as propagator
and all other terms as vertices. An internal line in a diagram
corresponds to the
contraction
\be
\langle \hat \eta (q,t) \eta(-q, t') \rangle =
         \left\{ \begin{array}{ll}
                 \Gamma_0 \exp\{-\Gamma_0(q^2 -\chi^{-1}/J)(t'-t)\} &
        \mbox{for $t' > t$} \\
                 0 & \mbox{for $t'\le t$}
                 \end{array}
         \right.
\ee
with $q$ in the infinitesimal momentum shell $\Lambda/b <q < \Lambda$
$(b>1)$ over which is integrated. This expression can be
approximated by
$ \delta(t-t')$ in the low frequency
approximation \cite{NarayanI}.
Note, however, that causality must be obeyed, {\it i.e.} $t'>t$.
Figure \ref{fig:feynman1} (c) shows an example of a diagram that
violates causality and
is therefore forbidden.
%
%XXXXXXXX figure feynman1 XXXXXXXXXXXX
%
External (loose) ends in a diagram correspond to operators that are
not integrated out, {\it i.e.} modes of momentum $q
<\Lambda/b$ outside of the momentum shell.
Each internal line carrying momentum contributes a factor
\be
1/(q^2 - \chi^{-1}/J) \, .
\ee
The entire loop in diagram \ref{fig:feynman1} (a)
contributes the integral
\be
I_1 = \int_{\Lambda/b}^{\Lambda} d^dq/(2 \pi)^d 1/(q^2 -
\chi^{-1}/J)^2
\ee
(integration over time is already performed).
Similarly the loop diagram in figure \ref{fig:feynman1} (b)
yields the integral
\be
I_2= \int_{\Lambda/b}^{\Lambda} d^dq/(2 \pi)^d 1/(q^2 -
\chi^{-1}/J)^3 \, .
\ee
After each integration step we also have to rescale momenta,
frequencies and fields.
Thus the recursion relations for $\chi^{-1}/J$ and $u$ become
\be
(\chi^{-1}/J)' = b^2 \left(\chi^{-1}/J + \frac{u_{2,0}}{2!} \frac{u}{3!}
6 I_1 \right)
\ee
and
\be
u' = b^\epsilon \left(\frac{u}{3!} + \frac{u_{2,0}}{2!}
\left[\frac{u}{3!}\right]^2
36 I_2 \right) \, .
\ee
(There are no loop corrections to $u_{2,0}$ at this order, so
$u_{2,0}' = u_{2,0}$.)
The integrals $I_1$ and $I_2$ have to be computed in $6-\epsilon$
dimensions. To that end
one uses the relation \cite{Goldenfeld}
\be
I=\int {\textstyle {d^dq \over (2 \pi)^d }}f(q^2)
= {\textstyle {S^d \over (2 \pi)^d}} \int dq q^{d-1} f(q^2)
\, ,
\ee
where $S_d$ is the surface area of a unit sphere in $d$ dimensions:
\be
S_d = 2 \pi^{d/2}/\Gamma(d/2) \, .
\ee
In performing this analytic continuation in dimension we make the
very strong assumption that the physical properties vary smoothly
with dimension. It
has been justified by its great success in many problems
(although it gives an asymptotic expansion in $\epsilon$ which is
not well behaved, see appendix \ref{ap:Borel}).
Disregarding potential complications we then
expand \cite{Goldenfeld}
\be
I_{1,2} = I_{1,2}(0) + \epsilon I_{1,2}'(0) + O(\epsilon^2) \, .
\ee
Since both $I_1$ and $I_2$ are multiplied by $u$ in the recursion
relations, which will be of $O(\epsilon)$ at the fixed point, we only
need $I_{1,2}(0)$ for an $O(\epsilon)$ calculation.
With $K_d= S_d/(2\pi)^d$ we obtain \cite{Goldenfeld}
\bea
I_1 &=& \int_{\Lambda/b}^{\Lambda} d^dq/(2 \pi)^d 1/(q^2 -
\chi^{-1}/J)^2
\nonumber \\
&=& \int_{\Lambda/b}^{\Lambda} d^dq/(2 \pi)^d
(1/q^4)(1+2(\chi^{-1}/J)/q^2 + O((\chi^{-1}/J)^2))
\nonumber \\
&=& K_6 \Lambda^2 (1-1/b^2)/2 + 2 K_6 (\chi^{-1}/J) \ln b +
O((\chi^{-1}/J)^2, \epsilon)
\, .
\eea
Similarly
\be
I_2 =
K_6 \ln b + O((\chi^{-1}/J)^2, \epsilon) \, .
\ee

\subsubsection{Recursion relations to $O(\epsilon)$}

With $v\equiv u_{2,0} u$ and $b^\epsilon = e^{\epsilon \ln b}
= 1 + \epsilon \ln b + O(\epsilon^2)$ the recursion relations then
become
\be
\label{chi-recursion}
(\chi^{-1}/J)^{\prime}=b^2\Bigl(\chi^{-1}/J+ v
/(4 \pi)^3 \Lambda^2(1-1/b^2)/4 + v /(4 \pi)^3 (\chi^{-1}/J)
\ln b \Bigr)
\ee
and
\be
\label{v-recursion}
v'=v+v[\epsilon+3 v/(4 \pi)^3 v] \ln b \, .
\ee
There are two fixed points of these relations.
The Gaussian fixed point $(\chi^{-1}/J)^* = 0$, $v^*=0$,
and a new, nontrivial fixed point at
\be
(\chi^{-1}/J)^* = -\epsilon \Lambda^2/12
\ee
and
\be
v^* = - (4 \pi)^3 \epsilon/3 \, ,
\ee
which is sometimes called the Wilson-Fisher (WF) fixed point.
We linearize the recursion relations around these fixed points to
calculate the critical exponents.
At the Gaussian fixed point the corresponding matrix is
\be
    \left( \begin{array}{cc}
   \partial \left( \frac{\chi^{-1}}{J}\right)'/\partial
\left(\chi^{-1}/J \right) & \partial\left(
\chi^{-1}/J \right)'/\partial v \\
\partial v'/\partial \left(\chi^{-1}/J \right) &
\partial v'/\partial v
                \end{array}  \right)
=
        \left( \begin{array}{cc}
                b^2 & \frac{\Lambda^2}{4} (b^2-1)/(4 \pi)^3 \\
                0   &  b^\epsilon
                \end{array}  \right)
\ee
At the Wilson-Fisher fixed point it is
\be
        \left( \begin{array}{cc}
                b^{(2-\epsilon/3)} & \frac{3 \Lambda^2}{2} (b^2-1)/(4
\pi)^3 \\
                0   & b^{-\epsilon}
                \end{array}  \right)
\ee
The eigenvalues of the transformation linearized around the Gaussian
fixed point are given by $e_{\chi^{-1}}=b^{y_t}$
with $y_t=2$ and
$e_2=b^\epsilon$ with eigendirections $e_{\chi^{-1}}=(1,0)$ and
$e_2=(\Lambda^2/(4(4\pi)^3),1)$ in the $(\chi^{-1}/J,v)$ plane.

The eigendirections of the transformation linearized around the
Wilson-Fisher fixed point are $e_{\chi^{-1}}=(1,0)$ and
$e_2=(-\epsilon \Lambda^2/12,- (4\pi)^3 \epsilon/3)$.
The corresponding eigenvalues,
called $\Lambda_{(\chi^{-1})}$ and
$\Lambda_2$ are given by
\be
\label{WFeigenvalues1}
\Lambda_{(\chi^{-1})}= b^{y_t}
\ee
with $y_t=2-\epsilon/3$ and
\be
\label{WFeigenvalues2}
\Lambda_2= b^{-\epsilon}\, .
\ee

\subsubsection{Flows and exponents}
\paragraph{For $\epsilon<0$, i.e. $d>6$:} the Gaussian fixed point is
unstable along the $e_{\chi^{-1}}$ direction, but stable
along the $e_2$ direction. Thus $v$ (or equivalently $u$) is an irrelevant
variable.
The correlation length scales as $\xi\left(\chi^{-1}/J\right) = b
\xi\left((\chi^{-1}/J)'\right)$
since under coarse graining the coordinates scale as $x'=bx$.
{}From the appendices \ref{ap:mean-field} and \ref{ap:RG-details}
we know that $w=0$ implies that $h=0$ and $\chi^{-1} \sim
r$.\footnote{See appendix \protect{\ref{ap:mean-field}},
\protect{\eq{t-scaling1}} in particular,
keeping in mind that the parameter $t$ given there
is proportional to $\chi^{-1}$, apart from some irrelevant
adjustments for the soft-spin model.}
One obtains (along $h=0$)
\be
\xi(r) \sim b \xi(b^{y_t} r) \, .
\ee
Choosing $b = r^{-1/y_t}$ it follows that
\be
\xi = r^{-1/y_t} \xi(1)\, .
\ee
Since $\xi \sim r^{-\nu}$ along $h=0$, one obtains
\be
\nu = 1/y_t \, .
\ee
Therefore $\nu = 1/2$ for $d>6$.

\paragraph{For $\epsilon>0$, i.e. $d<6$:} the Gaussian fixed
point becomes unstable along both $e_{\chi^{-1}}$ and $e_2$.
Along the $e_2$ direction the flow is towards the attractive WF
fixed point, which will govern the critical behavior of all systems
that do not initially sit at the Gaussian fixed point
$u=J \rho''(-JM -H)=0$.
For those systems one then obtains
(see \eq{WFeigenvalues1})
\be
\label{exponent-nu}
\nu = 1/{y_t} = 1/2 + \epsilon/12
\ee
The corresponding flows in the $(\chi^{-1}/J, v)$ plane are shown in
figure \ref{fig:flows}.

%XXXXXXXXXX figure fig:flows XXXXXXXX

As in the RG treatment of the standard Ising model, the anomalous
dimension $\eta$ is equal to zero to first order in $\epsilon$,
since the field rescalings acquire no loop corrections to first
order in $\epsilon$ \cite{Goldenfeld}.
(The coefficient of the $q^2$-term in the propagator and the
$u_{2,0}$-term only receive $O(\epsilon^2)$ and higher order loop
corrections.) Since the $i \omega$ term also receives no corrections
to $O(\epsilon)$, we have $z=2+O(\epsilon^2)$.

\subsubsection{Equation of state}
\label{subsub:equation-of-state}

We can also calculate $O(\epsilon)$ corrections to the
equation of state \cite{Domb6}, and
consequently for the entire scaling {\it function} of the
magnetization in \eq{magn-scaling}.
Following Wallace \cite{Domb6} we replace the field $\tilde \eta(x)$
in the Hamiltonian by a new field $L(x)$ with zero expectation value
\be
\tilde \eta(x) = L(x) + M
\ee
with
\be
\langle L \rangle =0\, .
\ee
In other words, in contrast to the preceding analysis, here we
expand the action around
the {\it true}, not mean-field, magnetization $M$ at the
considered order. At tree level
$[L(x) = \eta(x)]_{\it tree-level}$ and $[M=\eta^0]_{\it tree-level}$
and $[u_{1,0}(M)=0]_{\it tree-level}$.
At higher order however, unlike in an expansion around mean-field
theory, $u_{1,0}(M)$ is no longer zero.
Its role is to cancel off all tadpole graphs \cite{Domb6,NarayanI},
so that $\langle L \rangle =0$ at any given order. The
lowest order tadpole correction is shown in figure \ref{fig:tadpole}.
%
%XXXXXXXX figure fig:tadpole XXXX
%
Here we follow Wallace's calculation in \cite{Domb6}. Instead of
performing an
iterative integration over successive momentum shells, as we have
done before, all momentum modes are taken into account in one step
in which the integration extends
from $0$ up to the cutoff $\Lambda$ \cite{Amit}.
The analytic expression to be added to $u_{1,0}$, corresponding
to the lowest order correction of figure \ref{fig:tadpole}, is
then given by
\be
\label{tadpole-w}
(1/2) u_{2,0} w \int_0^\Lambda 1/(q^2-\chi^{-1})^2 d^dq \, .
\ee
As we have explained, the true magnetization
$M$ is chosen such that
\be
\label{defining-M}
u_{1,0}(M) + (1/2) u_{2,0} w \int_0^\Lambda 1/(q^2-\chi^{-1})^2 d^dq +
\cdot \cdot \cdot =0
\ee
where $\cdot \cdot \cdot$ stands for all higher order corrections.
\Eq{defining-M} then constitutes the equation of state at the given
order.
In an expansion around $\eta^0=M$ (where $M$ is now the true
magnetization),
$u_{1,0}(M)$ is given by
\eq{u-mn}, \eq{response-in-local-mft} and section
\ref{sub:Implementing-the-history} to be
\be
\label{mean-field-equation-of-state2}
u_{1,0}(M) = (k +H)/k - (k-J)M /k - 2
\int_{-\infty}^{-JM-H+k} \rho(h) dh \,.
\ee
Close to the critical point the integral can be expanded around
$-JM-H+k = -Jm -h = 0$, where $m=M-M_c$ is the deviation of the
magnetization $M$ from the mean-field critical value $M_c=+1$, and
$h=H-H_c(R_c)$, with the mean-field value $H_c(R_c)=k-J$,
and $r=(R_c-R)/R$ with $R_c=2 kJ/(\sqrt{2 \pi}(k-J))$.
One obtains
\be
\label{expanded-mean-field-equation-of-state}
u_{1,0} = \frac{h}{k} + \frac{\chi^{-1}}{J} rm + \frac{2
J^3}{3!}
\rho''(0) m^3 + \cdot \cdot \cdot \,,
\ee
where $\rho''(h)$ denotes the second derivative of the
distribution of random fields with respect to its argument
and $\cdot \cdot \cdot$ stands for higher order corrections.
To calculate the equation of state at one loop order we insert into
\eq{defining-M} the expressions
$\chi^{-1} = -2 J r/(\sqrt{2 \pi} R) + 1/2 u m^2 + \cdot \cdot
\cdot$ (see \eq{chi}),
$w=u m + \cdot \cdot \cdot$ (see \eq{w})
and $u=2J^3\rho''(0) +
\cdot \cdot
\cdot$ (see \eq{u}),
where $\cdot \cdot \cdot$ denotes higher orders in $m$ and $h$.
One then finds that the calculation
is analogous to the one done for the Ising model in two lower
dimensions.
For details on solving the integral etc.
we refer the reader to the article
on the equation of state for the Ising model in $4-\epsilon$
dimensions by D.J. Wallace in reference \cite{Domb6}, especially
equation (3.35).
(In fact,
with the following formal identifications, the resulting equations
of state in the two systems can be mapped onto each other:
$h/k=-h_w$, $2 J r/(\sqrt{2 \pi} R) = -t_w$, $2J^3\rho''(0)=-(u_0)_w$,
$m=-m_w$.
We have denoted the quantities in Wallace's article by an index
``w''.)

We find for the equation of state to one loop order:
\bea
\label{equation-of-state-to-one-loop-order}
u_{1,0} \equiv 0 &=& \frac{h}{k}  +2 J /(\sqrt{2 \pi} R) rm + \frac{2
J^3}{3!}\rho''(0) m^3 + \frac{1}{6} \epsilon m (2 J /(\sqrt{2 \pi}
R) r
\nonumber \\
& & +
J^3 \rho''(0) m^2 ) \ln (-2 J /(\sqrt{2 \pi} R) r - J^3 \rho''(0)
m^2) \,.
\eea
Here $\epsilon = 6-d$ and $r$ and $m$ are the reduced randomness
and magnetization respectively.

\subsubsection{The exponents $\beta$ and $\delta$}
This equation allows us to extract the critical exponents $\beta$
and $\delta$ by taking the corresponding limits:
The exponent $\delta$ can be obtained by setting $r=0$.
That leaves one with
\be
\label{equation-of-state-for-delta}
0 \equiv u_{1,0} = h/k + 1/3 J^3 \rho''(0) m^{3+\epsilon} \,.
\ee
Since by definition of $\delta$ one has $|h| \sim |m|^\delta$,
we find
\be
\label{delta}
\delta = 3+\epsilon \,.
\ee
Similarly
one sets $h=0$ to get $\beta$ and finds from $|r|\sim
|m|^{1/\beta}$
(for $R<R_c$)
that
\be
\label{beta}
\beta = \frac{1}{2} - \frac{\epsilon}{6} \,.
\ee

\subsubsection{The entire scaling function}
{}From \eqs{equation-of-state-to-one-loop-order},
(\ref{delta}) and (\ref{beta}) one can construct the entire scaling
function $\tilde f(r/m^{1/\beta})$ to $O(\epsilon)$ in
\be
|h/k| = |m|^\delta \tilde f(r/|m|^{1/\beta})
\ee
directly, as shown in reference \cite{Domb6}.
With the redefinitions
\be
m_{new}^2 = - (1/3) J^3 \rho''(0) m^2
\ee
and
\be
x= - (2Jr)/(\sqrt{2 \pi} R |m_{new}|^{1/\beta})
\ee
(for $r>0$),
one obtains
\be
f(x) = x + 1 + {\textstyle {\epsilon \over 6}} ((x+3) \ln(x+3)
-3(x+1) \ln(3)+ 2x \ln(2)) + O(\epsilon^2)) \, ,
\ee
in
\be
|h/k| = |m_{new}|^{\delta} f(x) \, .
\ee
In the next section we discuss a formal mapping of our $\epsilon$
expansion
to the $\epsilon$ expansion of the pure Ising model in two lower
dimensions, which allows us to copy the $O(\epsilon^2)$ correction
to the equation of state from previous calculations for the pure Ising
model \cite{equ-of-state}.
%Figure \ref{fig:M(H)-eps-numerics} shows a comparison
%of the $O(\epsilon^2)$ result for the scaling function
%to the scaling function obtained from numerical
%simulations in 5 dimensions.

\section{Mapping to the Thermal Random Field Ising Model}
\label{sec:mapping}
\subsection{Perturbative mapping and dimensional reduction}
We notice that the results found for $\nu$, $\beta$ and $\delta$
in $6-\epsilon$ dimensions
(\eqs{exponent-nu}, (\ref{beta}), and (\ref{delta}))
are the same as those for the regular
equilibrium Ising model in $4-\epsilon$ dimensions.
In fact, this equivalence actually extends to {\it all}
orders in $\epsilon$:
We will show that the $\epsilon$-expansion for our
model is the same
as the $\epsilon$-expansion for the {\it equilibrium} random field Ising
model to all orders in $\epsilon$ \cite{oopsI}. Once this equivalence is
established, we can use that the $(6-\epsilon)$-expansion of the
{\it equilibrium} random field Ising
has been mapped to {\it all} orders in $\epsilon$ to the
corresponding expansion of the regular Ising model in two
lower dimensions \cite{mapping,Parisi}.

The easiest way to recognize that the $\epsilon$-expansion for our
model and for the {\it equilibrium} RFIM should really be the same is by
comparing the corresponding effective actions. In a dynamical
description of the {\it equilibrium} RFIM at zero external magnetic field
the following effective Langevin equation of motion for the
spin-field $\phi(r,t)$ was used \cite{Krey}
\be
\label{effective-equation-of-motion}
\partial_t \phi(x,t) = -\Gamma_0 (-\nabla^2 \phi(x,t) + r_0 \phi(x,t)
+ 1/6 g_0\phi^3(x,t) - h_R(x)) \, .
\ee
$h_R(x)$
represents spatially uncorrelated
quenched random fields distributed according to a
Gaussian of width $\Delta_0$ and mean zero.
$h_T(r,t)$ is the thermal noise field, which is taken to be gaussian
with vanishing mean value and the variance
\be
\langle h_T(r,t) h_T(r',t') \rangle = 2 k T/\Gamma_0 \delta(r-r')
\delta(t-t') \, .
\ee
The corresponding Martin Siggia Rose generating functional is
\bea
\label{Z-eff-H0}
Z_{H_0}^{\it thermal} &=& \int [d\hat\phi] \int[d\phi] \exp\{ -\int d^dq \int
dt (\hat\phi(-q,t) (-\partial_t/\Gamma_0 +q^2 +r_0) \phi(q,t))
\nonumber \\
\nonumber
 & & +
\int d^dx \int dt (\hat\phi(x,t) (-1/6 g_0) \phi(x,t)^3) \\
 & & +
\int d^dx \int dt (\hat\phi(x,t)( h_R(x)+ h_T(r,t)) \} \, .
\eea
Since again $Z_{H_0}^{\it thermal}=1$, we can average the partition function
directly over the random fields $h_R$ and the thermal noise $h_T$:
\be
{\bar Z}_{H_0}^{\it thermal} = \langle Z_{H_0}^{\it thermal} \rangle_{h_R}.
\ee
The average over the random fields at each $x$
and over the thermal noise fields at each $x$ and $t$ yields, (after
completing the square):
\bea
\nonumber
        {\bar Z}_{H_0}^{\it thermal}
&=&
        \int [d\hat\phi] \int[d\phi] exp \left\{
        -\int d^dq \int
        dt \;
        \hat\phi(-q,t) (-\partial_t/\Gamma_0 +q^2 +r_0) \phi(q,t)
        \right.
\\
\nonumber
 & &
        + \int  d^dx  \int dt \;
        \hat\phi(x,t) (-
        {\textstyle {1 \over 6}} g_0) \phi^3(x,t)
\\
 & &
        + \int d^dx \int dt_1 \int dt_2
        \hat\phi(x,t_1)\hat\phi(x,t_2)
        \Delta^2/2
\nonumber
\\
 & &
\label{Z-bar-eff-H0}
        \left.
        + \int d^dx \int dt \hat\phi^2(x,t) (2 k T)/\Gamma_0
        \right\} \, .
\eea

With the identifications $r_0=-\chi^{-1}(H_0)$, $u=-1/6 g_0$,
$u_{2,0} = \Delta^2$ and $T=0$,
we see that the argument of the exponential function
is the same action as the effective action for our
zero temperature, nonequilibrium model in \eq{S-H}.
Setting $T$ to zero in the action for the equilibrium RFIM does not
change the expansion for the static behavior, since it
turns out that corrections involving temperature are negligible
compared to those involving the random magnetic
field \cite{Parisi,Krey,Tadic} --- the {\it temperature}
dependence is irrelevant in the {\it thermal} RFIM and the
{\it time} dependence is irrelevant in our zero--temperature
{\it dynamical} RFIM, leaving us with the same starting point for
the calculation!
This equivalence implies that in $6-\epsilon$ dimensions
we should obtain the same critical
exponents to all orders in $\epsilon$
for our model, as were calculated for the
thermal random field Ising model, which in turn are the same as
those of the pure equilibrium Ising model in $4-\epsilon$
dimensions \cite{mapping,Parisi}.

This observation is rather convenient, since it provides us with
results from the regular Ising model to
$O(\epsilon^5)$ for free. In $6-\epsilon$ dimensions
we read off \cite{Kleinert}
\bea
1/\nu &=& 2-\epsilon/3 - 0.1173 \epsilon^2 + 0.1245 \epsilon^3 -0.307
\epsilon^4 + 0.951\epsilon^5 + O(\epsilon^6) \\
\eta &=& 0.0185185 \epsilon^2 + 0.01869 \epsilon^3 - 0.00832876
\epsilon^4 +  0.02566 \epsilon^5 + O(\epsilon^6) \\
\nonumber
\beta &=& 1/2 -\epsilon/6 + 0.00617685\epsilon^2 - 0.035198 \epsilon^3
+ 0.0795387 \epsilon^4 \\
& & - 0.246111 \epsilon^5 + O(\epsilon^6) \\
\nonumber
\beta\delta &=& 3/2 + 0.0833454 \epsilon^2 - 0.0841566 \epsilon^3
+ 0.223194 \epsilon^4 - 0.69259 \epsilon^5 \\
& & + O(\epsilon^6)\, .
\eea
$\beta$ and $\delta$ have been
calculated from $\eta$ and $\nu$ using the perturbative relations:
$\beta = (\nu/2) (d-4+{\bar\eta})$ and
$\delta=(d-2\eta+{\bar\eta})/(d-4+{\bar\eta})$
\cite{appendix-beta-delta}, with
$\eta={\bar\eta}$
to all orders in $\epsilon$ \cite{Krey}.

By the same mapping we obtain the universal scaling function for the
magnetization to $O(\epsilon^2)$ \cite{equ-of-state,Domb6}.
We set
\be
h= m^\delta f\left(x= r/m^{1/\beta}\right)
\ee
in which the normalizations of $x$ and the universal scaling
function $f$ are chosen such that
\be
f(0)=1\, , \qquad f(-1) =0\,.
\ee
The expansion to second order in $\epsilon$ is then
\be
\label{f-x-scaling-fct}
f(x) = 1 + x + \epsilon f_1(x) + \epsilon^2 f_2(x)
\ee
with
\be
f_1(x) = {\textstyle{1 \over 6}} \left[  (x+3) \ln(x+3) -3(x+1) \ln(3)
+ 2x \ln(2) \right]
\ee
and
\bea
\nonumber
f_2(x) &=& \left[{\textstyle{1 \over 18}}\right]^2
\left\{  [6 \ln 2 -9 \ln 3 ]
 [ 3 (x+3) \ln(x+3)
+ 6 x \ln 2 - 9 (x+1) \ln 3 ] \right.\\
\nonumber
& & + \textstyle{9 \over 2}
(x+1) [\ln^2 (x+3) - \ln^2 3]
\nonumber \\
\nonumber
& & + 36 [\ln^2(x+3) - (x+1) \ln^2 3 + x \ln^2 2 ] \\
\nonumber
& &  - 54 \ln 2
[\ln(x+3)
+ x \ln 2 - (x+1) \ln 3 ] \\
& & \left. + 25 [(x+3) \ln(x+3) + 2 x
\ln 2 - 3 (x+1) \ln 3 ] \right\}  \, .
\eea

The scaling function $f(x)$ has actually been calculated
up to order $\epsilon^3$ \cite{Zinn-Justin}.
As it stands the expression (\ref{f-x-scaling-fct}) meets
the Griffith analyticity requirements \cite{Griffiths} only
within the framework of the $\epsilon$-expansion, but not
explicitly. These subtleties can be avoided by writing it in a
parametric form \cite{equ-of-state}, which can then be
compared directly with our numerical results for the universal
scaling function of $dM/dH$ in 5, 4, 3, and 2
dimensions. We will present the results in a forthcoming paper
\cite{Perkovic}.

%Figure \ref{fig:M(H)-eps-numerics} shows a comparison
%of the $O(\epsilon^2)$ result for the scaling function
%to the scaling function obtained from numerical
%simulations in 5 dimensions.

%XXXXXXXXXX fig:M(H)-eps-numerics XXXXXXX

The dynamic exponent $z$ cannot be extracted from the mapping
to the regular Ising model.
It was calculated separately to $O(\epsilon^3)$ for the equilibrium
RFIM \cite{Krey} and found
to be given to this order by\footnote{\Eq{z-result2} is only a
	perturbative result for $z$
	which does not reveal
	the presence of diverging barrier heights that lead to the observed
	slow relaxation towards equilibrium
\cite{RFIM,Nattermann-review,dynamics-RFIM,dyn-scal-RFIM,Fisher}.
	Nonperturbative corrections are expected to be important
	in the equilibrium random field Ising model.}\footnote{While
	we expect the $6-\epsilon$ results for the static exponents
	$\beta$, $\delta$, $\nu$, $\eta$, $\bar\eta$, $\tau$, $\sigma$, etc.
	to agree with our hard-spin simulation results close to 6 dimensions,
	this is not necessarily so for the dynamical exponent $z$. There
	are precedences for the dynamics being sensitive to the exact shape
	of the potential, at least in mean-field theory.
	Such differences have been found for example in the case of charge
	density waves \cite{FisherCDW,NarayanI,NarayanII,ParisiCDW}, if
	studied for smooth potentials, for cusp-potentials, or for sawtooth
	potentials.}
\be
\label{z-result2}
z=2+2\eta = 2 + 0.037037 \epsilon^2 + 0.03738 \epsilon^3 +
O(\epsilon^4) \, .
\ee
Because of the perturbative mapping of our model
to the equilibrium RFIM,
\eq{z-result2} also gives the result for $z$ in our nonequilibrium
hysteretic system.

We have performed a Borel resummation
(see also appendix \ref{ap:Borel}) of the
corrections to
$O(\epsilon^5)$ for $\eta$, $\nu$, and the
derived \cite{appendix-beta-delta} exponent
$\beta\delta$. The exponent $\beta$ is then given
through $\beta=\beta\delta - (2-\eta) \nu$ (\eq{fluct-diss}).
Figure \ref{fig:Borel-vs-numerics}
shows a comparison with our numerical
results in 3, 4, and 5 dimensions.

The agreement is rather good near 6 dimensions. However
the apparent dimensional reduction through the perturbative
mapping to the Ising exponents in 2 lower dimensions gradually
loses its validity at lower and lower dimensions. It is after all
only due to the equivalence of two asymptotic series, both of
which have radius of convergence zero.
Table \ref{tab:equil-RFIM} shows a comparison between the
numerical exponents for our model and for the equilibrium RFIM in
three dimensions.

%XXXXXX figure fig:Borel-vs-numerics XXXXXXX

%XXXXXXX tab:equil-RFIM XXXXXXX

\subsection{Nonperturbative corrections}
\label{sub:honest-dim-reduction}
\label{subsub:nonperturb-corrections}

The mapping of the $\epsilon$-expansion for the thermal random field
Ising model
to the expansion for the Ising model in two lower dimensions has
caused much controversy when first discovered.
The problem was that it had to break down
at the lower critical dimension, where the transition
disappears.
There is no transition in
the pure Ising model in $d=1$,
but the equilibrium RFIM is known rigorously to have a transition in
$d=3$ \cite{Imbrie,Kupiainen}.
The same is true for our model:
numerical simulations indicate \cite{Perkovic,hysterIII}
that the lower critical dimension
is lower than three --- probably equal to two.

In the case of the equilibrium random field Ising model
it was finally agreed \cite{Nattermann-review,Parisi}
that this breakdown
might be due to nonperturbative corrections.
The point is that proving a relation to all orders in $\epsilon$
does not make it true.
In the equilibrium RFIM there are at least two sources of
nonperturbative
corrections:

\paragraph{The ``embarrassing'' correction:}
It was found that there was a calculational error in the
$(6-\epsilon)$-expansion for the RFIM. The perturbation series was
tracing
over many unphysical metastable states of the system, instead of just
taking into account the ground state, which the system occupies in
equilibrium.
There were indications that this error leads to nonperturbative
corrections, which would destroy the dimensional
reduction outside of perturbation theory \cite{Parisi}.

In our calculation we have avoided
the embarrassing source of nonperturbative corrections found
in the equilibrium random field Ising problem.
Given the initial conditions and a history $H(t)$,
the set of coupled equations of motion for all spins will have only
one solution. In the Martin Siggia Rose formalism, the physical
state is selected as the only solution that obeys causality,
there are no unphysical metastable states coming in.
Therefore we believe our results should also apply to systems below
the critical randomness,
at least before the onset of the infinite avalanche.

\paragraph{Instanton corrections:}
Even without the embarrassing correction, there is no reason
why a perturbative mapping of the expansions about the
upper critical dimensions
should lead to a mapping of the lower
critical dimensions also. The $\epsilon$-expansion is only an
asymptotic expansion --- it has radius of convergence zero.
As we discuss in appendix \ref{ap:Borel}, there is no known
reason to assume that
the $\epsilon$-expansion uniquely determines an underlying function.
It leaves room for functions subdominant to the asymptotic
power series: If the series $\sum_0^\infty f_k z^k$ is asymptotic
to some function $f(z)$ in the complex plane as $z \rightarrow 0$,
then it is also asymptotic to any function which differs from
$f(z)$ by a function $g(z)$ that tends to zero
more rapidly than all powers of $z$ as
$z \rightarrow 0$ \cite{Bender}.
An example of such a subdominant function
would be $g(z)=\exp(-1/z)$. While some asymptotic expansions can be
proven to uniquely define the underlying function, this has not been
shown for the $\epsilon$-expansion (appendix \ref{ap:Borel}) --
not for our problem, nor for the equilibrium pure Ising model,
nor for the equilibrium thermal random field Ising model.
\vspace{2mm}

At this point, the $\epsilon$-expansion for our model is on no worse
formal footing than that for the ordinary Ising model. We believe,
the asymptotic expansion is valid for both models, despite the fact
that their critical exponents are different: the exponents for the
Ising model in $\epsilon=4-d$ and the exponents for our model in
$\epsilon=6-d$ are different analytic functions with the same
asymptotic expansion. The
$\epsilon$-expansion can not be used to
decide whether the lower critical dimension is at $\epsilon=3$ or
at $\epsilon=4$.

We conclude that because of instanton corrections
the dimensional reduction breaks down for the equilibrium
RFIM as well as for our nonequilibrium, deterministic zero
temperature RFIM. In addition there is another ``embarrassing''
source of non-perturbative corrections in the equilibrium
RFIM, which we do not have in our problem.
There is no reason to expect our exponents to be the same as those
of the
equilibrium RFIM \cite{Banavar},
though the perturbation series can be mapped.
There might actually be three different underlying functions
for the same $\epsilon$-expansion for any exponent:
one for the pure Ising model, one for the equilibrium random field
Ising model, and one for our model, so that the exponents in all
three
models would still be different although their $\epsilon$-expansions
are the same.

\section{$\epsilon$-Expansion for the avalanche exponents}
\label{sec:epsilon-expansion}

The exponents whose $\epsilon$-expansion we have
calculated so far using the mapping to the equilibrium
RFIM are $\nu$, $\eta$, $\bar\eta$, $\beta$, $\beta\delta$,
$\tilde\theta$ and $z$.
Unfortunately, we cannot extract the avalanche exponents $\tau$,
$1/\sigma$, and $\theta$ from this mapping.
The two exponent relations involving these exponents
\be
\tau-2 = \sigma\beta(1-\delta)
\ee
and
\be
1/\sigma = (d-\theta) \nu - \beta
\ee
are not enough to determine all three exponents from the information
already obtained.

In the following we will compute $\tau$ and $\sigma$
directly in an $\epsilon$-expansion. The method employed makes use
of the scaling of the higher moments of the avalanche size
distribution. They are being calculated using $n$ replicas of the
system for the $n$'th moment.

\subsection{The second moment of the avalanche size distribution}
\label{sub:s-2}

When calculating $\eta$ and $\nu$ we have already used all information
from the scaling behavior of the first moment $\langle S \rangle$
of the avalanche size distribution:
In \cite{eps-big2,thesis} we show that $\langle S
\rangle$ scales as the spatial integral over the
avalanche-response-correlation function, which in turn scales as the
``upward susceptibility'' $dM/dh$ calculated consistently with the
history of the system.

In the Martin-Siggia-Rose formalism it is given
by\footnote{As we explain in section \ref{sub:formalism} and in
\cite{eps-big2,thesis}
the expression
\be
\int dt_0 \langle \hat s(t_0,x_0) s(t,x) \rangle_f
\ee
gives the random-field-averaged static response of the system
at time $t$ and position $x$ to a positive $\theta$-function pulse
applied at position $x_0$  an infinitely long time before $t$
(so that all transients have died away).
The integral over all space will then give the {\it total} static
random-field averaged response of the system to a $\theta$-function
pulse applied at site $x_0$.
This should scale in the same way as the first moment of the
avalanche size distribution, {\it i.e.} as the average avalanche size.}
\be
\label{first-mom}
\langle S \rangle \sim \int dt_0 \int d^dx \langle \hat s(t_0,x_0)
s(t,x) \rangle_f = \int dt_0 \int d^dx \langle \delta s(t,x)/\delta
\epsilon(t_0,x_0) \rangle_f \, .
\ee

The second moment $\langle S^2 \rangle$ of the avalanche size
distribution is the random field average of the squared
avalanche response.
Note that it is not simply the square the expression
in \eq{first-mom} for the
first moment --- the product rule for taking derivatives gets in the
way:
A quantity such as
\be
\langle
 \hat{s}(t_0,x_0) s(t_1,x_1) \hat{s}(t_2,x_0) s(t_3,x_3)
\rangle_f \equiv A + B
\ee
not only contains the term which we need, namely
\be
A=\langle \frac{\delta s(t_1,x_1)}{ \delta \epsilon(t_0,x_0)}
\frac{ \delta s(t_3,x_3)}{\delta \epsilon(t_2,x_0)} \rangle_f
\ee
but also the terms
\be
B=\langle \frac{\delta^2 s(t_1,x_1)}{ \delta \epsilon(t_0,x_0)
\delta \epsilon(t_2,x_0)}
s(t_3,x_3)\rangle_f + \langle s(t_1,x_1) \frac{\delta^2 s(t_3,x_3)}{
\delta \epsilon(t_0,x_0)
\delta \epsilon(t_2,x_0)} \rangle_f \, ,
\ee
which are not related to $\langle S^2 \rangle$.

How can we separate $A$ from $B$?

\noindent
One possibility is to introduce a second replica of the system with
the identical configuration of random fields, the same initial
conditions, and
the same history of the external magnetic field.
One can then calculate the response in each of the two replicas
separately, multiply the results and afterwards take the average
over the random fields.
Denoting the quantities in the first replica with superscript
$\alpha$ and those in the second replica with superscript $\beta$
one obtains

\bea
A_2 &\equiv& \langle \hat{s^\alpha}(t_0,x_0) s^\alpha(t_1,x_1)
\hat{s}^\beta(t_2,x_0)
s^\beta(t_3,x_3)\rangle_f
\nonumber \\
 &=& \langle \frac{\delta s^{\alpha}(t_1,x_1)}{ \delta
\epsilon^{\alpha}(t_0,x_0)}
\frac{ \delta s^{\beta}(t_3,x_3)}{\delta \epsilon^{\beta}(t_2,x_0)}
\rangle_f \, ,
\eea
since
\be
\frac{\delta s^{\alpha}}{ \delta \epsilon^{\beta}}=0 \, ,
\ee
and
\be
\frac{\delta s^{\beta}}{ \delta \epsilon^{\alpha}}=0 \, .
\ee

Similarly for the $n$'th moment $\langle S^n \rangle$ of the
avalanche size distribution one would use $n$ replicas of the system.
In appendix \ref{ap:RG-details-avalanches} we make this argument more
precise and derive the scaling relation between $\langle S^2 \rangle$
and $A_2$
\be
\label{s2-main1}
\langle S^2 \rangle_f \sim \int dt_1 \int dt_\alpha dt_\beta
d^dx_\alpha
d^dx_\beta
\langle
 \hat{s}^\alpha(t_\alpha,x_0) s^\alpha(t_0,x_\alpha)
\hat{s}^\beta(t_\beta,x_0)
s^\beta(t_1,x_\beta)
\rangle_f \, .
\ee
In the following we generalize the RG treatment from previous
sections to the case of two replicas, and extract the
scaling behavior of $\langle S^2 \rangle$ from
\eq{s2-main1} near the critical point.
We will
compare the result to
the scaling relation
\bea
\label{s2-scaling-integral}
\langle S^2 \rangle &=& \int S^2 D(S,r,h) dS \sim \int S^2/S^\tau
{\cal D}_\pm(S r^{1/\sigma}, h/r^{\beta\delta}) dS
\nonumber \\
&\sim& r^{(\tau-3)/\sigma} {\cal S}_\pm^{(2)}
(h/r^{\beta\delta}) \, ,
\eea
where ${\cal S}_\pm^{(2)}$ is the corresponding scaling function,
and obtain the missing information to compute the
exponents $\tau$ and $\sigma$.

\subsection{Formalism for two replicas}
\label{sec:form-2-repl}

The generalization of the MSR generating functional to two replicas is
rather straightforward.
The equation of motion for each spin is the same in both replicas.
\be
\partial_t s^\alpha_i/\Gamma_0 - \delta {\cal H}(s^\alpha)/\delta
s^\alpha_i = 0
\ee
and
\be
\partial_t s^\beta_i/\Gamma_0 - \delta {\cal H}(s^\beta)/\delta
s^\beta_i = 0
\ee
where the Hamiltonian ${\cal H}$ is given by
\eq{model}.

The new generating functional is a double path integral
over two $\delta$-functions which impose the equations of motion for
both replicas. Again we can write the $\delta$-functions in their
``Fourier'' representation by introducing two auxiliary fields
$\hat s^\alpha$ and $\hat s^\beta$.

One obtains simply the square of the generating functional
from \eq{Z}, expressed in terms of two replicas:
\bea
\label{Z-alpha-beta-before-averaging}
Z^{\alpha\beta} &=&  \int \int [ds^{\alpha}] [d{\hat s}^\alpha]
\int \int [ds^\beta] [d{\hat s}^\beta] J[s^\alpha] J[s^\beta]
\nonumber \\
& & \exp ( i \sum_j \int dt {\hat s}^\alpha_j(t)
(\partial_t s^\alpha_j(t)/\Gamma_0-\delta {\cal H}(s^\alpha)/
\delta s^\alpha_j(t)))
\nonumber \\
& &
\exp{ ( i \sum_j \int dt {\hat s}^\beta_j(t)
(\partial_t s^\beta_j(t)/\Gamma_0-\delta {\cal H}(s^\beta)/
\delta s^\beta_j(t)))} \,.
\eea

We note that the two replicas do not interact before the average
over the random fields is taken. Since $Z=1$ we can again average
$Z$ directly over the random fields.

We rewrite the action using the same kinds of transformations
to the local fields ${\tilde \eta}^\alpha$, $\hat{{\tilde \eta}}^\alpha$,
${\tilde \eta}^\beta$ and $\hat{{\tilde \eta}}^\beta$ which we introduced
previously (see \eq{Z-as-a-function-of-eta})
{\it i.e.}
\bea
\label{Z_bar_as_a_function_of_eta_alpha_beta}
\lefteqn{Z^{\alpha\beta} = \int [d{\tilde \eta}^\alpha]
[d\hat{\tilde \eta}^\alpha]
[d{\tilde \eta}^\beta] [d\hat{\tilde \eta}^\beta] \prod_j \bar
Z_j[{\tilde \eta}^\alpha_j,{\hat{\tilde \eta}}^\alpha_j,
{\tilde \eta}^\beta_j,{\hat{\tilde \eta}}^\beta_j] } \\
\nonumber
& & \exp\left\{-\int dt \sum_j {\hat{\tilde \eta}}^\alpha_j(t)
              \left(\sum_\ell J_{j\ell}^{-1}
               J{\tilde \eta}^\alpha_\ell(t)\right)
                         -\int dt \sum_j {\hat{\tilde
\eta}}^\beta_j(t)
              \left(\sum_\ell J_{j\ell}^{-1}
               J{\tilde \eta}^\beta_\ell(t)\right)
\right\}
\, ,
\eea
where
$\bar{Z}_j[{\tilde \eta}^\alpha_j,{\hat{\tilde \eta}}^\alpha_j,{\tilde
\eta}^\beta_j,{\hat{\tilde \eta}}
^\beta_j]$ is a local functional
\be
\label{Zbar-j-alpha_beta}
\bar{Z}_j[{\tilde \eta}^\alpha_j,\hat{\tilde \eta}_j^\alpha,
{\tilde \eta}^\beta_j,\hat{\tilde \eta}_j^\beta] =
\int [ds^\alpha] [d\hat{s}^\alpha]
[ds^\beta] [d\hat{s}^\beta]\langle\exp {\tilde S}^{\alpha\beta}_{{\it
eff}_j}\rangle_f
\, ,
\ee
and
\bea
\label{S-j-alpha-beta}
\nonumber
{\tilde S}^{\alpha\beta}_{{\it eff}_j} &=&\frac{1}{J} \int dt
\left\{
J\hat{{\tilde \eta}}^\alpha_j(t)
s^\alpha_j(t) +
i \hat{s}^\alpha_j(t)\left(\partial_t s^\alpha_j(t) - J{\tilde
\eta}^\alpha_j
- H - f_j + \frac{\delta V^\alpha}{\delta s^\alpha_j}\right)
\right\} \\
\nonumber
& & + \frac{1}{J} \int dt \left\{
J\hat{\tilde \eta}^\beta_j(t)s^\beta_j(t) \right.\\
& & + \left.
i \hat{s}^\beta_j(t)\left(\partial_t s^\beta_j(t) - J{\tilde
\eta}^\beta_j
- H - f_j + \frac{\delta V^\beta}{\delta s^\beta_j}\right)
\right\}
 \, .
\eea
Here $V^\alpha$ and $V^\beta$ are given by the linear cusp potential
$V$ defined in \eq{V-def}, to be understood as a
function of $s^\alpha$ and $s^\beta$ respectively.

Again, we expand the action around its stationary point.
It is specified by four coupled equations, which turn out to be
solved self consistently by the replica symmetric mean field
solution, which we found earlier when studying just one replica:

\be
\label{replica_symmetric_mean_field}
\hat{\tilde \eta}^\alpha_0=0\,,
\ee
\be
\hat {\tilde \eta}^\beta_0=0\,,
\ee
\be
{\tilde \eta}^\alpha_0=\langle s^\alpha \rangle_f\,,
\ee
\be
{\tilde \eta}^\beta_0= \langle s^\beta \rangle_f\,.
\ee

Analogously to before  \cite{hysterII} we will now expand around
the mean--field solution ${\tilde \eta}^\alpha_0$, ${\tilde \eta}^\beta_0$
($J {\tilde \eta}^\alpha_0$ and $J {\tilde \eta}^\beta_0$ denote
the local field configurations about which the log of the
integrand in equation~(\ref{Z_bar_as_a_function_of_eta_alpha_beta})
is stationary).
Introducing shifted fields $\eta^\alpha \equiv {\tilde \eta}^\alpha -
{\tilde \eta}^\alpha_0$ so that
$\langle \eta^\alpha \rangle_f = 0$, and $\hat\eta^\alpha \equiv \hat {\tilde
\eta}^\alpha$
(and correspondingly for $\eta^\beta$, and $\hat \eta^\beta$),
leaves one
with the generating functional
\begin{equation}
\bar{Z} = \int [d\eta^\alpha] [d\hat\eta^\alpha]
[d\eta^\beta] [d\hat\eta^\beta]\exp({S}^{\alpha\beta})
\end{equation}
with an effective action
\bea
\label{tilde_S_alpha_beta}
\nonumber
{S}^{\alpha\beta} &=&
- \sum_{j,l} \int dt J_{jl}^{-1} J\hat{\eta}^\alpha_j(t)
\eta^\alpha_l(t)
- \sum_{j,l} \int dt J_{jl}^{-1} J\hat{\eta}^\beta_j(t)
\eta^\beta_l(t) \\
& &
\nonumber
+\sum_j \sum_{m,n,p,q=0}^{\infty} \frac{1}{m! n! p! q!}
\int dt_1\cdot \cdot \cdot dt_{m+n+p+q}
u^{\alpha\beta}_{mnpq}(t_1,...,t_{m+n+p+q}) \\
\nonumber
& &\hat{\eta}^\alpha_j(t_1)\cdot \cdot \cdot \hat{\eta}^\alpha_j(t_m)
\eta^\alpha_j(t_{m+1}) \cdot \cdot \cdot \eta^\alpha_j(t_{m+n}) \\
& &\hat{\eta}^\beta_j(t_{m+n+1})\cdot \cdot \cdot \hat{\eta}^
\beta_j(t_{m+n+p})\eta^\beta_j(t_{m+n+p+1}) \cdot \cdot \cdot
\eta^\beta_j
(t_{m+n+p+q}) \, .
%\nonumber
\eea
Here, the $u^{\alpha\beta}_{mnpq}$ are the derivatives of
$\ln \bar{Z}^{\alpha\beta}_j$
with
respect to the fields $\hat{\eta}^\alpha_j$, $\eta^\alpha_j$,
$\hat{\eta}^\beta_j$ and $\eta^\beta_j$
and thus are
equal to
the local, connected responses and correlations in mean-field theory:
\bea
\label{u-mnpq}
u^{\alpha\beta}_{mnpq} &=& \frac{\partial}{\partial
\epsilon^\alpha(t_{m+1})}
\cdot \cdot \cdot
\frac{\partial}{\partial \epsilon^\alpha(t_{m+n})}
\frac{\partial}{\partial \epsilon^\beta(t_{m+n+p+1})}
\cdot \cdot \cdot
\frac{\partial}{\partial \epsilon^\beta(t_{m+n+p+q})}
\nonumber \\
& & \langle s^\alpha(t_1) \cdot \cdot \cdot
s^\alpha(t_m) s^\beta (t_{m+n+1}) \cdot \cdot \cdot
s^\beta (t_{m+n+p})\rangle_{f,l,c}\,.
\eea
As before, local \cite{NarayanI} ($l$) means that we do not vary the
local field
$(\eta^\alpha_0)_j$ in the mean--field equation
\begin{equation}
\label{mean_field_alpha}
\partial_t s^\alpha_j(t) = J (\eta^\alpha_0)_j(t)+H+f_j-
\frac{\delta V^\alpha}{\delta s^\alpha_j(t)}+ J \epsilon^\alpha (t)
\end{equation}
when we perturb the replica $\alpha$ with the infinitesimal force
$J \epsilon^\alpha(t)$ (and correspondingly for replica $\beta$).
The index $c$ to the average in \eq{u-mnpq} is a reminder that these
are {\it connected} correlation and response functions.
In the same way as we discussed in section \ref{sub:calculating-u-mn}
the force $J\epsilon^\alpha(t)$
is only allowed to {\it increase} with time
consistently with the history we have chosen.
{}From \eq{u-mnpq} one sees that $u_{0npq}=0$ if $n\neq 0$,
$u_{mn0q}=0$ if $q\neq 0$, and $u_{0n0q}=0$, just as we had
$u_{0n}=0$ in our earlier calculation for just one replica.

\subsection{Coarse graining transformation}

The coarse graining transformation is defined in the same way as in
the single replica case. In the appendix
\ref{ap:RG-details-avalanches} we give the
Feynman rules for loop corrections, and derive the canonical
dimensions of the various operators in the action.

\subsection{The scaling of the second moment of the avalanche
size distribution}
\label{scaling-of-s2-first}

In appendix \ref{ap:RG-details-avalanches} we derive the scaling
expression (\eq{s2-main1})
for the second moment of the avalanche size distribution
\be
\label{s2}
\langle S^2 \rangle \sim \int dt_1 \int dt_\alpha dt_\beta
d^dx_\alpha
d^dx_\beta
\langle
 \hat{s}^\alpha(t_\alpha,x_0) s^\alpha(t_0,x_\alpha)
\hat{s}^\beta(t_\beta,x_0)
s^\beta(t_1,x_\beta)
\rangle_f \, .
\ee
In order to find the scaling dimension of
 $\langle S^2 \rangle$
we need to know how
\be
\langle
 \hat{s}^\alpha(t_\alpha,x_0) s^\alpha(t_0,x_\alpha)
\hat{s}^\beta(t_\beta,x_0)
s^\beta(t_1,x_\beta)
\rangle_f
\ee
scales under coarse graining.
The topology of the diagrams permits no $O(\epsilon)$ loop
corrections to the corresponding vertex function.
Since the anomalous dimensions of the external legs ({\it i.e.}
Greens functions) in the two replicas are also zero at $O(\epsilon)$
it is sufficient to use the plain field rescalings
to extract the scaling behavior of $\langle \hat s^\alpha {\hat
s}^\beta s^\alpha s^\beta \rangle$ under coarse graining.
As shown in appendix \ref{ap:RG-details-avalanches}
one obtains
\be
\langle \hat s^\alpha {\hat s}^\beta s^\alpha s^\beta \rangle \sim
\Lambda^{(2 (d+z) -4)}
\ee
where $\Lambda$ is the cutoff in the momentum shell integrals.

Inserting this result into \eq{s2} along with the canonical
dimensions of the various times
$[t] \sim \Lambda^{-z}$ and coordinates
$[x] \sim \Lambda^{-1}$, one obtains
\be
\label{s2-units}
\langle S^2 \rangle \sim \Lambda^{-(4+z)}
\ee
(Formally including the anomalous dimensions
$\eta=\bar\eta=0 + O(\epsilon^2)$,
one obtains (to first order in $\epsilon$)
$\langle S^2 \rangle \sim \Lambda^{-(z+(2-\eta)2)}$.
Similarly, one finds for the higher moments
$\langle S^n \rangle \sim \Lambda^{-((n-1)z+(2-\eta)n)}$.)

On the other hand, from \eq{s2-scaling-integral}
we know that
$\langle S^2 \rangle \sim r^{(\tau-3)/\sigma}
{\cal S}_\pm^{(2)}(h/r^{\beta\delta})$.
If we use that $r$ has scaling units
$\Lambda^{1/\nu}$, and that $\tau-2=\sigma\beta(1-\delta)$
(see section \ref{sec:exponent-equalities}, and
\cite{eps-big2,thesis}),
we find by comparison with \eq{s2-units}
that
$1/\sigma=z \nu + (2-\eta) \nu$ to first order in $\epsilon$.
One gets the same relation from comparing the dimensions
for the $n$'th moment, which scales as
$\langle S^n\rangle \sim r^{(\tau-(n+1))/\sigma}
{\cal S}_\pm^{(n)}(h/r^{\beta\delta})$.

\subsection{Results}
We've seen that
\be
1/\sigma =z \nu + (2-\eta)\nu = 2 + \epsilon/3 + O(\epsilon^2).
\ee
If one inserts the result for $1/\sigma = z \nu + (2-\eta) \nu
= 2 + \epsilon/3 + O(\epsilon^2)$
into the relation $\tau-2=\sigma\beta(1-\delta)$, one obtains
\be
\tau = 3/2 + O(\epsilon^2) \, .
\ee
{}From the violated hyperscaling relation $1/\sigma = (d-\theta) \nu
-\beta$ one finds
\be
\theta\nu = 1/2 -\epsilon/6 + O(\epsilon^2)\, .
\ee

This concludes the perturbative approach to the problem.

\section{Comparison with numerical simulations in 2, 3, 4, and 5
dimensions}
\label{sec:numerics-and-experiment}

Figure \ref{fig:eps-vs-numerics}
shows a comparison between the theoretical predictions
for various exponents and their values as obtained from numerical
simulations in 2, 3, 4, and 5 dimensions.
These prepublication results are courtesy of Olga Perkovi\'c.
A complete list of the
numerical exponents that were measured in the simulations,
and a detailled description of the algorithm that allowed to
simulate systems with up to $1000^3$ spins
is given in a forthcoming publication \cite{Perkovic}.
A quantitative comparison of the results to experiments
can be found in \cite{thesis,eps-big}.
Some first results and conjectures about the behavior in
in two dimensions, which is likely the
lower critical dimension of our critical point,
are presented elsewhere \cite{hysterIII}.
As is seen in the figure, the agreement between the numerics and
the results from the $\epsilon$ expansion
is surprisingly good, even down to $\epsilon=3$.

%XXXXXXX figure fig:eps-vs-numerics XXXXXX

%XXXXXXX tab:simul-exponents  XXXXX

The numerical values in 3 dimensions for $\beta$, $\beta\delta$,
$\nu$ and $\eta$ seem to have overlapping error bars with the
corresponding exponents of the equilibrium RFIM.
Maritan {\it et al} \cite{Banavar} conjectured that the exponents
might be equal in a comment to our first publication \cite{hysterI}
on this system.
Why this should be is by no means obvious. The physical states
probed by the two systems are very different. While the equilibrium
RFIM will be in the lowest free energy state, our system will be
in a history dependent metastable state.
Nevertheless, as we have seen, the perturbation expansions for the
critical exponents can be mapped onto another to all orders in
$\epsilon$. In appendix \ref{ap:temperature} we discuss possible
connections between the two models that might become clear if
temperature fluctuations are introduced in our zero-temperature
avalanche model.

\vspace{1cm}
{\bf Acknowledgements}

We acknowledge the support of DOE Grant \#DE-FG02-88-ER45364 and NSF
Grant \#DMR-9118065. We would like to thank Olga Perkovi\'c very
much for providing the beautiful numerical results in
two, three, four, and five dimensions to which the analytical
results of this paper compare favorably.
Furthermore we thank
Andreas Berger, Lincoln Chayes, Daniel Fisher, Stuart Field,
Sivan Kartha, Bill Klein, Eugene Kolomeisky, James A. Krumhansl,
Onuttom Narayan, Mark Newman, Jordi Ort\'in, Antoni Planes,
Mark Robbins, Bruce Roberts, Jean Souletie,
Uwe Taeuber, Eduard Vives, and Jan von Delft for helpful conversations,
and NORDITA where this project was started.
This research was conducted using the resources of the Cornell Theory
Center, which receives major funding from the National Science
Foundation (NSF) and New York State.  Additional funding comes from
the Advanced Research Projects Agency (ARPA), the National
Institutes of Health (NIH), IBM Corporation, and other members of
the center's Corporate Research Institute. Further pedagogical
information using Mosaic is available \cite{ToCome}.

\appendix

\section{Hard-Spin and Soft-Spin Mean-Field Theory}
\label{ap:mean-field}

\subsection{Hard-Spin Mean-Field Theory}
\label{ap:hard-spin-mean-field}

In this appendix we derive the scaling forms near the critical point
for the magnetization and the avalanche size distribution in the
hard-spin mean-field theory. At the end we briefly discuss changes
of nonuniversal quantities for the soft-spin mean-field theory.

We start from the hard-spin mean-field Hamiltonian:
\be
{\cal H} = -\sum_i (J M + H + f_i ) s_i
\ee
where the interaction with the nearest neighbors from the short
range model \eq{model1} has been replaced by an interaction with the
average spin value or magnetization $M=(\sum s_i)/N$.
This would be the correct Hamiltonian if every spin
would interact equally strongly with every other spin in the lattice
{\it i.e.} for infinite range interactions.

\subsection{Mean-field magnetization curve}

Initially, at $H=-\infty$,
all spins are pointing down. The field is
slowly increased to some finite value $H$.
Each spin $s_i$ flips, when it gains energy by doing so, {\it i.e.}
when its local effective field
$h^{\it eff}_i= J M + H + f_i$ changes sign.
At any given field $H$ all spins with $h^{\it eff}_i<0 \Leftrightarrow
f_i< - JM -H$ will still be pointing down, and all spins with
$h^{\it eff}_i >0 \Leftrightarrow f_i > - JM -H$ will be pointing up.
Self-consistency requires that $M=\int \rho(f) s_i df$.
Therefore
\be
\label{mft-sc}
M=(-1) \int_{-\infty}^{-J M -H} \rho(f) df + \int_{-J M
-H}^\infty \rho(f) df =
1 - 2 \int_{-\infty}^{-J M -H} \rho(f) df
\ee
is the self-consistency equation from which we can extract the
mean-field magnetization as a function of $H$.
As in the main text $\rho(f) = \exp(-f^2/2 R^2)/(\sqrt{2 \pi} R)$
is the distribution of random fields.

For $R \geq \sqrt{2/\pi} J \equiv R_c $ the solution
$M(H)$ of \eq{mft-sc} is analytic at all values of $H$.
For $R=R_c$ there is a critical magnetic field $H_c(R_c)=0$
where the
magnetization curve $M(H)$ has diverging slope.
For $R < R_c$ the solution $M(H)$ is unique only for $H$ outside a certain
interval $[H_c^l(R), H_c^u(R)]$.
For $H$ in the range between the two ``coercive fields''
$H_c^l(R)$ and $H_c^u(R)$, the equation has three solutions,
two that are stable and one that is unstable.
Unlike equilibrium systems, which will always occupy the
solution with the lowest overall free energy, our nonequilibrium
(zero temperature) system is forced by the {\it local} dynamics
to stay in the current local energy minimum until it is destabilized
by the external magnetic field. For increasing external magnetic
field this implies that the system will always occupy the
metastable state with the lowest possible magnetization.
Conversely, for decreasing external magnetic field, the system will
occupy the metastable state with highest possible magnetization.
One obtains a hysteresis loop
for $M(H)$ with a jump, or ``infinite avalanche''
at the upper and lower coercive fields
$H_c^u(R)$ and $H_c^l(R)$ respectively (see figure
\ref{fig:MFT-magnetization}).

{}From \eq{mft-sc} follows that $dM/dH=2 \rho(x)/(-t(x))$ with
$x=-JM-H$ and
\be
\label{t-hard-def}
t(x)=2J \rho(-JM -H) -1 \, .
\ee
$dM/dH$ diverges if $t(x_c) =0$, which defines $x_c$.
For $R>R_c$ the slope $dM/dH$ is
always finite, and $t<0$ at all $H \in [-\infty, +\infty]$. For $R
\leq R_c$ however the condition $t(x_c) = 0$ is fulfilled at the
critical field $H_c(R)$. To obtain potential scaling behavior near
this point, we expand $t(x)$ around $x_c =-JM(H_c(R)) - H_c(R)$
\be
\label{t-expand}
t=2J\rho(x) -1 = 2 J (\rho(x_c)-1) +
\rho'(x_c) (x-x_c) + 1/2 \rho''(x_c) (x-x_c)^2 + \cdot \cdot \cdot
\, ,
\ee
where
\be
\label{rho-c}
2 J \rho(x_c) -1 = 0 \, .
\ee
Then
\be
\label{dM-dH-xc}
\chi \equiv dM/dH  = (-\rho(x_c))/(J (\rho'(x_c) (x-x_c) +
1/2 \rho''(x_c) (x-x_c)^2
 + \cdot \cdot \cdot)) \, .
\ee
For a general analytic distribution of random fields $\rho(x)$
 with one
maximum with nonvanishing second derivative ($\rho''(x_c) <0$)
this suggests two different scaling behaviors corresponding to the
cases $\rho'(x_c) = 0$ and $\rho'(x_c) \neq 0$.

Let us consider the case $\rho'(x_c)=0$ first.
For a Gaussian distribution of width $R\equiv R_c$
with zero mean this implies that
$x_c=-JM(H_c)-H_c=0$ and consequently
$\rho(x_c) = 1/(\sqrt{2 \pi} R_c)$ and
$\rho''(x_c) = 1/(2 \pi R_c^3)$.
With
\eq{rho-c} one obtains
$R_c = \sqrt{2/\pi} J$.
This is in fact the largest possible value of $R$ for which $M(H)$
has a point of diverging slope.

Integrating \eq{dM-dH-xc} leads to a cubic equation for $M$ and the
leading order scaling behavior
\be
\label{mean-field-scaling-M-H}
M(r,h) \sim |r|^\beta {\cal M}_\pm(h/|r|^{\beta \delta}) \, ,
\ee
for small $h=H-H_(R_c)$ and $r=(R_c-R)/R$.
In mean-field theory
$\delta = 3$ and ${\cal M}_\pm$ is given by the smallest real root
$g_\pm (y)$ of the cubic equation
\be
\label{cubic-equation}
g^3 \mp {12 \over \pi } g-{ 12 \sqrt 2 \over \pi^{3/2} R_c } y = 0\,,
\ee
where here and throughout  $\pm$ refers to the sign of $r$.

The other case ($\rho(x_c) =1/(2J)$ and $\rho'(x_c) \neq 0$) is
found for distributions with $R<R_c$. Integrating \eq{dM-dH-xc}
with $x_c=-JM(H_c(R))-H_c(R)$
yields a quadratic equation for the
magnetization and the scaling behavior
\be
\label{mean-field-scaling-M-H-R<Rc}
M-M(H_c(R)) \sim (H-H_c(R))^\zeta
\ee
with $\zeta = 1/2$ for $H$ close to $H_c(R)$.
{}From \eq{mft-sc} and \eq{rho-c} one finds $H_c(R_c)=0$, $H_c(R)\sim
r^{\beta \delta}$ for small $r>0$, with $\beta \delta = 3/2$
and $H_c(R=0)=J$.
The corresponding phase diagram was shown in figure
\ref{fig:MFT-phase-diagram}.

Note that the scaling results for $R$ close to $R_c$
as given in \eq{mean-field-scaling-M-H}
remind one of the scaling results of the Curie-Weiss mean-field theory
for the equilibrium Ising model near the
Curie-temperature ($T=T_c$). For $T<T_c$ however,
the equilibrium model has a discontinuity in the magnetization at
$H=0$, while
for $R<R_c$ our model displays a jump in the magnetization
at a (history dependent) nonzero magnetic field $H_c(R)$,
where the corresponding metastable solution becomes unstable.
Our infinite avalanche line $H_c(R)$
is in fact similar to the spinodal
line in spinodal decomposition \cite{Goldenfeld}.

Note also that this mean-field theory does not show any hysteresis
for $R\geq R_c$ (see figure \ref{fig:MFT-magnetization}).
This is only an artifact of its particularly simple
structure and not a universal feature.
For example, the continuous spin model, which we use for the
$RG$ description in section \ref{sec:analytic-description}
has the same exponents in mean-field
theory, but shows hysteresis for {\it all} disorders $R$, even for
$R>R_c$. In that case there are two critical fields $H_c^u(R_c)$ and
$H_c^l(R_c)$ --- one for each branch of the hysteresis curve
(see figure \ref{fig:MFT-magn-soft-spin}, and figure
\ref{fig:MFT-soft-spin-phase-diagram}).

\subsection{Mean-field avalanche-size distribution}

As we have already discussed in the main text, one finds avalanches
of spin flips as the external field is raised adiabatically.
Due to the ferromagnetic interaction a flipping spin may cause some
of its nearest neighbors
to flip also, which may in turn trigger some of
their neighbors, and so on.
In mean-field theory, where all spins act as nearest neighbors with
coupling $J/N$, a spin flip from $-1$ to $+1$ changes the effective
field of {\it all} other spins by $2 J/N$.
For large N, the average number of secondary
spins that will be triggered to
flip in response to this change in the effective local field is then
given by
$n_{trig}\equiv(2 J/N) N\rho(-J M -H) = 2 J \rho(-JM -H)$.
If $n_{trig}<1$, any avalanche will eventually peter out,
and even in an infinite system all avalanches will only be
of finite size.
If $n_{trig}=1$, the avalanche will be able to sweep the
whole system, since each flipping spin triggers on average
one other spin. This happens when the magnetic field $H$ takes a
value at the infinite avalanche line $H=H_c(R)$, with $R\leq R_c$,
since $n_{\it trig}=1$ is equivalent to $t=0$ (see \eq{t-hard-def}).

Considering all possible configurations of random fields,
there is a probability distribution for
the number $S$ of spins that flip in an avalanche.
It can be estimated for avalanches in large
systems, i.e. for $S<<N$:
For an avalanche of size $S$ to happen, given that the primary spin has
random field $f_i$, it is {\it necessary}
that there are exactly
$S-1$ secondary spins with their random fields in the interval
$[f_i, f_i+2 (J/N)S]$.
Assuming that the probability density of random fields is
approximately
constant over this interval, the probability $P(S)$ for
a corresponding configuration of random fields
is given by the Poisson distribution, with the average value
$\lambda =2 J S \rho(-JM-H)= S(t+1)$, where
$t\equiv 2J\rho(-JM-H)-1$:

\be
\label{Ps}
P(S) = {\lambda^{(S-1)}\over{(S-1)!}} \exp(-\lambda) \, .
\ee
This includes cases in which the random fields of the $s$
spins are arranged in the interval $[f_i, f_i+2S (J/N)]$
in such a way
that they do not flip in one big avalanche, but rather in two
separate avalanches triggered at slightly different external
magnetic fields.
Imposing periodic boundary conditions on the interval $[f_i, f_i+2S
(J/N)]$ one can see that for any arrangement of the random fields
in the interval there is exactly one spin which can trigger the
rest in one big avalanche.
In $1/S$ of the cases,
the random field of this particular spin to trigger the avalanche
will be the one with the lowest random field, as desired.
Therefore we need to multiply $P(S)$ by $1/S$ to obtain the
probability $D(S,t)$ for an avalanche of size $S$ starting with a
spin flip at random field $f_i = -JM -H$
\be
\label{exact-Ds}
D(S,t) = S^{(S-2)}/(S-1)!(t+1)^{(S-1)} e^{-S(t+1)} \, .
\ee
With Stirling's formula we find for large $S$ the scaling form
\be
\label{Ds-t}
D(S,t) \sim {1\over\sqrt{2\pi}S^{3/2}}\exp(-S t^2/2) \, .
\ee
To obtain the scaling behavior
near the two different critical points, we will insert into the
expression in \eq{Ds-t}
the expansion
of $t(x)$ around $x_c$ from \eq{t-expand}.

\subsection{Avalanches near the critical endpoint}
\label{aval_near_crit_endpoint}
Near the critical point $(R_c, H_c(R_c))$, where $x_c=0$ and
$\rho'(x_c)=0$ we obtain $t= 2 J (\rho(0)-1) + J \rho''(0) (-JM
-H)^2$, which (by equation \eq{mean-field-scaling-M-H}) leads to
the scaling form
\be
\label{t-scaling1}
t\sim r[1 \mp 1/4 \pi
g_{\pm}(h/|r|^{\beta\delta})^2] \,.
\ee
$g_{\pm}$ was defined in \eq{cubic-equation}, and $\pm$ again
refers to
the sign of $r=(R_c-R)/R$.
Inserting the result into \eq{Ds-t} one obtains
\be
\label{scaling-Ds}
D(S,r,h) \sim S^{-\tau}{\cal D}_\pm(S/|r|^{-1/\sigma},
h/|r|^{\beta\delta}) \, ,
\ee
with the mean--field results $\tau =3/2$, $\sigma=1/2$,
$\beta\delta=3/2$,
and the mean-field scaling function
\be
{\cal D}_\pm(x,y)={1\over\sqrt{2\pi}}e^{-x{\left[ 1 \mp {\pi \over
4 }
g_\pm (y)^2 \right]^2/2}} \, .
\ee

\subsection {Mean-field avalanche size distribution near the
$\infty$-avalanche line (``spinodal
line'')}

For $R<R_c$ one has $\rho'(x_c) \neq 0$, so that the expansion
for $t$ becomes
\bea
t&=&2J(\rho(x_c) -1 ) + 2J \rho'(x_c) (x-x_c) + \cdot \cdot \cdot
= 2 J \rho'(x_c) (x-x_c) \\
\nonumber
&=& 2 J \rho'(x_c) (-J(M-M(H_c(R))) -(H-H_c(R)))
\eea
Following the steps that led to \eq{mean-field-scaling-M-H-R<Rc}
we arrive at
\be
t= -2 \sqrt{ J \rho'(x_c) (H-H_c(R))} + {\it higher
\,\,orders\,\,in\,\,} (H-H_c(R))
\ee
so that for $H$ close to the onset to infinite avalanche
(with $H \leq H_c^u(R)$ for increasing field $H$ and $H>H_c^l(R)$
for decreasing field)
\be
\label{spinodal-Ds}
D(S, (H-H_c(R))) \sim {1\over\sqrt{2\pi} S^{3/2}}
\exp{\{-2 [\rho'(-JM-H) J]
S |H-H_c(R)| \}} \, .
\ee
or, written more generally,
\be
\label{scaling-spinodal-Ds}
D(S,H-H_c(R)) \sim 1/S^\tau \bar{\cal F} (S |H-H_c(R)|^{1/\kappa})
\ee
with $\kappa=1$ and $\tau =3/2$ in mean-field theory,
and $\bar{\cal F}$ the
corresponding mean-field scaling function.

\subsection{Modifications for the soft-spin mean-field theory}
\label{ap:eta0}

In section \ref{sec:analytic-description} we have for calculational
convenience switched from the hard-spin model, where each spin $s_i$
could only take the values $\pm 1$, to a soft-spin model, where $s_i$
can take any value between $-\infty$ and $+\infty$.
In realistic systems these soft-spins can be considered as
coarse grained versions of the elementary spins.
The corresponding Hamiltonian with the newly introduced double-well
potential
\vspace{.3cm}
\be
V(s_i) = \left\{ \begin{array}{ll}
                            k/2~(s_i+1)^2 & \mbox{for $s<0$} \\
                            k/2~(s_i-1)^2 & \mbox{for $s>0$}
                            \end{array}
                   \right.
\ee
to mimic the two spin states of the hard-spin model, was given in
\eq{model}.
In the mean-field approximation, where the coupling term $-J_{ij}
s_i s_j$ is replaced by $-\sum_i J M s_i$ with $M=\sum_j s_j/N$, we
obtain
\be
{\cal H} = -\sum_i \left\{(JM + H + f_i) s_i - V(s_i) \right \} \, .
\ee
For adiabatically increasing external magnetic field the local
dynamics introduced earlier implies that each spin will be negative
so long as the lower well Hamiltonian
\be
{\cal H}_{-} \equiv k/2~(s_i+1)^2 - (H+JM+f_i) s_i
\ee
does have a local minimum with $\delta {\cal H}/\delta s =0$
for negative $s_i$. This implies that $s_i < 0 $ if
\be
\frac{\delta}{\delta s_i} [k/2~(s_i+1)^2 - (H + f_i + JM) s_i]_{s=0}
\geq
0 \, ,
\ee
else $s_i$ will be stable only at the bottom of the positive potential well,
where
\be
\frac{\delta}{\delta s_i} {\cal H}_{+}
= \frac{\delta}{\delta s_i} [k/2~(s_i-1)^2 - (H+JM+f_i) s_i] =
0 \, .
\ee
We conclude that for the given history
\be
 \left\{ \begin{array}{ll}
                            s_i \leq 0 & \mbox{for $f_i \leq - JM -
H + k$} \\
                            s_i >0 & \mbox{for $f_i > - JM - H + k$}
\,.
                            \end{array}
                   \right.
\ee
\noindent
{}From the self-consistency condition
\be
\langle s_i \rangle \equiv \int \rho(f_i) s_i df_i = M
\ee
we derive the mean-field self-consistency equation for the
{\it soft-spin} magnetization $M_u$ (for {\it increasing} external
magnetic field):
\be
\label{soft-sc-2}
M_u(H) = (k+H)/(k-J) - 2k/(k-J) \int_{-\infty}^{-JM-H+k} \rho(f) df \, .
\ee
Correspondingly one finds for the branch of {\it decreasing} external
magnetic field:
\be
\label{soft-sc-3}
M_l(H) = (k+H)/(k-J) - 2k/(k-J) \int_{-\infty}^{-JM-H-k} \rho(f) df \, .
\ee
Figure \ref{fig:MFT-magn-soft-spin} shows the corresponding hysteresis
loops in
the three disorder regimes $R<R_c=\sqrt{2/\pi} J (k/(k-J))$, where
the hysteresis loop has a jump,
$R=R_c$, where the jump has shrunk to a single point of infinite
slope $dM/dH$, and $R>R_c$, where the hysteresis loop is smooth.
In contrast to the hard spin model, this model displays hysteresis even for
$R \geq R_c$.

The critical magnetic fields $H_c^u(R)$ and $H_c^l(R)$
at which the slope of the static
magnetization curve diverges are found by differentiating
\eqs{soft-sc-2} and (\ref{soft-sc-3})
with respect to $H$ and by solving for $dM_{\it stat}/dH$. One
finds (for increasing external magnetic field) that
$dM_{\it stat}/dH \sim 1/\tau$ with
\be
\label{t-def-soft-spin-mft}
\tau = (2 k/(k-J)) J \rho(-JM_{\it stat} -H +k) -1 \, .
\ee
$\tau$ is defined analogously to the paramter $t$ in
\eq{t-hard-def}.
(It is worth mentioning that $t=\chi^{-1} k/(J(k-J))$,
where $\chi^{-1}$ is the constant term in the propagator
of the RG treatment, which was introduced in \eq{chi} in the
main text.) The critical field $H_c(R)$ is given by the solution to
the condition $\tau=0$.
To find the scaling
behavior near the critical point one can expand \eq{soft-sc-2}
around $H_c^u(R)$, and correspondingly \eq{soft-sc-3} around
$H_c^l(R)$.
For increasing external magnetic field
the critical point $R=R_c$, $H=H_c^u(R_c)$ and
$M=M_c\equiv M_u(H_c^u(R_c))$
is characterized by the equation $\tau=0$ and $\rho'(-JM_c-H_c +k)
=0$, {\it i.e.} $-JM_c-H_c +k=0$.
It follows that $R_c=\frac{1}{\sqrt{2\pi}} \frac{2kJ}{k-J}$.
Inserting these
results into \eq{soft-sc-2} one obtains $M_c^u=1$ and
$H_c^u(R_c) = k-J$.
Similarly for a decreasing external magnetic field one finds
$H_c^l(R_c) = -(k-J)$ and $M_c^l = M_l(H_c^l(R_c)) =-1$
The corresponding modified phase diagram is depicted
in figure \ref{fig:MFT-soft-spin-phase-diagram},
with $H_c^u(R=0) = +k$ and $H_c^l(R=0) = -k$.

In the same way as discussed in appendix
\ref{ap:mean-field} for the (static)
hard-spin model, expanding \eqs{soft-sc-2} and (\ref{soft-sc-3})
around $M_c$, $H_c$ and
$R_c$ yields cubic equations for the magnetization and one obtains
the scaling behavior near the critical point
\be
m \sim h^{1/3}
\ee
for $m\equiv M-M_c$ and $h\equiv H-H_c(R_c)$.
Consequently the slope of the magnetization as a function
of $h$ diverges at the
critical point
\be
dm/dh \sim h^{-2/3} \, .
\ee

It turns out that in fact none of the universal scaling features we
discussed for the hard spin model is changed.
The mean-field critical exponents
$\beta$, $\delta$, $\tau$ and $\sigma$, and the scaling forms near
the critical point are the same as in the hard-spin model.
(A ``spin-flip'' in the hard spin model corresponds to a spin moving
from the lower to the upper potential well in the soft-spin model.)
Apart from modifying some nonuniversal constants,
the new parameter $k>J$ in
the definition of the soft-spin potential does not appear to change
the calculation in any important way.

\subsubsection{Soft-Spin Mean-Field Theory at Finite Sweeping
Frequency $\Omega$}
\label{ap:eta0-timedependence}

In section \ref{sub:formalism}, \eq{response-in-local-mft},
we have derived the following equation of motion for
each spin in the dynamical soft-spin mean-field theory, as the
external magnetic field $H(t) = H_0 + \Omega t$ is slowly increased
\begin{equation}
\label{response-in-local-mft-2}
\frac{1}{\Gamma_0} \partial_t s_j(t) = J \eta^0_j(t)+H+f_j-
\frac{\delta V}{\delta s_j(t)}+ J \epsilon (t) \, .
\end{equation}
With the definition of the potential $V$ from \eq{V-def}
this becomes
\be
\label{response-in-local-mft-3}
        1/(\Gamma_0 k) \partial_t s_j(t) = - s_j(t) + J
\eta_j^0(t)/k +
        H(t)/k + f_j/k
        + sgn(s_j) + J \epsilon(t)/k \, .
\ee
{}From \eq{saddle-point2} we know that $\eta^0(t)=\langle s \rangle
\equiv
M(t)$ is the time dependent mean-field magnetization of the system.
It can be calculated
by taking the random-field average of \eq{response-in-local-mft-3}
and solving the resulting equation of motion
for $\eta^0(t)$.
One can show \cite{thesis} that
for driving rate $\Omega/k$ small compared to the
relaxation rate $k\Gamma_0$ of the system, for
all values of $H_0$ the solution
$\eta^0(t)$ can be expanded in terms of $(\Omega/\Gamma_0)$ in the
form
\be
\label{eta0-expand}
\eta^0(t) \equiv M(t) = M_{\it stat}(H_0) + (\Omega/\Gamma_0)^{p_1}
f_1(H_0) t + (\Omega/\Gamma_0)^{p_2} f_2(H_0) t^2 + \cdot \cdot \cdot
\ee
with $0<p_1 < p_2 < \cdot \cdot \cdot$. The $p_i$ depend on whether
$R<R_c$ or $R=R_c$.
$M_{stat}(H_0)$ is the
solution of the static mean-field theory equation (\ref{soft-sc-2})
for the given history.
If the series converges for $\Omega \rightarrow 0$, it follows that
$\eta^0(t)$ approaches the {\it constant} magnetization
$M_{\it stat}(H_0)$ in the adiabatic limit.
This is certainly expected
for $H_0$ {\it away} from the critical field $H_c(R)$,
where the static magnetization is non-singular:
as $\Omega$ tends to zero the time dependent
magnetization $M(t)$ simply lags less and less behind the static
value $M_{\it stat}(H(t))$.
The magnetization $M(t)$ can be expanded as $M(t)=M_{stat}(H_0) +
[dM/dH]_{H_0} \Omega t + \cdot \cdot \cdot$
and converges towards $M_{stat}(H_0)$ as $\Omega \rightarrow 0$, as
long as all derivatives
$[d^n M_{\it stat}/dH^n]_{H_0}$ are well defined and finite. This
argument
however does not obviously hold at the critical fields $H_0=H_c(R)$
with $R \leq R_c$, where $dM_{\it stat}/dH$ and all higher
derivatives diverge. Using boundary
layer theory one can show \cite{thesis}
that even at these singular points
$M(t)$ converges toward
its static limit $M(H_c(R))$ as $\Omega \rightarrow 0$, though with
power laws smaller than one in $\Omega$, as indicated in
\eq{eta0-expand}.
% XXXX \footnote{For
%$R<R_c$ in mean-field theory
%the duration of the infinite avalanche remains finite as $\Omega
%\rightarrow 0$, even for an infinite system. This is surely an
%unphysical artifact of the infinite range interactions. In a
%physical system with finite range coupling the
%infinite avalanche should rather be studied in terms of interface
%propagation as has been done in previous work
%\cite{Nattermann,NarayanIII,Robbins}
%(see also appendix \ref{ap:inf-aval-line}).
%In fact, the formalism used here to expand around mean-field theory
%describes only low-frequency behavior rather than fast events that
%take a time of $O(1/\Gamma_0)$. The formalism should apply, however,
%at least
%up to the onset point of the infinite avalanche $H_c(R)$ for
%$R<R_c$.}
Since we use $M_{stat}(H_0)$ as the foundation for our
$\epsilon$-expansion, this is reassuring.

%%%%%%%%%%%%%%%%%%%%%%%%%%%%%%%%%%%%%%%%%%%%%%%%%%%%%%%%%%%%%%%%%%%%%%%
%% This is a REVTEX file, which hopefully will run on your system too.%
%%%%%%%%%%%%%%%%%%%%%%%%%%%%%%%%%%%%%%%%%%%%%%%%%%%%%%%%%%%%%%%%%%%%%%%

\section{Tilting of the Scaling axes}
\label{ap:turning-of-axes}

In the appendix on mean-field theory we derived scaling forms for
the magnetization and the avalanche size distribution, which
depended on the (mean-field) scaling fields $r=(R-R_c)/R$ and
$h=H-H_c$. In finite dimensions however, the corresponding scaling
forms may depend not on $r$ and $h$, but on rotated variables
\be
\label{r'}
r'= r+ a h
\ee
and
\be
\label{h'}
h' = h + b r \, .
\ee
The amount by which the scaling axes $r'=0$ and $h'=0$ are tilted
relative to the $(r,0)$ and $(0,h)$ direction in the $(r,h)$ plane
is a nonuniversal quantity and has no effect on the critical
exponents. Nevertheless it can be important in the data analysis.
(The numerical results in 3, 4, and 5 dimensions do indeed seem to
indicate a slight tilting \cite{Perkovic}.)

\subsection{Extracting $\beta$ and $\beta\delta$ from the magnetization
curves}
With $m(r',h')=M(r',h')-M_c$, where $M_c=M(0,0)$ is
the magnetization at the critical point, we obtain the scaling
form
\be
\label{m'-def}
m \sim (r')^\beta {\cal M}_\pm(h'/(r')^{\beta \delta}) \, .
\ee
In simulations and experiments however the magnetization is given as
a function of $r$ and $h$ rather than $r'$ and $h'$. We rewrite
\eq{m'-def} in terms of $r$ and $h$ by inserting the definitions
of $r'$ and $h'$ from \eq{r'} and \eq{h'} and use the fact that
$\beta \delta >1$ in the cases we are considering. One obtains
for the leading order scaling behavior
\be
M-(M_c +c r) \sim r^\beta {\tilde{\cal
M}}_\pm((h+br)/r^{\beta\delta})
\ee
(where $c$ is a constant and ${\tilde{\cal M}}_\pm$ is the
appropriate
function). Equivalently this can be written as
\be
dM/dH \sim r^{\beta-\beta\delta} {\tilde{\cal
M}}_\pm((h+br)/r^{\beta\delta}) \, .
\ee

To extract the critical exponents $\beta$ and $\beta\delta$ one can
then use collapses of $dM/dH$ in the same way as if the
mean-field scaling forms were valid in finite dimensions,
except for the presence of a new tilting parameter $b$, that has
to be varied to its correct value simultaneously with $\beta$ and
$\beta \delta$ to find the best collapse of the data curves.
(A more detailed description of the procedure is given in reference
\cite{Perkovic}).

\subsection{Extracting the correlation length exponents $\nu$ and
$\nu/(\beta \delta)$}

Similarly to \eq{m'-def} the scaling for the correlation length in
finite dimensions takes the form
\be
\xi \sim (r')^{-\nu} {\cal Y}_\pm (h'/(r')^{\beta\delta}) \, ,
\ee

Thus $\xi \sim (h')^{-\nu/(\beta\delta)}$ along $r'=0$ and
$\xi \sim (r')^{-\nu}$ along $h'=0$.
Figure \ref{fig:contour-lines} shows contour lines
in the $(r',h')$ plane for the correlation length.
Since $\nu/(\beta\delta)<\nu$, $\xi$ changes faster in the $(0,h')$
direction than in the $(r',0)$ direction.
This implies that the correlation length diverges with the dominant
exponent $\nu/(\beta\delta)$ when the critical
point is
approached along any direction other than $h'=0$.
This can be used to extract $\nu/(\beta\delta)$ from collapses of
numerical or experimental curves for the correlation funcion
$G(x,h,r)$ measured as a
function of $h$ at fixed $R=R_c$:
\be
G(x,h,r=0) \sim 1/x^{(d-2+\eta)} {\tilde{\cal G}}_\pm(x
h^{-\nu/(\beta\delta)}, 0)
\ee
with the appropriate scaling function ${\tilde{\cal G}}_\pm$.
(Even in 3 dimensions $\nu/(\beta\delta) < \nu$ is still correct and
the
tilting is small.)

%XXXXXXXX figure fig:contour-lines XXXXXXX
%\begin{figure}
%\vbox to 5in{
% \vfil
% \special{psfile=contour.xfig.ps
%          hoffset=-400
%          voffset=-100
%          hscale=70
%          vscale=70}
%}
%\caption
%[Contour lines for the correlation length in the $(R-R_c,H-H_c)$
%plane (schematic).]
%{{\bf Contour lines for the correlation length}
%in the $(r',h')$ plane (schematic).
%The tilted coordinate axes indicate the physical directions
%$(r,h)$. Since $\nu/(\beta\delta)<\nu$, the
%correlation length $\xi$ changes faster in the
%$(0,h')$ direction than in the $(r',0)$ direction.
%\label{fig:contour-lines}}
% \end{figure}

On the other hand it seems rather difficult to find the weaker
scaling direction $(r',0)$ accurately enough to be able to
extract $\nu$ by approaching the critical point along this line.
If, instead, we integrate the
correlation function or the avalanche size distribution over $h'$,
we obtain a scaling form that depends only on $r'$.
{}From collapses of the integrated functions, the exponent $\nu$ can
then be extracted, even without knowledge of the exact
size of the tilting
angle.
Practically, in the analysis of our simulation data we actually
integrate over the magnetic field $H$ rather than $h'$.
On long length scales this is equivalent to an integration over $h'$:
The integration path $r=const\equiv c$ is written in the $(h',r')$
plane through $r'=a_1 c + b_1 h$ and $h'=a_2 c + b_2 h$.
Under coarse graining the
$h'$ component grows faster than the $r'$ component, so that
after several coarse graining steps
the integration path is deformed
into a straight line parallel to $h'$.
This procedure yields the dominant
contribution to an integrated scaling function
$G_{\it int}(x,r) \equiv \int dh G(x,r,h)$
in the region near the critical point.
It allows us to treat $r$ as a scaling field and to extract the
weaker exponent $\nu$: we have
$\xi \sim r^{-\nu}$ after integration over the hysteresis loop.
For the integrated avalanche correlation function, for example,
this means
$G_{\it int}(x,r) \sim 1/x^{d+\beta/\nu} {\cal G}^{\it
int}_\pm(x r^{-\nu})$, with the appropriate scaling function
${\cal G}^{\it int}_\pm$.

The rotation of the scaling axes (see also \cite{Perkovic})
is not apparent from the
$\epsilon$-expansion. For the renormalization group calculation
we have linearized the coarse graining transformation around the
fixed point. The rotation of the scaling axes is due to to
nonlinear corrections introduced during the flow to the fixed point.

{}From the mapping of the perturbation series for the critical
exponents in our model to the perturbation series for the critical
exponents in the equilibrium random field
Ising model, one might expect that for $R \geq R_c$ the
$H \rightarrow (-H)$ symmetry of the fixed point in the
thermal random field Ising model is mapped
to an $(h') \rightarrow (-h')$ symmetry in our problem.
(Nonperturbative corrections might destroy this of course.)
Note that this symmetry only emerges on long
length scales, near
the critical point, where the mapping to the thermal random field
Ising model holds.

%%%%%%%%%%%%%%%%%%%%%%%%%%%%%%%%%%%%%%%%%%%%%%%%%%%%%%%%%%%%%%%%%%%%%%%
%% This is a REVTEX file, which hopefully will run on your system too.%
%%%%%%%%%%%%%%%%%%%%%%%%%%%%%%%%%%%%%%%%%%%%%%%%%%%%%%%%%%%%%%%%%%%%%%%

\section{Some details of the RG calculation}
\label{ap:RG-details}
\subsection{Calculating some $u_{mn}$ coefficients}
\label{sub:calculating-some-u-mn}

In section \ref{sub:formalism} in \eq{u-mn}
we have given an expression for the coefficients $u_{mn}$ in the
expansion around mean-field theory:
\be
\label{u-mn2}
u_{m,n}= \frac{\partial}{\partial \epsilon(t_{m+1})} \cdot \cdot
\cdot
\frac{\partial}{\partial \epsilon(t_{m+n})} \langle
(s(t_1)-\eta^0(t_1)) \cdot \cdot \cdot
(s(t_m)-\eta^0(t_m)) \rangle_{l,\hat{\eta}^0, \eta^0} \, ,
\ee
where $s_j(t)$ is the solution of the {\it local} mean-field
equation
\be
\label{response-in-local-mft-4}
        1/(\Gamma_0 k) \partial_t s_j(t) = - s_j(t) + J
\eta_j^0(t)/k +
        H(t)/k + f_j/k
        + sgn(s_j) + J \epsilon(t)/k \, .
\ee
We need to insert the solution for $\eta^0(t)$ from appendix \ref{ap:eta0}
into \eq{response-in-local-mft-4} to calculate the higher
response and correlation functions $u_{mn}$ as given in \eq{u-mn2}.

\subsubsection{The coefficient $u_{1,0}$}

The vertex $u_{1,0}$ is defined as
\be
u_{1,0}(t) = \langle s_j(t) \rangle_l - \eta^0(t) \,.
\ee
{}From the stationary phase equation \eq{saddle-point2}
we have $\eta^0(t) = \langle s_i(t) \rangle_l$. Therefore
(by construction)
\be
u_{1,0}(t) = 0 \, .
\ee

\subsubsection{The coefficient $u_{1,1}(t_1,t_2)$}

As is shown in appendix \ref{ap:eta0}, $\eta^0(t)$ can be expanded in terms
of $\Omega$ (at least for $R\geq R_c$ and for $R<R_c$ before the
jump up to $H_c(R)$):
\be
        \eta^0(t) = M(H_0) + \Omega^p t + \cdot \cdot \cdot \, ,
\ee
where $M(H_0)$ is the static magnetization, $p>0$, and
$\cdot \cdot \cdot$ implies higher orders in $\Omega$.
Inserting this expansion into \eq{response-in-local-mft-4}
and \eq{u-mn2} allows us to calculate the coefficients $u_{mn}$
perturbatively in $\Omega$. Only the lowest order remains as
$\Omega \rightarrow 0$.
The vertex function $u_{11}(t_1,t_2)$ is then given by
\be
        lim_{\Omega \rightarrow 0}[\partial_t(
        lim_{\epsilon \rightarrow 0} \langle s(t_2)|_{H(t_2) + J
\epsilon
        \theta(t_2-t_1)} - s(t_2)|_{H(t_2)} \rangle_f /\epsilon)] \, .
\ee

To evaluate $s(t_2)|_{H(t_2) + J \epsilon \theta(t_2-t_1)}$ from the
equation of motion \eq{response-in-local-mft-4} we have to consider
three cases:
\vspace{2mm}

1. Neither $s_j(t)|_{H(t)}$ nor $s(t)|_{H(t) + J \epsilon
\theta(t-t_1)}$ flip at any time $t$ with $t_1 < t < t_2$.
To lowest order in $\Omega$
the solution of \eq{response-in-local-mft-4} is then given
by simple relaxation:
\be
\label{simple-relaxation}
        s_j^{no\, flip} (t) |_{H(t)+J\epsilon \theta(t-t_1)} -
	s_j(t)|_{H(t)}
        = (J \epsilon/k) (1 - \exp{[-k \Gamma_0(t-t_1)]})
	\theta(t-t_1) \, .
\ee

\vspace{2mm}
2. The unperturbed $s_j(t)|_{H(t)}$ flips at some time $t_J$ with
$t_1\leq t_J<t_2$. The fraction of spins for which this is the case
is proportional to $\Omega$. Thus, for $\Omega \rightarrow 0$
these spins will yield no contribution to $u_{11}$ for finite
$t_2-t_1$.( This is true even if one chooses to keep $H(t_2) - H(t_1)
\equiv \Delta H_0 \neq 0$ fixed as $\Omega \rightarrow 0$.
One finds that the resulting contribution to $u_{11}$ involves
terms of the form $\exp[-k \Gamma_0 \Delta H/\Omega]$
which clearly vanish as $\Omega \rightarrow 0$.)

3. The unperturbed spin $s_j(t)|_{H(t)}$ does not flip at any time
between $t_1$ and $t_2$, but the perturbed spin $s(t)|_{H(t) + J
\epsilon \theta(t-t_1)}$ does flip at time $t_J$ with $t_1 \leq t_J
<t_2$.
For fixed, finite $t_2-t_1$ we can again expand the contribution
$\Delta s$ to
$u_{11}$ in terms of $\Omega$. In the adiabatic limit at fixed,
finite $t_2-t_1$ only the lowest order term survives,
which is independent of $\Omega$.
It is calculated at constant $\eta^0(t_2) =\eta^0(t_1)$ and
$H(t_2) = H(t_1) \equiv H$, as done in the following paragraph:

The time $t_J$ with $t_1\leq t_J <t_2$
at which a spin moves from the lower to the upper
well is given by
\be
\label{flip}
s(t_J)|_{H + J\epsilon \theta(t_J-t_1)} = 0 \, .
\ee
Before the perturbation is switched on, the static solution
for the spin in the lower potential well is
\be
\label{static-initial}
        s_i(t_1) = J \eta^0/k + H/k + f_i/k -1 \, .
\ee
As soon as the perturbation is switched on, the spin starts relaxing
into its new equilibrium position according to
\eq{simple-relaxation}.
Its value will remain negative until time $t_J$, when it will
flip to a positive value in the upper potential well.
Thus (using \eq{static-initial} and \eq{simple-relaxation})
\be
\label{tJ-definition}
        0=s_i(t_J) = (1-\exp(-k(t_J-t_1)))J\epsilon/k + (J\eta^0/k +
H/k
        +f_i/k-1)\, ,
\ee
or
\be
\label{t-J}
        t_J = t_1 - 1/k \Gamma_0 \ln ( 1+ (J\eta^0/k + H/k +f_i/k -
1)
        k/J\epsilon) \, .
\ee

The shift in $s_J(t)$ at time $t_2 > t_J$ will therefore not only
consist of a contribution due to simple relaxation as
given by \eq{simple-relaxation} but also a contribution due to the
spin flip. It is
proportional to the distance of the equilibrium point of the upper
well to the equilibrium point of the lower well,
which is at all times two in our model.
Solving \eq{response-in-local-mft-4} with $sgn(s_i(t<t_J))=-1$ and
$sgn(s_i(t>t_J))=+1$, we find in this case
\bea
\label{s-flip}
\nonumber
\lefteqn{s_i^{flip}(t_2)|_{H+J\epsilon(t_2)} - s_i(t_2)|_H
= J\epsilon/k (1-exp(-k \Gamma_0(t_2-t_1)))\Theta(t_2-t_1) } \\
& & + 2 (1-exp(-k\Gamma_0(t_2-t_J))) \Theta(t_2-t_J)
        \Theta(t_J-t_1) \, .
\eea

The first of the two terms on the right hand side of the equation
also appears in \eq{simple-relaxation}.
We can therefore separate the contribution
solely due to the spin flip from the relaxational part:

{}From \eq{simple-relaxation} and \eq{s-flip} one obtains
\be
\label{delta-s}
\Delta s \equiv s_i^{flip}(t_2)-s_i^{noflip}(t_2) = 2 (1-\exp(-k
\Gamma_0(t_2-t_J))) \, .
\ee

The disorder averaged shift in the local magnetization is then
\bea
\label{average-shift-in-m}
\lefteqn{\langle s(t_2) |_{H+J \epsilon(t)} - s(t_2)|_{H} \rangle_f
\sim
(s_i^{no flip}(t_2)|_{H+\epsilon(t)} -
s_i(t_2)|_{H})} \\
\nonumber
& & + \int_{t_1}^{t_2} dt_J dN(t_J)/dt_J 2 (1 - \exp(-k
\Gamma_0(t_2-t_J))) \Theta(t_2-t_J) \, .
\eea
$dN(t_J)$ is the number of spins that are going to flip in the time
interval $[t_J, t_J + dt_J]$ with $t_1 < t_J < t_2$.
{}From \eq{tJ-definition}
we find a relation between the random field $f_i$
and $t_J$:
\be
f_i = - J \eta^0 - H +k - J\epsilon( 1- exp (-k \Gamma_0(t_J-t_1)))
\, .
\ee
Therefore
\begin{eqnarray}
\lefteqn{dN(t_J)=} \\
\nonumber
& & d t_J |df_i/dt_J| \rho(f_i=- J \eta^0 - H +k -
J\epsilon(1- \exp (-k \Gamma_0(t_J-t_1)))) \\
\nonumber
& &=d t_J( J\epsilon k \Gamma_0 exp( - k \Gamma_0 (t_J - t_1)) \\
\nonumber
& & \quad\rho(f_i=- J \eta^0 - H +k -
J\epsilon( 1- \exp (-k \Gamma_0(t_J-t_1)))) \\
\nonumber
& &= dt_J (J \epsilon k\Gamma_0 \exp{-k \Gamma_0 (t_J -t_1)}
\rho(f_i=-J \eta^0 - H +k)) + O(\epsilon^2).
\end{eqnarray}

\Eq{simple-relaxation}, \eq{s-flip}, and \eq{average-shift-in-m}
then yield for the
average shift (for $t_2 > t_1$)
divided by $\epsilon$, in the limit of small $\epsilon$
\begin{eqnarray}
\label{total-response-before-time-derivative}
\lefteqn{{\it lim}_{\epsilon \rightarrow 0}
[\langle s_i(t_2)|_{H+J\epsilon(t_2)}
- s_i(t_2)|_H \rangle]/\epsilon } \\
\nonumber
&& = [J/k + 2 J \rho(f_i=-J\eta^0 - H + k) + \\
\nonumber
& &\quad \exp (-k \Gamma_0 (t_2-t_1))(-J/k - 2 J (1+
k\Gamma_0(t_2-t_1))
\\
\nonumber
& &\quad \rho(f_i=-J\eta^0 - H + k ))]
\Theta(t_2-t_1) \, ,
\nonumber
\eea
which correctly goes to zero for $t_2 \rightarrow t_1$.
(The terms proportional to $\rho$ are due to spin flips and the
others stem from simple relaxation.)
For $(t_2-t_1) \rightarrow \infty$ it approaches the static value
$J/k + 2 J \rho(f_i=J\eta^0 - H + k )$, which is
the static response to a small perturbation.
The $u_{1,1}(t_1,t_2)$ vertex is given by the negative time
derivative
of \eq{total-response-before-time-derivative}
with respect to $t_1$, i.e.
\begin{eqnarray}
\label{u-11}
\lefteqn{u_{1,1}(t_1,t_2) = (J/k +2 J \rho(f_i=-J\eta^0 - H + k ))
\delta(t_2-t_1)}\\
\nonumber
& &+ \delta(t_2-t_1) [ exp (-k \Gamma_0(t_2-t_1)) (-J/k - 2 J (1+ k
\Gamma_0 (t_2-t_1)) \\
\nonumber
&& \quad \rho(f_i=-J\eta^0 - H+ k ))]\\
\nonumber
&& + \Theta(t_2-t_1) \partial_{t_1} [exp (-k \Gamma_0
(t_2-t_1))(-J/k - 2 J (1+ k \Gamma_0 (t_2-t_1)) \\
\nonumber
&& \quad \rho(f_i=-J\eta^0 - H
+ k ))] \, .
\end{eqnarray}
In the corresponding term in the action this is multiplied by
$\eta(t_2)$ and $\hat\eta(t_1)$ and integrated over $dt_1$ and
$dt_2$.
The terms multiplied by $\delta(t_2-t_1)$ in \eq{u-11} cancel
exactly,
so that there is only left the term multiplied by
$\theta(t_2-t_1)$. We perform an integration by parts in $t_1$
and obtain two terms: The boundary term which has
a purely static integrand and leaves only one time integral,
and a time dependent part with two time integrals.

\noindent
The static term contributing to the action is then
\be
\label{static-part-of-u-11}
-\int_{-\infty}^{+\infty}dt_1 {\hat \eta}(t_1) \eta(t_1) (-J/k- 2
J\rho(f_i=-J\eta^0 - H+ k )) \, .
\ee
The dynamical part can be written as
\begin{eqnarray}
\label{dynamical-part-of-u-11}
\nonumber
\lefteqn{\int_{-\infty}^{+\infty}dt_2 \int_{-\infty}^{+\infty}dt_1
\Theta(t_2-t_1) {\hat \eta}(t_2) (\partial_{t_1}\eta(t_1))
\exp(-k \Gamma_0(t_2-t_1))} \\
& & (-J/k -2 J (1+k\Gamma_0 (t_2-t_1))) \rho(-J \eta^0 - H +k) \, .
\end{eqnarray}
In the low frequency approximation this becomes
\bea
\label{i-omega-contrib}
\label{dyn-part-low-freq-of-u-11}
\nonumber
\lefteqn{\int_{-\infty}^\infty dt_1 {\hat \eta}(t_1) \partial_{t_1}
\eta(t_1)[ -J/k - 4 J \rho(-J \eta^0 - H +k)]/(\Gamma_0 k)} \\
& & = -\int_{-\infty}^\infty dt_1 {\hat \eta}(t_1) \partial_{t_1}
\eta(t_1) a/\Gamma_0
\, ,
\eea
with
\be
a = [J/k + 4 J \rho(-J \eta^0 - H +k)]/k \, ,
\ee
as can be seen using that
\be
e^{-k \Gamma_0 (t_2 - t_1) } \Theta(t_2-t_1) =
{\textstyle {1 \over 2 \pi}} \int d\omega
e^{ i \omega (t_2-t_1)} /(k\Gamma_0 + i \omega)\, ,
\ee
and
\be
k \Gamma_0 (t_2-t_1) e^{-k \Gamma_0 (t_2-t_1)} \Theta(t_2-t_1)
= {\textstyle {1 \over 2 \pi}} \int d\omega e^{ i \omega (t_2-t_1)}
{k \Gamma_0 \over (k \Gamma_0 + i \omega)^2} \, .
\ee

\Eq{i-omega-contrib} contributes
to the ``$i\omega$'' term in the propagator
expressed in frequency space.

We have performed the above calculation for the case $H(t_1)=H(t_2)$.
If instead one keeps $H(t_2)-H(t_1)= \Delta H \neq 0$ fixed as
$\Omega \rightarrow 0$ ({\it i.e.} $t_2 -t_1 \rightarrow \infty$),
one obtains
\be
lim_{\epsilon \rightarrow 0} \langle s_i(t_2)|_{H+\epsilon(t_2)}
- s_i(t_2)|_H \rangle_f/\epsilon = J/k + 2 J \rho(-J\eta^0(t_2) -H
+k)
\ee
up to dynamical corrections of the form
$(\exp(-\Delta H \Gamma_0/\Omega))$, which
are
negligible as $\Omega \rightarrow 0$.
Consequently, the derivative with respect to $(-t_1)$
yields zero in this limit. Therefore there is no contribution to the
action from these cases and the result converges to the expressions
in \eq{static-part-of-u-11} and \eq{dynamical-part-of-u-11}
as $\Omega \rightarrow 0$.

\subsubsection{The coefficients $u_{1,2}$, $u_{1,3}$ and $u_{2,0}$}
The coefficients $u_{1,2}$ and $u_{1,3}$ at field $H$
are calculated similarly.
One obtains for the
terms in the action corresponding to $u_{1,2}$ in the adiabatic
limit:
\bea
\label{u-12-timedependence}
\int &d^dx& \int_{-\infty}^{+\infty} dt {\hat\eta}(x,t) \left( w
[\eta(x,t)]^2
+ \int_{-\infty}^t dt_2 a(t,t_2,t_2) \eta(x,t_2) \partial_{t_2}
\eta(x,t_2) \nonumber \right.\\
&+& \left. \int_{-\infty}^t dt_2 \int_{-\infty}^{t_2} dt_1
a(t,t_1,t_2) \partial_{t_2} \eta(x,t_2) \partial_{t_1} \eta(x,t_1)
\right) \, .
\eea
Here, $w=-2 J^2 \rho'(f_i = -J \eta^0 -H +k)$, and
$a(t,t_1,t_2)$ is a transient function due to the
relaxational dynamics of the system. It consists of
terms proportional to
$\exp\{-\Gamma_0(t-t_1)\}$ or $\exp\{-\Gamma_0(t-t_2)\}$.
The transient terms proportional to $a(t,t_1,t_2)$ turn out to be
irrelevant for the critical behavior observed on long length scales.

The static and the transient terms in the action contributed by
$u_{1,3}$, are calculated similarly. Again, only the static part
turns out to be relevant for the
calculation of the exponents below the upper critical
dimension. It is given by
\be
\label{static-u-13}
\int d^dx \int_{-\infty}^{+\infty} dt u {\hat\eta}(x,t)
[\eta(x,t)]^3 \, ,
\ee
with
$ u= 2 J^3 \rho''(f_i = -J \eta^0 -H +k)$.

Finally, the vertex $u_{2,0}(t_1,t_2)=\langle s_i(t_1)
s_i(t_2)\rangle$
is calculated similarly. It is a local correlation
function instead of a local response function. Therefore the times
$t_1$ and $t_2$ can be infinitely far apart, i.e.
even for $H(t_1) \neq H(t_2)$ the vertex $u_{2,0}$ is still nonzero.
One obtains:
\bea
\label{u-20-t1-t2}
\nonumber
\lefteqn{u_{2,0}(t_1,t_2) = R^2/k^2 +
\left(\int_{-\infty}^{-H(t_2)-\eta^0(t_2)+k} \rho(h) dh \right)} \\
\nonumber
& &\left(4-4
\int_{-H(t_2)-\eta^0(t_2)+k}^{-H(t_1)-\eta^0(t_1)+k} \rho(h)
dh\right)
 \\
& & - 4 \left(\int_{-\infty}^{-H(t_2)-\eta^0(t_2)+k} \rho(h) dh
\right)^2
- 4 \left( \int_{-\infty}^{-H(t_2)-\eta^0(t_2)+k} (h/k) \rho(h) dh \right)
\nonumber \\
& &-2 \left(\int_{-H(t_2)-\eta^0(t_2)+k}^{-H(t_1)-\eta^0(t_1)+k}
(h/k)
\rho(h) dh\right) \, ,
\eea
which is positive (or zero) for any normalized distribution
$\rho(f)$.

\subsection{Feynman Rules}
\label{feynman-rules}

In the following discussion we denote with $u_{mn}$ the
static part of $u_{mn}(t_1...t_{m+n})$,
i.e. the part which, (for $n \ne 0$, after taking the time derivative
and integrating by parts) is not multiplied by any time
derivative
of the fields. This is usually the only part of the vertex which is
not irrelevant under coarse graining (except for the propagator
term, which also has a contribution proportional to $i\omega$).
We formulate the Feynman rules in general terms, although for our
discussion only $u_{1,1}=\chi^{-1}/J + 1$, $u_{1,2} = w$, $u_{1,3}=u$
and $u_{2,0}$ come into play.

We introduce the following notation for the Feynman diagrams, which
we will use in the perturbation expansion :
A vertex $u_{1,n}$ is a dot with $m$ outgoing arrows (one for each
$\hat\eta$ operator)  and $n$ incoming arrows (one for each
$\eta$ operator.)
Figure \ref{fig:feynman} (a) shows the graph for $u$ as an example.
The propagator can only connect $\hat\eta(t)$ and $\eta(t')$
operators. It is important to remember that causality must be obeyed,
i.e. $t'>t$. Figure \ref{fig:feynman} (b)
shows an example of a diagram which violates
causality.

The vertex $u_{2,0}$ is denoted as shown in figure \ref{fig:feynman}
(c). The
black ellipse connects the two parts of the vertex that are taken at
different times.

The symmetry factors for each diagram are obtained in the usual way
 \cite{Ryder,Ramond} by drawing all topologically distinct graphs
and counting their multiplicities. One must not forget to include
the factors $1/(m! n!)$ that appear in the expansion (see
\eq{effective-action-final}).

In each diagram, momentum conservation requires that vertices should
be connected by loops rather than a single, dead end propagator line.
Figure \ref{fig:feynman} (d) shows an example of a diagram that is
zero \cite{Wilson}.

%XXXXXXXXXXX figure fig:feynman XXXXXXX

%\begin{figure}
%\vbox to 3in{
% \vfil
% \special{psfile=feynman.xfig.ps
%          hoffset=-340
%          voffset=-150
%          hscale=60
%          vscale=60}
%}
%\caption
%[Illustration of the Feynman rules through Feynman diagrams.]
%{{\bf Feynman diagrams.} The perturbative expansion about mean--field
%theory is presented here by Feynman diagrams.
%(a) Graph for the vertex $u$. Incoming arrows denote $\eta$ fields,
%outgoing arrows denote $\hat \eta$ fields.
%(b) Example of a diagram which violates causality and is
%therefore forbidden.
%(c) Graph for the vertex $u_{2,0}$.
%(d) Example of a diagram that is
%zero due to momentum conservation {\protect \cite{Wilson}}.
%\label{fig:feynman}}
% \end{figure}

Each internal loop contributes an integral over the internal
momentum--shell  \cite{Goldenfeld} (see also below)
\be
\label{loop-integral}
        \int_{\Lambda/b}^{\Lambda} d^dq/(q^2-\chi^{-1}/J) \, ,
\ee
where $\Lambda$ is the cutoff and $b > 1$.

Integration over time is already performed in \eq{loop-integral}.
(In the low frequency approximation the propagator
can be approximately taken to be
$\delta(t-t')$, when we integrate out modes in the infinitesimal
momentum shell $\Lambda/b<q<\Lambda$ \cite{NarayanI}.)

\subsection{Implementation of the history}
\label{history}

As we have mentioned in section \ref{sub:Implementing-the-history}
it turns out that on long length scales different magnetic fields
decouple and the static critical exponents can be extracted from a
renormalization group analysis performed at a single, fixed value of
the external magnetic field $H_0$ due to a separation of time scales.
In the following paragraph we will show that this statement is
self-consistent using an argument by Narayan and Middleton in the
context of the CDW depinning transition \cite{NarayanII}.

An expansion around mean-field theory in the way performed here
corresponds to {\it first} increasing the magnetic field $H$ within
an infinite ranged model and {\it then} tuning the elastic coupling
to a short ranged form, while the actual physical behavior
corresponds to {\it first} tuning the elastic coupling to a short
ranged form and {\it then} increasing the force within the short
ranged model. The concern is that in the presence of many metastable
states the critical behavior of the two approaches might not be the
same. For example, spins might tend to flip backwards upon reduction
of the interaction range in the expansion around mean-field theory.
Although there will of course always be {\it some} spins for which this is
the case, no such effects are expected on long length scales since
the susceptibility is actually {\it more} divergent near the
critical point for $d<6$ than in mean-field theory:
\be
(dm/dh)_{h=0} \sim r^{-\gamma}
\ee
and
\be
(dm/dh)_{r=0} \sim h^{1/\delta -1}
\ee
with
\be
\gamma({\it in\,\,\,} d<6) > \gamma({\it in\,\,\,mean\,\,\,field\,\,\,theory})
\ee
and
\be
(1-1/\delta)({\it in\,\,\,} d<6) > (1-1/\delta) ({\it in\,\,\,
mean\,\,\,field\,\,\,theory})
\ee
as we learn from numerics and analytics.
This suggests that on long length scales spins tend to flip {\it
forward} rather than backward upon reduction of the coupling range.
One would therefore expect the expansion around mean-field theory
performed here to correctly describe the critical behavior.

Reassured by this self-consistency argument we now briefly discuss
for separated time scales that
the decoupling of the different magnetic fields
is consistent also within the RG description.
As discussed in appendix \ref{sub:calculating-some-u-mn}
the response functions $u_{1,n}(t_1,\cdot\cdot\cdot, t_{n+1})$
with fixed $H(t_1) \neq H(t_j)$ with $j\neq1$
tend to zero in the adiabatic limit.
However, the original action $S$ of \eq{effective-action-final}
also contains terms of the form $u_{2,0}(H_1,H_2)$ which do couple
different fields even as $\Omega \rightarrow 0$.
These ``multi-field'' vertices however do not contribute to the
renormalization of the vertices evaluated at a {\it single} value
of the external magnetic field, because the propagator does not
couple different field values. It turns out that the multi-field
vertices are also irrelevant on long
length scales:
setting up the RG in section \ref{sub:RG-analysis} we have treated
$u_{2,0}(H_0,H_0)$ as a marginal variable.
In fact, we have been choosing the rescaling $\gamma_{\hat\phi}$ of the
fields $\hat\eta$ in $\hat\eta(x,t) = b^{-d/2-z-\gamma_{\hat\phi}/2}
\hat\eta'(x,t)$ such that $u_{2,0}(H_0,H_0)$ remains marginal to
all orders in perturbation theory.
Similarly, the rescaling $\gamma_\phi$ of the fields $\eta$
in $\eta(x,t)= b^{-d/2+2-\gamma_\phi/2} \eta'(x,t)$ is chosen
such that the coefficient of the $q^2$ term remains marginal to all
orders also. (These choices are made so
that the rescaling of the response and the cluster
correlation function under coarse graining immediately gives their
respective power law dependence on momentum (see
section \ref{sub:Gaussian-response})).
To decide what this implies for $u_{2,0}(H_1,H_2)$ at two {\it
different} magnetic fields $H_1 \neq H_2$, we need consider two loop
order corrections for $u_{2,0}(H_1,H_2)$.
In figure \ref{fig:u-20-h1-h2} we have indicated the magnetic fields
corresponding to
the times at which the vertices are evaluated, taking into account
that the propagators do not couple different fields. One finds the
following recursion relation:
\bea
\label{u-20-h1-h2}
{\textstyle {1 \over 2}} u_{2,0}'(H_1,H_2) &=&
{\textstyle {1 \over 2}} u_{2,0}(H_1,H_2) \\
\nonumber
& & + {\textstyle {1 \over 2}}
u_{2,0}(H_1,H_2) ( u_{2,0}(H_1,H_2)^2 - u_{2,0}(H_1)
u_{2,0}(H_2)) \\
\nonumber
& & {\textstyle {1 \over 54}} K_6^2 (4 \pi)^6 \epsilon^2/(u_{2,0}(H_1)
u_{2,0}(H_2)) \ln b \, ,
\eea
with $u_{2,0}(H_1) \equiv u_{2,0}(H_1,H_1)$.
The term that is subtracted stems from the second order correction
to the rescaling of the fields $\hat\eta$ (from terms as
depicted in figure \ref{fig:u-20-h1-h2}
but with $H_1=H_2$).

%XXXXXX figure fig:u-20-h1-h2   XXXXXX
%\begin{figure}
%\vbox to 3in{
% \vfil
% \special{psfile=u20.xfig.ps
%          hoffset=-290
%          voffset=-170
%          hscale=60
%          vscale=60}
%}
%\caption
%[Feynman diagram for the lowest order correction to
%$u_{2,0}(H_1,H_2)$.]
%{Feynman diagram for the lowest order correction to
%$u_{2,0}(H_1,H_2)$. The magnetic fields corresponding to
%the times at which the vertices are evaluated are indicated.
%The propagators do not couple different fields.
%\label{fig:u-20-h1-h2}}
% \end{figure}

{}From \eq{u-20-t1-t2} it follows that $u_{2,0}(H_1,H_2) \geq 0$.
Furthermore, using the same equation one finds after some
algebra
that (for the special history considered here and
for any distribution $\rho$ of random fields, using the
Cauchy-Schwartz inequality)
\be
\label{u-20-difference}
u_{2,0}(H_1,H_2)^2 -u_{2,0}(H_1)u_{2,0}(H_2) \leq 0 \, .
\ee
In fact, for $H_1 \neq H_2$ ( which excludes also $H_1$ and $H_2$
being both equal to $\pm \infty$, with the same sign),
and $R \neq 0$ the expression in \eq{u-20-difference}
is actually negative.
This implies that for $R\neq 0$ the part in the quadratic action,
which was coupling different magnetic fields, will be irrelevant at
dimensions $d < 6$ under coarse graining. Its fixed point value is
zero. In contrast, $u_{2,0}(H)$ stays constant under coarse graining.

What does this mean for the coarse grained action?
If we leave out all the terms in the action that are zero or
irrelevant at $d=6-\epsilon$, the different magnetic fields are
completely decoupled, and the critical exponents (for $R \geq R_c$ at
least) can be extracted from coarse graining the following action
at fixed magnetic field $H_0$:
\bea
\label{action-H-0}
\tilde S_{H_0} &=& -\int d^dq \int dt \hat\eta(-q,t)
(-1/\Gamma_0\partial_t +q^2-\chi^{-1}(H_0)/J ) \eta(q,t) \\
\nonumber
& & + 1/6 \int d^dx \int dt \hat\eta(x,t) (\eta(x,t)^3 u(H_0) \\
\nonumber
& & + 1/2 \int d^dx \int dt_1 \int dt_2 u_{2,0}(H_0,H_0) \hat\eta(x,t_1)
\hat\eta(x,t_2) \, .
\eea
(The index and argument $H_0$ serve as a reminder that all
coefficients in this action are evaluated at the same magnetic field
$H_0$.
The time integrals extend from $-\infty$ to $\infty$.
The constant coefficients of the $\partial_t$-term and the
$q^2$-term have been rescaled to $1$
(see section \ref{sub:Gaussian-response}).

\section{Borel resummation}
\label{ap:Borel}
It is known \cite{Binney,Zinn-Justin} that the $\epsilon$-expansion
yields only asymptotic series for the critical exponents as functions
of dimension. It is important to note that asymptotic series
in general do not
define a {\it unique} underlying function.
In the following we will give a definition of an asymptotic series,
discuss several examples and give the special conditions under which
there does exist a unique underlying function. So far it has
not been possible
to show that the $\epsilon$-expansion would meet these conditions;
we therefore cannot assume that it uniquely determines
the critical exponents as functions of dimension.
Applying the methods of reference \cite{Vladimirov} to our problem
we Borel resum the perturbation series for our critical
exponents to $O(\epsilon^5)$. A comparison of the result to the
numerical values of the exponents in 3, 4, and 5 dimensions is given
in the main text.

\subsection{Definition of an asymptotic series}
Let the variable $z$ range over a sector $S$ in the complex plane
with the origin as a limit point (which may or may not belong to
$S$):
\be
\label{sector}
Arg(z)\leq \alpha/2 \,,\, |z|\leq |z_0| \, .
\ee
A power series $\sum_{k=0}^\infty f_k z^k$ is said to be asymptotic
to the numerical function $f(z)$ defined on $S$ as $z\rightarrow 0$
\be
f(z) \sim \sum_{k=0}^\infty f_k z^k
\ee
if the approximation afforded by the first few terms of the series
is better the closer $z$ is to its limiting value, which is zero
in this case \cite{Erdelyi}. Formally:
\be
\label{bound1}
 |f(z)- \sum_{k=0}^N f_k z^k | << |z|^N
\ee
as $z\rightarrow 0$ for every $N$
({\it i.e.} for any given $\epsilon > 0$ there exists a neighborhood
$U_{\epsilon}$ of the origin so that
\be
|f(z) - \sum_{k=0}^{N} f_k z^k| \leq \epsilon |z|^N
\ee
for all $z$ common to $U_{\epsilon}$ and $S$).

This is equivalent to the statement that the remainder of the sum
after summing the first $N$ terms is of the order of the first
neglected term and goes to zero rapidly as $z\rightarrow 0$:
\be
\label{bound2}
 \left(f(z)- \sum_{k=0}^N f_k z^k \right) \sim f_M z^M
\ee
for every $N$, where $f_M$ is the first nonzero
coefficient after $f_N$. As shown in \cite{Erdelyi} this implies
that there exists a constant $C(N+1)$ ({\it i.e.} a number independent of
$z$) and a neighborhood $U$ of the origin such that
\be
| \left(f(z)- \sum_{k=0}^N f_k z^k \right) | \leq |C(N+1) z^{(N+1)} |
\ee
for all $z$ common to $U$ and $S$.

Often the terms in the series at first decrease rapidly (the faster,
the closer $z$ approaches zero) but higher order terms start
increasing again.
If the series is truncated at the minimum term, one obtains usually
the best possible truncated estimate for $f(z)$ with a finite error
$E(z)$.
If the coefficients $C(k)$ are of the form
\be
\label{C(k)}
C(k) \sim C (k!)^\beta A^{(-k)}
\ee
with $A>0$, $\beta >0$, and $C>0$ real constants,
the error $E(z)$ can be estimated explicitly \cite{Zinn-Justin}
to
\be
E(z) = min_{N} C(k) |z|^k \sim \exp[ -\beta (A/|z|)^{(1/\beta)}] \, .
\ee
which decreases rapidly as $|z| \rightarrow 0$.
It determines a limit to the accuracy beyond which it is impossible
to penetrate by straightforward summation of a finite number of
terms of the series.

\subsubsection{Non-uniqueness of $f(z)$}
\label{uniqueness-theorem}

One sees that an asymptotic series does not in general define a
unique analytic function $f(z)$ over $S$.
Any other function $g(z)$ which is analytic in $S$
and smaller than $E(z)$ in the whole sector can be added to $f(z)$
such that in $S$ the series $\sum_{k=0}^\infty f_k z^k$ is
asymptotic also to $g(z) + f(z)$.
If however the sector is big enough, {\it i.e.} if
$\alpha \geq \pi \beta$ then according to a classical theorem
(Phragmen-Lindel\"of)
there exists no
nonvanishing function over $S$ that is analytic in $S$ and bounded
by $E(z)$ in the whole sector. In that case the underlying function
is uniquely determined by the series \cite{Zinn-Justin}.

\subsubsection{Examples}

Let $S_\Delta$ be the sector $0<|z|< \infty$, $\phi \equiv
|Arg(z)| < \pi/2 - \Delta$
with $\Delta >0$.

1. The function $\exp(-1/z)$ is asymptotic to
$\sum_{n=0}^\infty 0 \cdot z^n =
0$ over $S_\Delta$:
\be
|\exp(-1/z) -0| = \exp(-\cos(\phi)/|z|) < \exp(-\cos(\pi/2 -
\Delta)/|z|) << |z|^N
\ee
for any $N$ as $|z| \rightarrow 0$.

\vspace{2mm}

2. Similarly one can show that over $S_{\Delta}$ the series
\be
\sum_{k=0}^\infty (-1)^{(k-1)} z^k
\ee
is asymptotic to the functions $1/(1+z^{(-1)})$,
$(1+exp(-1/z))/(1+z^{(-1)})$ and $1/(1+\exp(-\sqrt{(1/z)})+z^{(-1)})$
as $z \rightarrow 0$ \cite{Erdelyi}.
(A similar nonuniqueness is expected for the underlying functions of
the $\epsilon$-expansion.)

\vspace{2mm}

3. The Stirling series is one of the oldest and most venerable of
asymptotic series. It expresses the asymptotic behavior of the
factorial function $n!$ for large values of $n$
\be
\Gamma(n) = (n-1)! \sim \left(\frac{2 \pi}{n}\right)^{\frac{1}{2}}
\left(\frac{n}{e} \right)^n (1 + \frac{A_1}{n} + \frac{A_2}{n^2} + \cdot
\cdot \cdot )
\ee
with $A_1=1/12$, $A_2=1/288$, etc.
It becomes the full asymptotic expansion of the Gamma function
$\Gamma(z)$ for complex argument $z$:
\be
|\Gamma(z) -  \left(\frac{2 \pi}{n}\right)^{\frac{1}{2}}
\left(\frac{z}{e}\right)^z
(1+ \sum_{j=1}^N A_j z^{-j}) | << |\left(\frac{z}{e}\right)^z
z^{-\frac{1}{2}-N}
\ee
for $z \rightarrow \infty$ and $|Arg(z)| < \pi$ \cite{Bender}.

\subsection{Borel resummation}

The Borel summation is a technique for expressing
an asymptotic series $\sum_{k=0}^\infty f_k z^k$ as the limit of a
convergent integral.
For an introduction to this
technique see reference \cite{Bender,Erdelyi}.
The idea is to modify the coefficients of the series such that
for sufficiently small $z$ it
converges to a function $B(z)$, which can
be calculated. In the simplest case
\be
B(z) \equiv \sum_{k=0}^\infty f_k/k! z^k
\ee
would be appropriate.
Suppose that also
\be
f(x) \equiv \int_0^\infty \exp(-t) B(x t) dt
\ee
exists.
Then, by expanding the integral
$\int_0^\infty \exp(-t/x) B(t) dt/x$
for small $x$ and integrating term by term, one
finds \cite{Bender} that
$f(x) = \sum_{k=0}^\infty f_k x^k$ as $x \rightarrow 0^+$.

\subsubsection{Example}
The series $\sum f_k z^k = \sum_{n=0}^{\infty} n!
(-x)^n$ diverges, but $B(x)\equiv \sum_{n=0}^{\infty} (-x)^n$ converges
for $|x| \leq 1$ to $1/(1+x)$. Thus the Borel sum of
$\sum_{n=0}^{\infty} n! (-x)^n$ is
$f(x) = \int_0^\infty \exp(-t)/(1+xt) dt$. For $z$ in the
complex plane, $f(z)$ is asymptotic to $\sum (-z)^n n!$ over the
sector $S$: $0<|z|<\infty$, $|Arg(z)| \leq \pi - \epsilon$, $\epsilon
>0$. Since $|f_n| \sim n!$ we have $\beta = 1$ from \eq{C(k)}, and
since $\alpha > \pi$ for small enough $\epsilon$, the uniqueness
theorem of section \ref{uniqueness-theorem} implies that $f(z)$ is
the unique underlying analytic function of the series over $S$.

\subsection{Borel resumming the $\epsilon$-expansion}

It is known \cite{Zinn-Justin}, that the $\epsilon$-expansion
series is an asymptotic series. It has a zero radius of convergence.
Lipatov, Br\'ezin, LeGouillou, and
Zinn-Justin have shown \cite{Large-order-perturbation}, that
the coefficients of higher orders of the $\epsilon$ expansion
$f(\epsilon) = \sum_k (-\epsilon)^k f_k$
(with $\epsilon=4-d$) have the asymptotic
form $f_k\sim_{k\rightarrow\infty} k! a^k k^b c$.
(The factorial growth of the coefficients reflects the zero
radius of convergence.) Here, $f$ stands for $\eta$ or $1/\nu$
(or $\omega$ or $g_0$, see \cite{Vladimirov}), and
$a = 1/3$, $b=7/2$ for $\eta$, and $a = 1/3$, $b=9/2$ for $1/\nu$.

Using the newest results for the coefficients to $O(\epsilon^5)$
as given in Kleinert {\it et al.} \cite{Kleinert} and the asymptotic
form given above,
we resum the coefficients to $O(\epsilon^5)$, using
the method discussed in reference \cite{Vladimirov,Kazakov}
and reference \cite{LeGuillou}. The method is based on a modified Borel
transformation:
\be
f(\epsilon) = \int_0^\infty dx/(\epsilon a)
\exp(-x/(\epsilon a)) (x/(\epsilon a))^{(b+3/2)} B(x)\, .
\ee
One obtains $B(x) = \sum_k (-x)^k B_k$, with $B_k=f_k/(a^k
\Gamma(k+b+5/2)) \sim_{k\rightarrow \infty} c/k^{3/2}$.

$B(x)$ is given by its Taylor series only for $|x|<1$.
It can be continued analytically
beyond the unity convergence circle in the complex plane, using the
conformal mapping \cite{Loeffel} $x \rightarrow w$ with
$w(x) = [(1+x)^{1/2}-1] / [(1+x)^{1/2} +1]$, or $x=4 w/(1-w)^2$.
The integration interval $[0,\infty)$ then goes over into the
interval
$[0,1]$ and the cut $(-\infty,-1]$ goes over in the boundary of the
unit disk. The expansion of $B(x(w))$ in terms of $w$ converges
to $B(x)$ in the entire region of integration $w \in [0,1)$.
The coefficient of $w^N$ is determined on the basis of the
coefficients $f_k$ for $k \leq N$ of the initial
$\epsilon$ expansion.
Neglecting the terms of $O(w^m)$, with $m>N$,
one obtains an approximation $f_N(\epsilon)$ which takes into
account the $O(\epsilon^N)$ corrections to $f$.

Since we are working with truncated rather than infinite series
there is an
arbitrariness in the reexpansion of $B(x)$ in terms of $w$
\cite{LeGuillou,Vladimirov}. Rather than expanding $B(x(w))$
directly, Vladimirov {\it et al.} actually expand
$((1-w)^2)^\lambda B(x(w))/4 \equiv B^\lambda(w) $ in terms of $w$,
so that (since $x=4 w/(1-w)^2$) the result for $B(x(w))$ is given
by
$B(x) \rightarrow x^\lambda B^\lambda(w) = (x/w)^\lambda \sum_k^N
B_{\lambda}^{(k)} w^k$.
$\lambda$ is a parameter that must be chosen correctly, such
that
$f_N(\epsilon)$ is matched to the asymptotic form $f(\epsilon)$
as
$\epsilon \rightarrow \infty$, given by:
$f(\epsilon) \sim_{\epsilon\rightarrow \infty}
(\epsilon)^\lambda$.
Since the asymptotic form of the exponents relative to $\epsilon$
is unknown, it is argued in \cite{Kazakov} and \cite{LeGuillou}
that $\lambda$ should be
fixed from the condition of fastest convergence of $f_N$, i.e. it
should minimize $\Delta_N =|1-f_N(1)/f_{N-1}(1)|$.
We have written a computer program that performs this procedure
for the exponents $1/\nu$ and $\eta$ using the following results
for the $\epsilon$-expansion to $O(\epsilon^5)$ from
reference \cite{Kleinert}:
\be
1/\nu=2-\epsilon/3 - 0.1173 \epsilon^2 + 0.1245 \epsilon^3 -0.307
\epsilon^4 + 0.951\epsilon^5 + O(\epsilon^6)\, ,
\ee
\be
\eta= 0.0185185 \epsilon^2 + 0.01869 \epsilon^3 - 0.00832876
\epsilon^4 +  0.02566 \epsilon^5 + O(\epsilon^6)\, .
\ee

Through the perturbative mapping of our problem to the pure Ising
model in two lower dimensions to all orders in $\epsilon$ these are
also the results for the exponents in our system in $6-\epsilon$
dimensions.
Figure \ref{fig:Borel-vs-numerics}
shows the comparison between the Borel resummed exponents
$1/\nu$, $\eta$, and
$\beta\delta = {\textstyle {\nu \over 2}}(d-2\eta + {\bar\eta})$,
and the corresponding numerical results at integer dimensions.
We have used that in perturbation theory $\bar\eta=\eta$,
as follows from the mapping to the equilibrium random field
Ising model \cite{Krey,Nattermann-review}.

The standard Borel resummation has been modified in various ways
specifically to resum
the $4-\epsilon$ expansion for the exponents of the pure Ising model.
Le Gouillou and Zinn-Justin \cite{LeGuillou} have developed a resummation
prescription that takes into account that the critical exponents
have a singularity at the lower critical dimension $d=1$
({\it i.e.} $\epsilon=3$).
In a more recent calculation \cite{LeGuillou},
they treated the dimension of the singularity
as a variational parameter to improve the apparent convergence of
the expansion. They also
imposed the exactly known values of the exponents in two dimensions.
Their results agree (partly by design) very well with the numerical
exponents of the pure equilibrium Ising model in two and three
dimensions. Below two dimensions the apparent errors
of their results are rapidly increasing.\footnote{We should note
that their work was based on a form for the fifth order
term which later turned out to be incorrect
\cite{Kleinert}.}

In our problem the lower critical dimension is probably 2
rather than 3, {\it i.e.} $\epsilon=4$ rather than $\epsilon=3$.
In our Borel-resummation of the $6-\epsilon$ expansion results for
our critical exponents
we have made no assumptions about a singularity for the critical
exponents, nor did we impose any other information.
Instead, we have applied the conventional Borel-resummation as adapted
by Vladimirov {\it et al.}, which shows good agreement with our
numerical data (see figure \ref{fig:Borel-vs-numerics}).
Our resummation of
corrections for $\nu$ up to order $O(\epsilon^5)$ has a pole
around 2.3 dimensions. (In reality $\nu$ is expected to diverge
at the lower critical dimension.)
The pole did slowly increase with the order
to which the resummation was performed. It is not clear however
whether this tendency persists at higher orders.

Figure \ref{fig:two-Borel} shows a comparison of our Borel
resummation
for $\nu$ with the results of
Le Guillou and Zinn-Justin from 1987 \cite{LeGuillou}.
It is believed that the "straightforward"
Borel-resummation according to
Vladimirov {\it et al.} \cite{Vladimirov,Kazakov} and
the resummation according to
LeGuillou and Zinn-Justin \cite{LeGuillou}
which treats the dimension of the singularity in the exponents
as a variational parameter,
should both at high enough orders converge to the
pure Ising exponents, although this has not been proven.
Unfortunately, the curve from LeGuillou and Zinn-Justin
\cite{LeGuillou}
shown in figure \ref{fig:two-Borel} has been obtained using
an epsilon expansion for the critical exponents, which later turned
out to be incorrect in the 5th order term \cite{Kleinert}.
Our own Borel-resummation however,
which is also shown in figure \ref{fig:two-Borel},
is based on
the newest, presumably correct result for the fifth
order term \cite{Kleinert}.
We are currently trying to duplicate the variable-pole analysis
of LeGuillou and Zinn-Justin with the
correct fifth order term.

As we have already mentioned, there is no known reason
to assume that the $\epsilon$-expansion is Borel-summable,
{\it i.e.} that the values of the exponents would be uniquely
determined as functions of dimension by the Borel summation
of their asymptotic $\epsilon$-expansion.
At this stage we do not know the size of the section of analyticity
for the various Borel sums \cite{LeGuillou}.
Hence we expect the $\epsilon$-expansion to give
the correct value of the exponents only
asymptotically as $\epsilon\rightarrow 0$, even after Borel
resummation.
There is no reason to expect that its extrapolation to
$\epsilon=3$
can be used to determine the lower critical dimension, i.e. the
dimension where $\nu$ has a pole, and where the transition
disappears or loses its universality
(see also section \ref{sub:honest-dim-reduction} in the text).

\section{Infinite avalanche line}
\label{ap:inf-aval-line}

In most of this paper we have focussed on the critical endpoint
$(R_c,H_c(R_c))$, in particular as it is approached from $R \geq
R_c$.
In this appendix we discuss the onset of the infinite avalanche for
$R<R_c$.
Our $\epsilon$-expansion can be applied to the entire line $H_c(R)$,
$R<R_c$ at which the infinite avalanche occurs (with some
reservations which we will discuss later).
In mean-field theory the approach to this line is continuous with a
power law divergence of the susceptibility $\chi \sim dM/dH$
and precursor avalanches on all scales (see appendix
\ref{ap:mean-field}).
{}From
\begin{eqnarray}
\label{S-H2}
\lefteqn{S_H = -\sum_{j,l} \int dt J_{jl}^{-1}J \hat\eta_j(t)
\eta_l(t)
-\sum_j \int dt\, \hat{\eta}_j(t)
        [-\partial_t/\Gamma_0 - u_{11}^{stat}]\, \eta_j(t)} \\
\nonumber
&+& \!\!\! \sum_j(1/6) \int dt\, \hat{\eta}_j(t) (\eta_j(t))^3 u
+ (1/2) \sum_j \int dt_1 \int dt_2\, \hat{\eta}_j(t_1)
\hat{\eta}_j(t_2)
u_{2,0} \, ,
\end{eqnarray}
quoted from in \eq{S-H}, and from the rescaling of the vertices
given in \eq{u-mn-rescaling}
\be
\label{u-mn-rescaling2}
u_{m,n}' = b^{[-(m+n)+2]d/2+2n} u_{m,n} \, .
\ee
one sees that on long length scales the effective action is purely
quadratic above 8 dimensions.
This suggests that there is a {\it continuous}
transition (as $H$ approaches $H_c(R)$)
with mean-field critical exponents and a diverging correlation
length $\xi(\chi^{-1})$ with the scaling
behavior $\xi(\chi^{-1}) = b \xi(b^2 \chi^{-1})$, {\it i.e.}
$\xi \sim (\chi^{-1})^{-1/2}$. Since $\chi^{-1} \sim \sqrt{|H-H_c(R)|}$
(see appendix \ref{ap:mean-field})
it follows that $\xi \sim |H-H_c(R)|^{-\nu_h}$ with $\nu_h = 1/4$ for
$d>8$.\footnote{If the Harris-criterion is
not violated through the presence of large
rare nonperturbative fluctuations in an infinite system, such as a
preexisting interface (for a discussion see appendix
\ref{ap:related}), {\it i.e.} if $\nu_h \geq 2/d$ is a valid
exponent-inequality,
then the mean-field critical exponents with $\nu_h = 1/4$ are only
correct for $d \geq 8$, which is consistent with our result from
perturbation theory.}

For $d=8-\tilde\epsilon$ ($\tilde\epsilon>0$) the vertex $w$ in the
action $S_H$ becomes relevant. In contrast to the critical endpoint
where $\chi^{-1}=0$ and $w=0$, the infinite avalanche line is
characterized by the ``bare values '' $\chi^{-1}=0$ and $w=-2
J^2\rho'(-JM(H_c(R)) -H_c(R) +k) \neq 0$.
Figure \ref{fig:feynman-w} shows the lowest order the correction to vertex $w$.
With the Feynman-rules of appendix
\ref{feynman-rules} and section \ref{subsub:loop-corrections}
the recursion relation to the same order becomes
\begin{equation}
\label{w-loop2}
w^{\prime} /2= b^{(-d/2+4)}\left\{w/2+(u_{2,0}/2) (w/2)^3 8 /(4
\pi)^4
\int_{\Lambda/b}^{\Lambda} dq \, 1/(q^2-\chi^{-1}/J)^4\right\} \, .
\end{equation}
Performing the integral over the momentum shell $\Lambda/b<q<\Lambda$
and writing $b^{(-d/2+4)}=b^{(\tilde{\epsilon}
/2)}=1+\tilde{\epsilon}/2 \,\log b$
we find
\begin{equation}
\label{w2}
w^{\prime} /2=w/2+(w/2)\Bigl(\tilde\epsilon/2+u_{2,0}(w/2)^2 4 /(4
\pi)^4
\log b \Bigr) \, .
\end{equation}
Since $u_{2,0}>0$, this equation has only
two fixed points $w^*$ with $w^{\prime}=w$ for
$\tilde \epsilon >0$: either $w^*=0$ or
$w^*=\infty$.
Any system with bare value $w \neq 0$ will have effectively larger
$w$ on longer length scales. The system flows to the strong
coupling limit. We interpret this as indication that in
perturbation theory the transition is of first-order type below 8
dimensions. Indeed, our numerical simulation suggests an abrupt
onset of the infinite avalanche in 2, 3, 4, 5 and 7
dimensions, but smooth in 9 dimensions \cite{Perkovic}.

As we discuss in appendix \ref{ap:related} there are some questions
as to whether in an infinite system the onset of the infinite
avalanche would be abrupt in {\it any} finite dimension due to large
rare preexisting clusters of flipped spins which provide a
preexisting interface that might be able to advance {\it before} the
perturbatively calculated critical field $H_c(R)$ is reached. These
large rare fluctuations might be  nonperturbative contributions
which are not taken into account by our $\epsilon$-expansion.
The progression of a preexisting interface has been studied
previously in the framework of depinning transitions
\cite{Nattermann,NarayanIII,Robbins}.
Our preliminary numerical simulation results in 9 dimensions
do however seem to show a continuous
transition at the onset of the infinite avalanche as predicted by
the RG calculation \cite{Perkovic}.

%XXXXXXXXXXX figure fig:feynman-w XXXXXX
%\begin{figure}
%\vbox to 1.5in{
% \vfil
% \special{psfile=tadpole.xfig.ps
%          hoffset=-380
%          voffset=-320
%          hscale=80
%          vscale=80}
%}
% \caption
%[Feynman diagram for the relevant correction to first order
%in $\tilde \epsilon=8-d$
%for systems with less than critical disorder $R<R_c$.]
%{{\bf Feynman diagram.}
%The correction to $O(\tilde{\epsilon})$ to the vertex w in
%an expansion
%about 8 dimensions, see eq.~(\protect\ref{w-loop2}) in the text.
% \label{fig:feynman-w}}
% \end{figure}

%%%%%%%%%%%%%%%%%%%%%%%%%%%%%%%%%%%%%%%%%%%%%%%%%%%%%%%%%%%%%%%%%%%%%%%
%% This is a REVTEX file, which hopefully will run on your system too.%
%%%%%%%%%%%%%%%%%%%%%%%%%%%%%%%%%%%%%%%%%%%%%%%%%%%%%%%%%%%%%%%%%%%%%%%

\section{Details for the $\epsilon$-expansion of the avalanche
exponents}
\label{ap:RG-details-avalanches}

\subsection{The second moment of the avalanche size distribution}
\label{sub:s-2-2}

In this section we show that the second moment $\langle S^2 \rangle_f$
of the avalanche size distribution $D(S,r,h)$
scales in the adiabatic limit as
\be
\label{s2-2}
\langle S^2 \rangle_f \sim \int dt_1 \int dt_\alpha dt_\beta
d^dx_\alpha
d^dx_\beta
\langle
 \hat{s}^\alpha(t_\alpha,x_0) s^\alpha(t_0,x_\alpha)
\hat{s}^\beta(t_\beta,x_0)
s^\beta(t_1,x_\beta)
\rangle_f \, .
\ee
where $\alpha$ and $\beta$ specify the corresponding replica
that have identical configurations of random fields and are exposed
to the same external magnetic field $H(t)=H_0+\Omega t$, with
$\Omega \rightarrow 0$.\footnote{A heuristic justification of this
was given in section \ref{sec:epsilon-expansion} together with an
explanation of why replicas are necessary.}

We start by computing the (not yet random field averaged) expression
\bea
\label{DeltaSaDeltaSb}
\lefteqn{\int dt_\alpha d^dx_\alpha dt_\beta d^dx_\beta
\{ \hat{s}^\alpha(t_\alpha,x_0) s^\alpha(t_0,x_\alpha)
\hat{s}^\beta(t_\beta,x_0)
s^\beta(t_1,x_\beta)
\}}\\
&=& \int dt_\alpha d^dx_\alpha \{\hat{s}^\alpha(t_\alpha,x_0)
s^\alpha(t_0,x_\alpha) \}
\int  dt_\beta d^dx_\beta \{ \hat{s}^\beta(t_\beta,x_0)
s^\beta(t_1,x_\beta)
\}
\nonumber
\eea
where $\{~\}$ stands for the path integral over the product with the
$\delta$-function weight in $Z$ that singles out the correct path
through the space of possible states for the given configuration
of random fields and the given history.
In \eq{DeltaSaDeltaSb} the two replicas are uncoupled since we have
not yet averaged over the random fields. As we have seen in appendix
\ref{ap:RG-details}
\bea
\label{DeltaSa}
\lefteqn{
(\Delta S/\Delta H)_\alpha \equiv
\int dt_\alpha d^dx_\alpha \{\hat{s}^\alpha(t_\alpha,x_0)
s^\alpha(t_0,x_\alpha) \}=} \\
\nonumber
& &\int dt_\alpha d^dx_\alpha
\frac{\partial}{\partial t}_\alpha
{\it lim}_{\Delta H \rightarrow 0} \\
\nonumber
& &\left(s^\alpha(t_0,x_\alpha)|_{H_{x_0}^a(t_0)=H(t_0)+\Delta H
\Theta(t_0-t_\alpha)}
-s(t_0,x_\alpha)|_{H_{x_0}^a(t_0)=H(t_0)}\right)/\Delta H
\eea
is the response of replica $\alpha$ to a perturbing pulse of
amplitude $\Delta H$ applied at field $H(t_\alpha)$ at site $x_0$
integrated over the entire system.

If no spin flips in response to the perturbation, the
total response will be
\be
(\Delta S/\Delta H) = \Delta S_{\it harmonic}/\Delta H = C_2
\ee
where $C_2$ is a constant that depends only on the parameters $k$,
$J$, and the coordination number $z$ of the lattice.

If, on the other hand the perturbation triggers an avalanche
of spin flips from the lower to the upper potential well,
$\Delta S =S_{\it flip}\equiv S_\alpha$ will be of the order
of the number of spins
participating in the avalanche (see also appendix
\ref{sub:calculating-some-u-mn}).\footnote{$S_\alpha$
is not exactly equal to the number of spins
flipping in the avalanche. It contains also the harmonic response
that each spin flip causes through the coupling to the neighboring
spins.
This harmonic response couples back to the original spin and
propagates to the next-nearest neighbors with
an amplitude damped by the factor $J_{ij}/k$ and
so on. Occasionally it may cause an avalanche
to continue which would otherwise (in the hard spin model)
have come to a halt. However
since this is a short-ranged effect, we do not expect it to be of
any relevance to the scaling behavior on long length scales. In
mean-field theory the harmonic response
only amounts to a constant factor relating
$S_\alpha$ to the number of spins participating in the avalanche.}

The expression in \eq{DeltaSaDeltaSb} is the product of the total
response to the same perturbation at site $x_0$ measured
in replica $\alpha$
at time $t_0$ and in replica $\beta$ at time $t_1$.
At finite sweeping rate $\Omega/k$ the corresponding values $((\Delta
S)_\alpha/\Delta H)$ and $((\Delta S)_\beta/\Delta H)$ do not have to
be the same, since the responses are measured at potentially
different values of the external magnetic field ($H_0 \equiv H(t_0)$ and
$H_1 \equiv H(t_1)$ respectively).
(We only consider the
adiabatic case, in which the sweeping rate $\Omega/k$ is small
compared to the relaxation rate $\Gamma_0 k$, so that the magnetic
field can be assumed to be constant during the course of an
avalanche.\footnote{
We take the adiabatic limit $\Omega \rightarrow 0$ at finite
correlation length $\xi$,
{\it before}
approaching the critical point of diverging avalanche size and
time to
avoid triggering a new avalanche before the previous one has come to
a halt.
This is consistent with our computer simulations at finite system
sizes where avalanches occur only sequentially.})

Without loss of generality let us assume that
$H(t_1) \geq H(t_0)$.
First we discuss the case that there is an avalanche $S_\alpha$
triggered by the perturbation of amplitude
$\Delta H$ in replica $\alpha$ at field $H_0$.
We further assume that $t_1$ is much bigger than $t_0$, such
that $H_1 \geq H_0 + \Delta H$.
In this case the response
to the pulse in replica $\beta$ will be substantially different from
the response $S_\alpha$ in replica $\alpha$. The spins
that are
pushed over the brink by the {\it perturbation} at field $H_0$ in
replica $\alpha$,
will in replica $\beta$ be triggered by the increased
external magnetic field {\it before} it reaches the bigger value
$H_1$ at which the response is measured.
For $\Omega/\Gamma_0$, $\Delta H$, and $H_1-H_0$ small enough, the
response in replica $\beta$ at field $H_1$ will then be just the
harmonic response $C_2$ or a {\it different} avalanche.
If it is the harmonic response,
the expression in \eq{DeltaSaDeltaSb} takes the form
\be
(\Delta S/\Delta H)_\alpha (\Delta S/\Delta H)_\beta = (S_\alpha/\Delta
H) C_2 \, .
\ee
Similarly one might imagine scenarios in which there is an
avalanche $S_\beta$ triggered only in replica $\beta$, {it i.e.}
\be
(\Delta S/\Delta H)_\alpha (\Delta S/\Delta H)_\beta = (S_\beta/\Delta
H) C_2 \, ,
\ee
or where there is no avalanche happening at either field value
\be
(\Delta S/\Delta H)_\alpha (\Delta S/\Delta H)_\beta =
(C_2)^2 \, .
\ee
It is also possible that
two {\it different} avalanches are triggered
in the two replicas:
\be
(\Delta S/\Delta H)_\alpha (\Delta S/\Delta H)_\beta =
(S_\alpha/\Delta H) (S_\beta/\Delta H)
\ee
with $S_\alpha \neq S_\beta$.

We are interested however in contributions
due to the {\it same avalanche} response in both replicas
\be
(\Delta S/\Delta H)_\alpha (\Delta S/\Delta H)_\beta =
(S_\alpha/\Delta H) (S_\beta/\Delta H)
\ee
with $S_\alpha = S_\beta$.
As we have seen,
a necessary condition is that $H_0-\Delta H \leq H_1 \leq
H_0 + \Delta H$.
We denote with $P_{\it flip} = c_0 \Delta H + o((\Delta H)^2)$
(with $c_0$ a constant in the critical regime)
the fraction of all possible configurations
of random fields in which a local perturbation
of amplitude $\Delta H$ at field $H$, applied at site $x_0$,
causes at least one spin to flip.
For $\Omega$ and $\Delta H$ small enough the fraction of all possible
configurations of random fields in which the local perturbation
will lead to the same initial spin flip triggering the same
avalanche $S$ in replica $\alpha$ and replica $\beta$,
is to leading order in $\Delta H$ proportional to the size of the
overlap $ P^{\it both}_{\it flip} $ of the two intervals
$[H_0, H_0 + \Delta H]$ and $[H_1, H_1 + \Delta H]$,
multiplied by $P_{\it flip}$, with
\be
\label{p-both-flip-def}
 P^{\it both}_{\it flip} =
(1 - \Theta(|H_1-H_0|-\Delta H)) (\Delta H
-|H_1-H_0|)/\Delta H  \,
\ee
(see figure \ref{fig:Pbothflip}).
We can now compute the
random-field average of the expression in \eq{DeltaSaDeltaSb},
denoted by $\langle~\rangle_f$ to leading order in $\Delta H$
\bea
\label{S2-response-2}
\nonumber
\langle \frac{\Delta S}{\Delta H_\alpha} \frac{\Delta S}{\Delta
H_\beta} \rangle_f &=& {\bar C}_1 \frac{\langle S^2 \rangle_f}{\Delta
H^2}
P^{\it both}_{\it flip} P_{\it flip} +
{\bar C}_2 \frac{\langle S \rangle_f}{\Delta
H} \left(1-P^{\it both}_{\it flip} \right) P_{\it flip} \\
& &+ (C_2)^2
+ \langle S_\alpha S_\beta \rangle/(\Delta H)^2 P_{\it
flip}^2 (1-P^{\it both}_{\it flip}) \, ,
\eea
where $\langle S^2  \rangle_f$ is the mean square avalanche size,
and $\langle S  \rangle_f$ is the mean avalanche size, and
${\bar C}_1$ and ${\bar C}_2$ are constants in the critical regime.
The last term accounts for cases in which two different avalanches
$S_\alpha \neq S_\beta$ are triggered in the two replicas.

The last three terms in \eq{S2-response-2}
approach a constant as $\Delta H
\rightarrow 0$, since $P_{\it flip} \sim \Delta H$.
We will now analyze the first term, which is proportional to
$\langle S^2 \rangle$ in more detail.
The function multiplying $\langle S^2 \rangle_f$ is sharply peaked
around $H_0=H_1$ (see figure \ref{fig:Pbothflip}).
Since $P_{\it flip} \sim \Delta H$ it is proportional to
$P^{\it both}_{\it flip}/ \Delta H $. From \eq{p-both-flip-def}
we have
\be
\int_{H_0-\Delta H}^{H_0+\Delta H} dH_1 P^{\it both}_{\it
flip}/ \Delta H  = 1
\ee
independent of $\Delta H$.
The same integral applied to the other terms in \eq{S2-response-2}
yields contributions of order $O(\Delta H)$ which are negligible
compared to the first term as $\Delta H$ is chosen small.
With $H_1= H_0 + \Omega t$ we can express the integral in terms
of time
\be
\int_{-\Delta H/\Omega}^{\Delta H/\Omega} \Omega dt_1 P^{\it
both}_{\it flip}/ \Delta H = 1 \, .
\ee
We then obtain
\be
{\it lim}_{\Delta H \rightarrow 0} {\it lim}_{\Omega \rightarrow 0}
\int_{-\Delta H/\Omega}^{\Delta H/\Omega} \Omega dt_1 \langle
\frac{\Delta S}{\Delta H_\alpha} \frac{\Delta S}{\Delta H_\beta}
\rangle_f = {\bar C}_1 \langle S^2  \rangle_f \, .
\ee
With \eq{DeltaSa} this leads to the scaling relation
\be
\langle S^2 \rangle_f \sim \int dt_1 \int dt_\alpha dt_\beta
d^dx_\alpha
d^dx_\beta
\langle
 \hat{s}^\alpha(t_\alpha,x_0) s^\alpha(t_0,x_\alpha)
\hat{s}^\beta(t_\beta,x_0)
s^\beta(t_1,x_\beta)
\rangle_f \, .
\ee
which was to be shown.
In this notation we have suppressed the factor $\Omega$ and
the various limits for clarity. The integrals over time extend
from $-\infty$ to $+\infty$ with an infinitesimal associated change
in magnetic field.

%XXXXXXXXXX figure fig:Pbothflip XXXXXXX
%\begin{figure}
%\vbox to 4in{
% \vfil
% \special{psfile=Pbothflip.ps
%          hoffset=-340
%          voffset=-150
%          hscale=60
%          vscale=60}
%}
%\vspace{2cm}
%\caption{
%The function $P^{\it both}_{\it flip}$ defined in equation
%({\protect{\ref{p-both-flip-def}}}), plotted as a function of $H_1$.
%In the figure, $dH$ denotes the amplitude which is called $\Delta H$
%in the text.
%\label{fig:Pbothflip}}
% \end{figure}

\subsection{Feynman rules for two replicas}
\label{sub:aval-renormalization}

We study the behavior of $S^{\alpha \beta}$ of
\eq{tilde_S_alpha_beta} under coarse graining
analogously to the calculation done before for just one replica,
with the
difference that instead of two, there are now four fields to be
considered (two for each replica).\footnote{In section
\protect{\ref{sec:form-2-repl}} we already derived the appropriate
partition function. In this appendix we use the notation introduced
there.}
In the following section we briefly describe the associated Feynman
rules. This section may be skipped by the reader uninterested in the
details, since it turns out that there are no loop corrections
to $O(\epsilon)$ to $\langle S^2 \rangle$.

In the Feynman graphs for the loop corrections,
the fields of the $\alpha$ replica are symbolized by arrows on full
lines, whereas those for the $\beta$ replica are symbolized by
arrows on dashed lines.
A vertex $u_{mnpq}$ has then $m$ outgoing arrows on full lines,
$n$ incoming arrows on full lines, $p$ outgoing arrows on dashed
lines and $q$ incoming arrows on dashed lines.
In this notation, the fact that $u_{0npq}=0$ if $n\neq 0$,
$u_{mn0q}=0$ if $q\neq 0$ and $u_{0n0q}=0$, which we discussed in
section \ref{sec:form-2-repl},
means that any vertex with incoming arrows of a certain replica
must have at least one outgoing arrow of the same type of replica,
i.e. there are no ``sinks'', with only incoming lines of a certain
replica.
Furthermore, since the spins from different replica do not interact
directly, and since $u_{0,1,1,0}=u_{1,0,0,1}=0$, there are only
two kinds of propagators, one for each replica. In other words,
in any diagram,
an outgoing line can be connected only to an incoming line of
the same replica.

Using the above rules and causality, one finds that
corrections to vertices
with lines of only one replica, can only receive corrections
from vertices of the {\it same} replica.
There are no contributions
from diagrams that also involve the other replicas. That means that
our results for the magnetization and other quantities that can be
calculated using only one replica, are unaffected by the
introduction of a second replica.

``Pure'' (or one-replica) vertices which depend on more than one
time usually have several different contributions. For example
the
vertex $u_{2,2}(t_1,t_2,t_3,t_4)$ has two main contributions that
are obtained by partial integration of the corresponding term in the
action as discussed in appendix \ref{ap:RG-details}. One contribution
is derived from $u_{2,2}(t_1,t_2,t_3,t_4)$ and has
$t_1=t_3^+$ and $t_2=t_4^+$. The other contribution is derived from
$u_{2,2}(t_1,t_2,t_3,t_4)$ and has $t_1=t_3^+$ and $t_1=t_4^+$.
If we have two replicas, there are {\it different} ``mixed'' vertices
each of which corresponds to one such
the individual contribution to a pure vertex.
Each of them has the same bare value as such a pure
counterpart, since both are obtained in the
same way from mean-field theory.
With ``corresponding'' we mean that the times associated with the
different legs of the mixed vertex, are assigned in the same way to
the legs of the corresponding part of a pure vertex.
The part of a pure vertex formally corresponding to
$u_{1,1,1,1}$, for example, is given by that contribution to
$u_{2,2}(t_1,t_2,t_3,t_4)$, which has $t_1=t_3^+$ and $t_2=t_4^+$.
Conversely the part of a pure vertex corresponding to $u_{1,2,1,0}$
is given
by that contribution to $u_{2,2}(t_1,t_2,t_3,t_4)$, which has
$t_1=t_3^+$ and $t_1=t_4^+$.
Notice, that in any mixed vertex all legs
carrying a certain time label (one outgoing and any number
of incoming arrows),
must belong to the same replica.
In contrast to the different contributions to one pure vertex,
the corresponding mixed vertices do not add up to a single mixed
vertex,
since they are multiplied by field from different replicas
$\eta^\alpha \neq \eta^\beta$.

The loop corrections to mixed vertices formally look the same as
those the corresponding parts of the pure vertices.
For each loop correction
to a mixed vertex there is a matching correction to the corresponding
{\it part} of the pure
vertex and vice versa. The combinatoric factors are also
the same. This implies
in particular that
choosing the same spin rescaling for both replica as we did before
in the case of only one replica, renders marginal
not only $u_{2,0}^{\alpha}$ and $u_{2,0}^{\beta}$,
but also $u_{1,0,1,0}$.

\subsection{Scaling of the second moment of the avalanche
size distribution}

We need to find the scaling behavior of the ``Greens function''
\be
\langle
 \hat{s}^\alpha(t_\alpha,x_0) s^\alpha(t_0,x_\alpha)
\hat{s}^\beta(t_\beta,x_0)
s^\beta(t_1,x_\beta)
\rangle_f
\ee
from its behavior under coarse graining.
The topology of the diagrams permits no $O(\epsilon)$ loop
corrections to the corresponding vertex function.

One finds the canonical dimensions of the fields \cite{Ryder}
(where ``dimension of'' is denoted by ``[~]'' and $\Lambda$ is the
upper cutoff in momentum):
$[\eta(p,\omega)] = \Lambda^{-d/2-2-z}$,
$[\hat \eta(p,\omega)]\sim \Lambda^{-d/2}$.

For calculating Greens functions one introduces source terms in the
action.
{}From the (functional) derivative with respect to the source fields,
one obtains the corresponding average correlation functions.
In the end the source fields are taken to zero again, since usually
they have no physical significance.
In our case the following three source terms are needed:
$\int d^dq \int d\omega L({q,\omega}) \eta(q,\omega)$,
$\int d^dq \int d\omega {\hat L}({q,\omega}) {\hat \eta}(q,\omega)$,
and the term needed for the calculation
of the (spacially) composite operator in $\langle S^2
\rangle_f$, given by
\be
\int d^dq \int d\omega_1 \int d\omega_2
L_2({q,\omega_1,\omega_2}) \int d^dq {\hat \eta}(q,\omega_1)
{\hat \eta}(p-q,\omega_2)\, .
\ee
$L$, ${\hat L}$ and $L_2$ are the respective source fields:
the corresponding canonical dimensions are
$[L(q,\omega)] \sim \Lambda^{-d/2 + 2}$,
and $[\delta/\delta L(q,\omega)] \sim \Lambda^{(d/2-2)}
\Lambda^{(-d-z)}
\sim \Lambda^{-d/2 -2 -z}$.
Similarly $[\hat L(q,\omega)] \sim \Lambda^{-d/2-z}$, and
$[\delta/\delta \hat L(q,\omega)] \sim \Lambda^{-d/2}$.
And also $[L_2(p,\Omega)] \sim \Lambda^{(d-2z)}$, and $[\delta/\delta
L_2(p,\Omega)] \sim \Lambda^0$.
{}From \eq{s2}
and the fact that Greens functions in the fields $\eta$
and $\hat\eta$ scale in the
same way as those in terms of $s$ and $\hat s$ (see section
\ref{source-terms}),
we then find (without loop corrections) that
$\langle S^2 \rangle_f \sim \Lambda^{-(4+z)}$.
Below the upper critical dimension, the canonical dimensions
of the fields $\eta(q,\omega)$ and $\hat\eta(q,\omega)$
are corrected by $\Lambda^{(\eta/2)}$ and
$\Lambda^{(\eta-\bar\eta/2)}$ respectively.
With $\eta=\bar\eta$ from the mapping to the pure
Ising model \cite{Krey}, one obtains (to $O(\epsilon)$)
$\langle S^2 \rangle_f \sim \Lambda^{-(z+(2-\eta)2)}$.
Similarly, one finds for the higher moments
$\langle S^n \rangle_f \sim \Lambda^{-((n-1)z+(2-\eta)n)}$
to $O(\epsilon)$.
In section \ref{scaling-of-s2-first} this result is compared
to the scaling behavior of $\langle S^2 \rangle$ as obtained
from the scaling form of the avalanche size distribution
\be
\langle S^2 \rangle \sim r^{(\tau-3)/\sigma}
{\cal S}_\pm^{(2)}(h/r^{\beta\delta})
\ee
(with the appropriate scaling function ${\cal S}_\pm$)
to extract the results for $1/sigma$ and $\tau$.
There we obtain
\be
1/\sigma = 2 + \epsilon/3 + O(\epsilon^2).
\ee
and
\be
\tau = 3/2 + O(\epsilon^2) \, .
\ee

%%%%%%%%%%%%%%%%%%%%%%%%%%%%%%%%%%%%%%%%%%%%%%%%%%%%%%%%%%%%%%%%%%%%%%%
%% This is a REVTEX file, which hopefully will run on your system too.%
%%%%%%%%%%%%%%%%%%%%%%%%%%%%%%%%%%%%%%%%%%%%%%%%%%%%%%%%%%%%%%%%%%%%%%%

\section{Related Problems}
\label{ap:related}

\subsection{Comparison with depinning transitions}

\subsubsection{Relation to fluid invasion in porous media}
\label{ap:Robbins}

Much progress has been made in the study of fluid invasion
of porous media
\cite{porous-media-Robbins-ref-5-17,fluid-invasion,Robbins}.
A preexisting fluid in the porous medium, for
example oil in rock, is displaced by an invading fluid (water),
which is
driven by an applied pressure $P$. The interface between the two
fluids is pinned at pressures lower than a threshold value $P_c$ and
advances continuously at higher pressures. The interface depinning
transition is accompanied by a diverging correlation length,
critical fluctuations and universal exponents. There are three
different universality classes for the associated critical
exponents, corresponding to low, intermediate and high disorder in
the system. At low disorder the marginally stable interface
at $P_c$ is faceted, at intermediate disorder it is self-affine and
at high disorder it is self-similar.

Already before we did our work on hysteresis,
Robbins, Cieplak, Ji and Koiller
have pointed out the analogy of fluid invasion to the physics of domain
wall motion in Ising ferromagnets
\cite{Robbins,Nattermann,NarayanIII}. There, the interface separates
regions of up and down spins. As the magnetic field is ramped, one
domain grows at the expense of the other --- the interface is pushed
forward. Quenched disorder may be due to random fields or random
bonds.
The authors study the zero temperature nonequilibrium RFIM with a
{\it rectangular distribution} of random fields of width $\Delta$ and a
preexisting flat interface separating up spins from down spins. They
increase the external magnetic field adiabatically and study
avalanches of spin flips that are triggered by flipping spins at the
interface. As in our system a spin in such an avalanche flips up
if its local effective field becomes positive. {\it The main difference is
that in their system no spins are allowed to flip ahead of the interface.}
Each spin flip can be interpreted as an advancement of the
pre-existing system-spanning
interface. The value of the magnetic field at which any part of the
interface reaches the other side of the system is the critical field
which corresponds to the threshold pressure $P_c$ in fluid invasion
in porous media. As in fluid invasion, Ji and Robbins find in 3
dimensions a faceted regime at low disorder $\Delta$,
a self-affine regime at intermediate $\Delta$ and a percolation
regime at large $\Delta$. At a critical width $\Delta_2^c=3.41 J$,
which separates the self-affine regime from the self-similar regime,
a diverging length scale in a bulk quantity (``fingerwidth'') is
observed. The critical exponents associated with the interface
depinning transition in the self affine regime have also been
calculated in an $\epsilon$-expansion \cite{Nattermann,NarayanIII}.
They are the same for random-field and random-bond
disorder \cite{NarayanIII}.

We expect that in our system it would be possible to observe the
same critical behavior at the onset of the infinite avalanche
in large enough systems with less than critical disorder ($R<R_c$).
If the system is big enough, there will certainly be
somewhere a rare large cluster of flipped spins, even at relatively
low magnetic fields. As the field is slowly raised, the surface of
such a cluster is expected to act as a preexisting interface
analogously to Ji and Robbins system. The small clusters that are
flipped ahead of the interface probably have no influence on
the long length scale behavior and
the critical exponents associated with interface progression such
as roughening exponents.
The onset field for the infinite avalanche in an infinite system
corresponds to the threshold field at which a preexisting interface
gets depinned in Ji and Robbins system.

Numerical and analytical
studies
%% FOLLOWING LINE CANNOT BE BROKEN BEFORE 80 CHAR
\cite{NarayanI,NarayanII,NarayanIII,Nattermann,Robbins,Myers,Middleton,Ertas,flux-lattice}
leave no doubt that
the interface depinning
transition
has an associated diverging height-height correlation length and
critical fluctuations. There are scaling forms for quantities
related to the shape of the interface and its progression, which
suggest that the interface gets depinned in a second order
transition.
Nevertheless in our system (in three dimensions)
we have called the onset of the infinite
avalanche in systems with less than critical disorder an abrupt
or ``{\it first} order'' transition. What's come over us ?
Our numerical simulations in three dimensions clearly
show a kink in the magnetization curve at the threshold field
$H_c(R)$,
rather than power law behavior as expected in continuous transitions.
This is not a contradiction. One has to be careful about which
quantities are measured:
The critical fluctuations at the interface yield no contribution to
bulk quantities ({\it e.g.} the magnetization), which are measured
per unit volume. In references \cite{thesis,eps-big}
we propose an experimental setup that would allow one to observe
bulk and interface fluctuations simultaneously.
In the conventional setup for measuring magnetic hysteresis
loops discussed in this paper however, the second order depinning transition
with a diverging height-height correlation function at the interface
discussed by Robbins {\it et al.} is buried invisibly inside the
line of abrupt (``first order'') transitions seen in our
magnetization curves in sufficiently low dimensions.
Note that this connection can only be established for large enough
systems, where rare large fluctuations provide a preexisting
interface of flipped spins. In smaller systems, disorder induced
nucleation effects will determine the size of the onset field for
the infinite avalanche in our hysteresis model.

At the critical disorder $\Delta_2^c$, separating the self-affine
regime from the self-similar regime, Ji and Robbins find a diverging
bulk length scale, the ``fingerwidth'' of regions of flipped spins.
The critical exponents associated with this transition are
different from the exponents seen at our critical endpoint $(R_c,
H_c(R_c))$. This is not too surprising: unlike the case for $R<R_c$,
near $R_c$ a large advancing interface in our problem runs into
pre-existing flipped regions {\it of all sizes} --- presumably
changing the universality class.
Also we have several infinite fronts at $R_c$.
Ji and Robbins report that for a bounded nonanalytic distribution of
random fields the corresponding
exponents are nonuniversal in the depinning problem, but depend
on the analytic form of
the edges of the distribution of random fields.
However, for an analytic distribution they presumably take universal
values \cite{Robbins-private}.
In our problem we use an analytic distribution and
$\nu$ is universal, as is shown by the RG calculation.
It has been shown \cite{Olami} however that the exponents are
different for a rectangular distribution of random fields in our
system also.
Perhaps it would not be universal in our problem either for rectangular
distributions of random fields \cite{Olami}
(or other distributions with singularities at their tails).
In section \ref{sub:other-models} this issue will be discussed
in more detail.

Above a critical amount of disorder, using single interface dynamics,
Ji and Robbins see a self-similar regime, where the interface
grows to a percolating cluster at a certain ``threshold field''
$H_c^{\it single}(R)$.
In our dynamics in an infinite system
such a percolating cluster might occur already at
a lower value of the external magnetic field,
because it is probably easier to connect preexisting clusters
of flipped spins to form a
percolating cluster than it is to push a domain wall through the
system all the way till it reaches the the opposite side.
One would expect the same argument to apply for $R \leq R_c$,
{\it i.e.} $H_c(R) \leq H_c^{\it single}(R)$.
In references \cite{thesis,eps-big2} we discuss four potential
experimental setups
for measuring Barkhausen noise (or avalanche size distributions
in general), explaining for each case whether the single interface
model of Ji and Robbins or our model with
many interfaces and domain nucleation is expected to apply.

\subsection{Other models}
\label{sub:other-models}

\subsubsection{Adiabatic models for hysteresis}

There are several numerical studies of related hysteresis models,
which we discuss in more detail in reference
\cite{thesis,eps-big}. Here we only list a few closely related
models.

Maslov and Olami \cite{Olami}
have simulated the same model as we have studied, but for a
rectangular distribution of random fields \cite{Olami} rather than
a Gaussian. The authors find different critical exponents than ours
in 3 dimensions and in mean-field theory. They report that
their numerical results speak in favor of the upper critical
dimensions being $d_c=6$. They claim that the model does not belong
to any known universality \label{page:foottails}
class.\footnote{Presumably in each dimension
there will be a critical power law for the tails of a bounded
distribution, so that distributions with a power law larger than
this critical value will lead to
different critical exponents. In this sense the tails of a
rectangular distribution can be thought of as an infinite power law,
and conceivably might lead to a different universality
class than ours in all dimensions.
\label{foot:tails}}

Koiller, Ji and Robbins \cite{Robbins} point out that for a
rectangular distribution at $R<R_c$ there is a close connection to
diffusion percolation and bootstrap percolation in $2$
and also in $3$ dimensions \cite{Robbins-private},
which are modified percolation models in which the occupation of a
site depends on its environment \cite{bootstrap-percolation}.

Coram, Jacobs, and Heinig \cite{Coram}
have studied the zero temperature nonequilibrium random bond Ising
models, both spin glasses (SG) and random ferromagnets (RFM) with
nearest neighbor interactions in one, two and three dimensions.
They report power law sensitivity to single-spin-flip perturbations
in 2 and 3 dimensional random ferromagnets, if, starting from the
spin state with all spins pointing down (ground state at $H=0$) the
field has been raised to a positive critical value.
The associated critical exponents are $\tau = 1.37$ in 2 dimensions,
and $\tau = 2.8$ in three dimensions.
It is likely that the value of the exponent $\tau$
would be different for a Gaussian distribution of random bonds
(see footnote \ref{foot:tails} on page \pageref{page:foottails}).

Vives {\it et al.} \cite{Ortin-RBIM}
have studied that case for a negative mean of the Gaussian
bond distribution. They  trigger avalanches
by a slowly increasing homogeneous external magnetic field (as in our
model).
For the avalanche size distribution integrated over one branch of the
hysteresis curve
they find $\tau+\sigma\beta\delta=2.0 \pm 0.2$
in three dimensions and $\tau+\sigma\beta\delta=1.45 \pm
0.1$ in two dimensions, which are rather close to the values of
the nonequilibrium RFIM (see section \ref{sub:universality}).
One would expect the value the exponent $\tau$
to be smaller than $\tau+\sigma\beta\delta$, {\it i.e.}
their result seems to
deviate from the number obtained by Coram {\it et al.}
for a rectangular distribution of ferromagnetic bonds with strengths
between 0 and 1.
That is not surprising if we consider the differences in
the two approaches.
It might be interesting to compare the shape of the
hysteresis loops of the two models at various values of the
relevant tunable parameters. A similar comparison of our
hysteresis loops to those of Maslov and Olami's RFIM with a
rectangular distribution of random fields \cite{Olami}
revealed marked differences of the two models. For example,
in our model at less than critical randomness, there were precursors
to the infinite avalanche, while in Maslov and Olami's model
there were none.

Bertotti and Pasquale \cite{BertottiI} have studied
hysteresis phenomena in
the Sherrington-Kirkpatrick spin glass model
\cite{SK} with $N$ Ising spins $s_i=\pm1$ on a lattice
with random infinite range interactions that are distributed
according to a Gaussian with mean zero.
They used a
slightly generalized dynamics that allows for temperature
like fluctuations at a finite sweeping
frequency $\Omega$ with $H=\Omega t$.
They report power law scaling of the power spectrum
due to avalanches of spin flips
near the central
part of the saturation hysteresis loop, where the rate of change
of the magnetization $dM/dt$ is approximately constant.
The authors note that this scaling behavior as well as the
associated shape $F(\omega)/\Omega$ very much resembles the behavior
observed in Barkhausen effect experiments in soft magnetic materials
\cite{Bertotti-experiment}.

For sweeping rate $\Omega \rightarrow 0$ we can think of their model
without random fields
as an infinite range mean-field theory
for the nonequilibrium random bond Ising
model, simulated by Vives {\it et al.}
\cite{Ortin-RBIM}.
Vives {\it et al.} found that
in two and three dimensions the nonequilibrium RFIM
does reveal a critical point of the kind
we found in the nonequilibrium RFIM. It would be interesting
to look for the same kind of critical point in Bertotti {\it et
al.}'s
nonequilibrium SK-model (with and without random fields), for
example by setting the mean $J_0$ of the distribution of random
bonds to a nonzero value
and tuning the widths of distributions of random fields and random
bonds.

Rudyak \cite{Rudyak} has suggested a theory for dielectric hysteresis in
ferroelectrics which leads to the same picture as our hard-spin
mean-field theory for $R<R_c$.

\subsubsection{Dynamical and other hysteresis models}

There are several studies of scaling behavior of the area of the
hysteresis loop as a function of driving frequency $\Omega$ and
amplitude $H_0$ of the of the external magnetic field $H(t) = H_0
\sin \Omega t$, and other dynamical effects
\cite{dynamic-hysteresis,BertottiI,Rao-Krishnamurthy-Pandit}.
Most of these model continuum and lattice spin
systems do incorporate temperature effects, but no quenched disorder.
Rao, Krishnamurthy and Pandit give a nice review and discussion of
previous (empirical) hysteresis models such as the Preisach model
and mean-field types of theories like the Stoner-Wohlfarth theory and
others \cite{Rao-Krishnamurthy-Pandit}.

\subsubsection{Conjectures about other models in the same universality
class}
\label{sub:universality}
Recently Vives, Goicoechea, Ort\'in, and Planes
found \cite{Ortin-RBIM} that the
numerical exponents $\nu$, $\beta$, $\tau$, and $z$
in the nonequilibrium zero temperature RFIM and the RBIM (random
bond Ising model) have very similar values in two and three
dimensions. In two dimensions, the exponents for
the RFBEG (random field Blume-Emery-Griffiths
model \cite{RFBEG}) seem to be similar also.
In this interesting paper the authors suggest that these
models might actually be in the same
universality class. Admiring their work, we however have some
concerns as to whether their critical exponents will remain
unchanged for larger system sizes:
they used systems of linear size
up to $L=100$; we used much larger systems, up to $7000^2$
and $800^3$ for the RFIM, and found
that finite size effects are actually quite prominent and lead
to shifted results for the exponents.
Although the {\it equilibrium} versions of these
models are not in the same universality class,
it is known \cite{Nattermann,NarayanIII}, however, that
the nonequilibrium {\it single interface depinning} transitions
of the RFIM and the RBIM do have the same critical
exponents.

In the following section we will discuss some symmetry arguments,
that
would indeed speak in favor of Vives {\it et al.}'s conjecture
and would suggest that the universality class of our model
extends even beyond just the RBIM. A large universality class
would also explain the surprisingly good agreement with experiments
discussed in section \ref{sec:numerics-and-experiment}.
Generally one may ask how robust the universality class of our model
is against the introduction of other kinds of disorder, other
symmetries for the order parameter,
long range interactions, different dimensions, and altered dynamics.
The variation with dimension has already been discussed at the
appropriate places in this paper (see for example
section \ref{sec:numerics-and-experiment}).

\vspace{2cm}
{\bf Other forms of disorder and symmetries}

\vspace{2mm}
\noindent
If a new kind of disorder in an otherwise unaltered system
changes neither the symmetries, nor the interaction range,
nor the dynamics, nor the relevant dimensions,
we may be hopeful that it does not lead to a different
universality class.

{\bf Random fields and random bonds:}
\label{par:RFBIM}
Uncorrelated fluctuations in the
nearest neighbor coupling strengths (random bonds)
in the presence of random field
disorder do not break any new symmetries.
Our random field Ising model fulfills two Harris criteria
$\nu/\beta\delta \geq 2/d$ and $\nu \geq 2/d$. Adding random bond
disorder cannot destroy the fixed point in the Harris criterion
sense through added statistical fluctuations, because the random field
disorder has already broken the relevant (translational) symmetry.
It then seems plausible that systems with random bonds and random
fields are in the same universality class as systems with random
fields only. The ultimate justification for this conjecture may be drawn
from the renormalization group picture. If the change in the
generating functional due to the added new disorder turns out to be
irrelevant under coarse graining, it will not affect the critical
behavior on long length scales. Some preliminary studies seem to
indicate that this would be case for random bonds in the presence of
random fields. Further elaborations on this issue will be presented
elsewhere \cite{eps-big}.

{\bf Random bonds only:}
\label{par:RBIM}
Similarly one might expect systems with random bonds only to be in
the same universality class also. Because the critical magnetization
$M_c \equiv M(H_c(R_c))$
is nonzero, the time reversal
invariance will be broken at the critical point,
just as it is broken in the case of
random fields. The symmetries of the
random field model and the random bond model would then be the same.
In a soft spin model the dynamics could also be defined in the same
way, using relaxational dynamics.
One would then expect to see the same critical behavior on
long length scales. In fact, in the random bond model
one may consider the spins that flip
outside the critical region to act as random fields for the spins
that participate in the large avalanches near the critical
point. We have already suggested that random bonds in the presence of
random fields do not change the critical behavior. It then seems
plausible that the random bond problem would be in the same
universality class also, as numerical simulations seem to confirm
\cite{Ortin-RBIM}.
As a warning to the enthusiastic reader we should mention that the
simplest mean field theory for this problem has some sicknesses that make it
difficult to verify this conjecture in the RG framework.
We believe, however, that these subtleties occur only in the infinite
range mean-field theory, and that they are unimportant
for the behavior in finite dimensions.
Some simulations of the infinite range model with random bonds
are reported in \cite{BertottiI,Bertotti-SK-model}.

{\bf Random anisotropies:}
Realistic models of Barkhausen noise in polycrystalline magnets
usually involve random anisotropies rather than random fields.
In the same way it appears plausible that a nonequilibrium $O(n)$
model with
random anisotropies \cite{Jiles,Chikazumi,Ashcroft-Mermin}
may be in the same universality class as the nonequilibrium RFIM.
The  external magnetic driving field breaks the rotational symmetry
and time reversal invariance, just as in the case of random bonds.
Again, spins that do not flip in the critical region may act as
random fields for the spins participating in avalanches near the
critical point, so that the essential features are the same as in
our model, and one may expect to see the same critical
exponents.\footnote{It may be that in some strong coupling limit the
system will lose the ability to avalanche and all spins will smoothly
rotate from down to up as the external magnetic field is increased.
Our discussion above refers to the case where the coupling is weaker
and avalanches do occur, as of course they do experimentally.}
The $O(n)$ model with random anisotropies is very similar to a
continuous scalar spin model with random couplings to the external
magnetic field (random ``g-factors''). The mean-field theory for
the random g-factor model turns out to have the same critical
exponents as our random-field Ising model. There are no new terms
generated in the RG description of this model either, it is
therefore expected to be in the same universality class as our
model. By symmetry we would neither expect any change in the
exponents if there was randomness added through a distribution
in the soft-spin potential well curvatures $k$ (see our definition
of the soft-spin potential $V(s_i)$ in \eq{V-def}),
nor if random bonds are added to the system, as may be the
case in real experimental systems.

The RG formalism developed in this paper can be used as a convenient
tool to verify these conjectures. One can
write down the most general generating functional and
verify for each of these models whether
on long length scales the
same kinds of terms become important or irrelevant as in our model.

{\bf Long range interactions:}

The question about the effect of long range interactions is of equal
importance. Depending on the sample shape, dipole-dipole interactions
can lead to long-range, antiferromagnetic interaction forces which
are the reason for the breakup of the magnetization into Weiss-domains
in conventional magnets \cite{Chikazumi,Jiles}.
In the case of martensites there are long range
antiferroelastic strain fields present \cite{Kartha,Martensites}.
In references \cite{thesis,eps-big2} we give an example
of a critical exponent in a system with long range forces
(from avalanche duration measurements in
martensites \cite{Ortin-martensite-aval}) that appears to be quite
different from the
corresponding exponent in our model.
On the other hand, measurements of Barkhausen-noise distributions in
magnets in the presence of long range demagnetizing fields seem to
yield a critical exponent quite close to the corresponding exponent in our
model \cite{Urbach}.

In a recent preprint \cite{Urbach} Urbach, Madison and Markert study
a model for a {\it single} moving domain wall without overhangs in the
presence of infinite range antiferromagnetic interactions
and quenched (random field) disorder. In an
infinite system their model self-organizes\footnote{This
self-organization to the
critical point is similar to the trivial
self-organization expected in an experiment
in the presence of a gradient field,
which we proposed in references \protect{\cite{thesis,eps-big2}}}
without necessary
parameter tuning to the same critical state seen in the
absence of the infinite range interactions right at the interface
depinning threshold \cite{Robbins}.
An analysis \cite{BWR} of our
ferromagnetic RFIM in the presence of infinite range
antiferromagnetic interactions leads to an unchanged critical
behavior except for a tilt of the entire magnetization curve
in the $(M,H)$ plane: here too it does not change the critical
properties.
It would be interesting to see how these results
would change for more physical long-range interactions.
Dipole-dipole interactions decaying
with distance as $1/x^3$, for example, might be appropriate.

\subsection{Thermal fluctuations}
\label{ap:temperature}
{\bf (a) The equilibrium random field Ising model:}

The equilibrium properties of the random field Ising model, in
particular the phase transition from paramagnetic to ferromagnetic
(long range ordered) behavior, have been the subject of much
controversy since the 1970s \cite{Nattermann-review}.
The reason is intriguing:
experimental and theoretical studies of the approach to equilibrium
show that near the critical temperature there seems to appear
a ``glassy'' regime where relaxation to equilibrium becomes very
slow. Activated by thermal fluctuations the system tumbles over free
energy barriers to lower and lower valleys in the free energy
landscape, until it has reached the lowest possible state, the
equilibrium or ground state. The higher those barriers are compared
to the typical energy of thermal fluctuations, the longer the
relaxation process takes.  At low temperatures, due to the
effect of disorder, some of these barriers are so large (diverging
in an infinite system), that the system gets stuck in some
metastable state and never reaches true equilibrium on measurement
time scales.  On long length scales (and experimental time scales)
thermal fluctuations become irrelevant and collective behavior
emerges. When driven by an external field, the system moves through
a local valley in the free energy landscape, and collective behavior
in the form of avalanches is found when the system reaches a
descending slope in the free energy surface.  The present state of
the system depends on its history --- a phenomenon commonly observed
as hysteresis.
\vspace{1mm}

\noindent
{\bf (b) The nonequilibrium random field Ising model:}

We have studied this hysteresis in the zero temperature random field
Ising model, far from equilibrium and in the absence of any thermal
fluctuations.  We found a critical point, at which the shape of the
hysteresis loops (magnetization versus magnetic field) changes
continuously from displaying a jump in the magnetization to a smooth
curve.  Interestingly, the nonequilibrium critical exponents
associated with the universal behavior near this point in $d=3$
dimensions seem to match those obtained from 3 dimensional
simulations of the equilibrium phase transition point
approximately within the error bars (see table \ref{tab:equil-RFIM}).
This is surprising, since the physical starting
points of the two systems are very different. Furthermore, our
perturbation expansion in $\epsilon=6-d$ for nonequilibrium
critical exponents can be mapped onto the expansion for the
equilibrium problem to all orders in $\epsilon$. Our expansion stems
from a dynamical systems description of a deterministic process,
which takes into account the history of the system and is designed to
single out the correct metastable state, while the calculation for
the equilibrium problem involves temperature fluctuations and no
history dependence at all.

\vspace{1mm}
\noindent
{\bf (c) The crossover:}

It would be interesting to see if there is actually a deeper
connection between the nonequilibrium and equilibrium critical
points, and whether the calculation for the nonequilibrium model
could be used to resolve long-standing difficulties with the
perturbation expansion for the equilibrium model.  The idea is to
introduce temperature fluctuations in the nonequilibrium
calculation, and at the same time a finite sweeping frequency for
the external driving force.  The lower the sweeping frequency
$\Omega$ at fixed temperature,
the more equilibrated the system and the
longer the length scale above which nonequilibrium behavior emerges.
Tuning $\Omega$ would allow one to explore the whole crossover
region between the two extreme cases that are found in the
literature (far from and close to equilibrium). Contrary to previous
treatments of relaxation, the history dependence that is so
essential in experimental realizations, emerges naturally from this
approach.  At fixed temperature,
but for progressively lower sweeping frequencies,
one expects to see smaller and smaller hysteresis loops,
asymptotically attaining a universal shape at low enough frequencies.
The tails of these hysteresis loops will match the equilibrium
magnetization curve. In the limit of zero frequency, the hysteresis
loop shrinks to a point, and equilibrium is expected at all values
of the external magnetic field.  On the other hand,
taking temperature to zero first,
should yield nonequilibrium behavior as seen in our recent work.
The prospect of relating equilibrium and nonequilibrium critical
behavior as two limits at opposite edges of the experimentally
relevant crossover regime is an exciting challenge.

%XXXXXXXXXXXXXXXXXXXXXXXXXXXXXXXXXXXXXXXXXXXXXXXXXX
%XXX  Figure Captions
%XXXXXXXXXXXXXXXXXXXXXXXXXXXXXXXXXXXXXXXXXXXXXXXXXX
\newpage
\tightenlines

\begin{figure}
\caption
[Experiment: Magnetic hysteresis loops in a thin Gd film
after annealing the sample at different temperatures]
{{\bf Experiment: Magnetic hysteresis loops of a 60 nm thick Gd film
for
various annealing temperatures} (as indicated next to each loop)
and constant annealing time (3 minutes each).
All measurements are performed at $200 \pm 5 K$. The sweeping
frequency of the external magnetic field is $0.5$ Hz (from A.
Berger, unpublished).
\label{fig:Berger}}
\end{figure}

\begin{figure}
\caption
[Experiment: Distribution of Barkhausen pulse areas in a $81 \%$
Ni-Fe wire
after annealing the sample at different temperatures]
{{\bf Experiment: Distribution of pulse areas ($p$), integrated over
the hysteresis
loop for $81 \%$ Ni-Fe wires after various heat treatments.}
The originally hard drawn wires have been subjected to a one-hour
heat treatment in high vacuum at temperatures of $240^\circ C$ or
$460^\circ C$ and cooled down in the furnace.
(From U. Lieneweg and W. Grosse-Nobis, {\it Int. J. Magn} {\bf 3},
11 (1972).)
\label{fig:Barkhausen-annealing}}
\end{figure}

\begin{figure}
\caption
[Equilibrium magnetization curve $M(H)$
for the pure Ising model at zero temperature.]
{{\bf Equilibrium magnetization curve $M(H)$}
for the pure Ising model at zero temperature.
\label{fig:equil-M(H)}}
\end{figure}

\begin{figure}
\caption
[Nonequilibrium magnetization curve $M(H)$ for the pure Ising model
at zero temperature with a local dynamics.]
{{\bf Nonequilibrium magnetization curve $M(H)$}
in the pure Ising model at zero temperature for the dynamics
defined in the text.
\label{fig:rectangle}}
\end{figure}

\begin{figure}
\caption
[Mean-field magnetization curves for the nonequilibrium
zero temperature random field Ising model at
various values of the disorder.]
{{\bf Mean-field magnetization curves} for the nonequilibrium
zero temperature random field Ising model at
various values of the disorder $R=0.6 J<R_c$ (a),
$R=R_c=\protect{\sqrt{(2/\pi)}} J=0.798 J$ (b), and $R=J>R_c$ (c).
\label{fig:MFT-magnetization}}
\end{figure}

\begin{figure}
\caption
[Mean-field magnetization curves for the soft-spin
version of the zero temperature random field Ising model
at various values of the disorder.]
{{\bf Mean-field magnetization curves for the soft-spin
version} of the zero temperature random field Ising model
at various values of the disorder $R=1.3 J< R_c$ (a),
$R=R_c= 2kJ/((k-J) \protect{\sqrt{2\pi}})=1.6 J$
(see appendix {\protect{\ref{ap:eta0}}}) (b), and $R= 2J>R_c$ (c).
\label{fig:MFT-magn-soft-spin}}
\end{figure}

\begin{figure}
\caption
[Mean-field phase diagram for the nonequilibrium
zero temperature random field Ising model.]
{{\bf Mean-field phase diagram} for the nonequilibrium
zero temperature random field Ising model. The critical point
studied in this paper is at $R=R_c$, $H=H_c(R_c)$, with $H_c(R_c)=0$
in the hard-spin mean field theory. There are two
relevant directions $r=(R_c-R)/R$ and $h=H- H_c(R_c)$ near this
critical point.
The bold line indicates the threshold field $H_c^u(R)$
for the onset of the infinite avalanche upon monotonically increasing
the external magnetic field.
The dashed line describes $H_c^l(R)$ for a decreasing external
magnetic field. The three dotted vertical lines marked (a), (b), and
(c) describe the paths in parameter space which lead to the
corresponding hysteresis loops shown in figure
\protect{\ref{fig:MFT-magnetization}}.
\label{fig:MFT-phase-diagram}}
\end{figure}

\begin{figure}
\caption
[Mean-field phase diagram for the soft-spin version of the
nonequilibrium zero temperature random field Ising model.]
{{\bf Mean-field phase diagram for the soft-spin version} of the
nonequilibrium zero temperature random field Ising model.
The diagram is plotted analogously to figure
\protect{\ref{fig:MFT-phase-diagram}}. Magnetic field sweeps along
the
lines (a), (b) and (c) lead to the corresponding soft-spin
hysteresis curves shown in figure
\protect{\ref{fig:MFT-magn-soft-spin}}.
Note that here, in contrast to the hard-spin model
the value of the
critical field, $H_c(R_c)$ does depend on the history of the
system: for monotonically increasing external magnetic field
$H_c(R_c) =
H_c^u(R_c) = k-J$, and for monotonically decreasing external magnetic
field $H_c(R_c) = H_c^l(R_c) = -(k-J)$ (see appendix
\protect{\ref{ap:eta0}}).
This implies that in contrast to the hard-spin mean-field theory
of figure \protect{\ref{fig:MFT-phase-diagram},}
the soft-spin mean-field theory displays hysteresis for {\it all}
finite
disorder values, {\it i.e.} even at $R \geq R_c$.
\label{fig:MFT-soft-spin-phase-diagram}}
\end{figure}

\begin{figure}
\caption
[Simulated hysteresis curves for two small realizations
of the nonequilibrium zero temperature random field Ising model
in three dimensions.]
{{\bf Simulated hysteresis curves for two small realizations
of the nonequilibrium RFIM in three dimensions.}
Each sample consists of only $5^3$ spins,
with periodic boundary conditions. The two systems have
different configurations of random fields that are taken from
the same distribution $\rho(f)$ with standard
deviation $R=5 J > R_c$.
(In 3 dimensions $R_c=(2.16 \pm 0.03) J$ \protect{\cite{Perkovic}}.)
Note that here, as in all plots of numerical simulation results in
finite dimensions $J$ denotes the strength of the nearest neighbor
coupling $J_{ij}$, which differs from its definition in the
analytical calculation by the coordination
coordination number $z$ -- see footnote \protect{\ref{foot:J}}
on page \protect{\pageref{page:footJ}}.
\label{fig:MFT-simul-m(h)}}
\end{figure}

\begin{figure}
\caption
[Mean-field avalanche size distribution for the nonequilibrium
random field Ising model.]
{{\bf Mean-field avalanche size distribution} integrated
over the
hysteresis loop for systems with 1000000 spins at various disorder
values $R>R_c=0.798 J$:
(a) $R=1.46 J$ (averaged over 10 different configurations of random
fields), (b) $R=1.069 J$ (averaged over 5 different configurations
of random fields), and (c) $R=0.912 J$ (averaged over 10 different
configurations of random fields). Each curve is a histogram
of all avalanche sizes found as the magnetic field is raised from
$-\infty$ to $+\infty$, normalized by the number of spins in the
system.
For small $|r|=|R_c-R|/R$ the distribution
roughly follows a power law
$D(S, r) \sim S^{-(\tau+\sigma\beta\delta)}$
up to a certain
cutoff size $S_{\it max} \sim |r|^{-1/\sigma}$ which scales to
infinity as $r$ is taken to zero.
The straight line above the three data curves in the figure
represents an extrapolation to the critical point $R=R_c$
in an infinite system,
where one expects
to see a pure power law distribution on all length scales
$D(S, r) \sim S^{-(\tau+\sigma\beta\delta)}$
with the mean field values of the corresponding exponents
$\tau+\sigma\beta\delta = 2.25$.
\label{fig:MFT-D(s)}}
 \end{figure}

\begin{figure}
\caption
[Feynman diagrams for the the relevant corrections to first
order in $\epsilon=6-d$.]
{{\bf Feynman diagrams.}
The relevant corrections
to first order in $\protect{\epsilon}=6-d$ for the constant part
$\protect{\chi^{-1}}/J$
in the propagator (a), and for the vertex $u$ (b).
Figure (c) shows an example of a diagram forbidden by causality.
\label{fig:feynman1}}
\end{figure}

\begin{figure}
 \caption
[Phase diagram and renormalization group flows (schematic).]
{{\bf Phase diagram and flows (schematic).} (a) The
vertical
axis is the
external field $H$, responsible for pulling the system from down to
up.
The horizontal axis is the width of the random--field distribution
$R$.
The bold line is $H_c(R)$, the location of the infinite avalanche
(assuming
an initial condition with all spins down and a slowly increasing
external
field).  The critical point we study is the end point of the infinite
avalanche line $(R_c, H_c(R_c))$.\hfill\break
\null\hskip0.2in
Using the analogy with the Ising model (see text) we also show the RG
flows around the critical point.  Here we ignore
the RG motion of the critical point itself: equivalently, the figure
can represent a section through the critical fixed point tangent to
the two unstable eigenvectors (labeled $h$ and $r$).  Two systems on
the same RG trajectory (dashed thin lines) have the same
long--wavelength
properties (correlation functions ...) except for an overall change
in
length scale, leading to the scaling collapse of
equation~(\protect\ref{magn-scaling}).
The $r$ eigendirection to the left extends along the infinite
avalanche
line; to the right, we speculate that it lies along the percolation
threshold
for up spins.
\hfill\break
\null\hskip0.2in
(b) $O(\epsilon)$ RG flows below 6 dimensions in the
$(\chi^{-1},u)$ plane (see text). Linearization around the
Wilson-Fisher (WF) fixed point yields the exponents given to
$O(\epsilon)$ in the table. In the vicinity of the repulsive
$u=0=\chi^{-1}$ (MFT) fixed points one obtains the old mean-field
exponents.
 \label{fig:flows}}
 \end{figure}

\begin{figure}
\caption
[Feynman diagram for the relevant correction to first order
in $\tilde \epsilon=8-d$
for systems with less than critical disorder $R<R_c$.]
{{\bf Feynman diagram.}
The correction to $O(\tilde{\epsilon})$ to the vertex w in
an expansion
about 8 dimensions, see eq.~(\protect\ref{tadpole-w}) in the text.
 \label{fig:tadpole}}
 \end{figure}

\begin{figure}
\caption
[Comparison of the
Borel resummed critical exponents (to $O(\epsilon^5)$)
with the simulation results in 3, 4,
and 5 dimensions.]
{{\bf Borel resummed critical exponents and simulation results.}
Shown are the numerical values of the exponents
$1/\nu$, $\eta$, and
$\beta \delta = \nu (d-\eta)/2$ (triangles, diamonds,
and circles respectively)
in $3$, $4$, and $5$ dimensions and in mean field theory (dimension
$6$ and higher).
The error bars denote systematic errors in finding the
exponents from collapses
of curves at different values of disorder $R$. Statistical errors
are smaller.
The dashed lines are the Borel sums to fifth order in $\epsilon$
for the same exponents (see text).
\label{fig:Borel-vs-numerics}}
\end{figure}

\begin{figure}
\caption{{\bf Comparison to numerical results.}
Numerical values (filled symbols) of the exponents $\tau +
\sigma\beta\delta$, $\tau$, $1/\nu$, $\sigma\nu z$, and $\sigma\nu$
(circles, diamond, triangles up, squares, and triangle left) in 2,
3,4,
and 5 dimensions. The empty symbols are values for these exponents in
mean field (dimension 6). Note that the value of $\tau$ in $2$d was
not
measured. The empty diamond represents the expected value
\protect{\cite{hysterIII,eps-big2}}. The numerical results are
courtesy of Olga Perkovi\'c \protect{\cite{Perkovic,hysterIII}}
from simulations of
sizes up to $7000^2$, $1000^3$, $80^4$, and $50^5$ spins, where
for $320^3$ for example, more than $700$ different random field
configurations were measured. The long-dashed lines are the
$\epsilon$
expansions to first order for the exponents $\tau +
\sigma\beta\delta$,
$\tau$, $\sigma\nu z$, and $\sigma\nu$.
They are: $\tau+\sigma \beta
\delta = {9
\over 4} - {\epsilon \over 8}$, $\tau = {3 \over 2} +
O({\epsilon}^2)$,
and $\sigma\nu z= {1 \over 2} + O(\epsilon^2)$,
and $\sigma\nu = {1 \over 4} + O(\epsilon^2)$
where $\epsilon = 6 - d$ and $d$ is the
dimension.
The short-dashed line is the
Borel sum for $1/\nu$ to fifth order in $\epsilon$.
The other exponents can be obtained from exponent equalities (see
section \protect{\ref{sec:exponent-equalities}} in the text).
The error bars denote systematic errors in finding the exponents from
collapses of curves at different values of disorder $R$. Statistical
errors are smaller. \label{fig:eps-vs-numerics}}
\end{figure}

% XXX in tilting.appendix.tex XXXXXX

\begin{figure}
\caption
[Contour lines for the correlation length in the $(R-R_c,H-H_c)$
plane (schematic).]
{{\bf Contour lines for the correlation length}
in the $(r',h')$ plane (schematic).
The tilted coordinate axes indicate the physical directions
$(r,h)$. Since $\nu/(\beta\delta)<\nu$, the
correlation length $\xi$ changes faster in the
$(0,h')$ direction than in the $(r',0)$ direction.
\label{fig:contour-lines}}
 \end{figure}

%XXXXXX details-of-RG.app.tex XXXXXXX

\begin{figure}
\caption
[Illustration of the Feynman rules through Feynman diagrams.]
{{\bf Feynman diagrams.} The perturbative expansion about mean--field
theory is presented here by Feynman diagrams.
(a) Graph for the vertex $u$. Incoming arrows denote $\eta$ fields,
outgoing arrows denote $\hat \eta$ fields.
(b) Example of a diagram which violates causality and is
therefore forbidden.
(c) Graph for the vertex $u_{2,0}$.
(d) Example of a diagram that is
zero due to momentum conservation {\protect \cite{Wilson}}.
\label{fig:feynman}}
 \end{figure}

\begin{figure}
\caption
[Feynman diagram for the lowest order correction to
$u_{2,0}(H_1,H_2)$.]
{{\bf Feynman diagram} for the lowest order correction to
$u_{2,0}(H_1,H_2)$. The magnetic fields corresponding to
the times at which the vertices are evaluated are indicated.
The propagators do not couple different fields.
\label{fig:u-20-h1-h2}}
 \end{figure}

% XXXXX borel.app.tex XXXXX

\begin{figure}
\caption
[Results of two different ways to Borel resum the
perturbation expansion for the exponent $\nu$ to $O(\epsilon^5)$.]
{{\bf Different Borel-resummations for $\nu$.}
Borel resummation of the perturbation series for $\protect{\nu}$
to $O({\protect{\epsilon^5}})$ in
$\protect{6-\epsilon}$ dimensions using (a) the method
of Vladimirov, Kazakov, and
Tarasov {\protect {\cite{Vladimirov}}}, which does not impose a pole
for $\nu$ or any other independent information, and (b) the
results of Le Guillou and Zinn-Justin
\protect{\cite{LeGuillou}}, which are obtained by explicitly
assuming a singularity for $\nu$ at a (variationally determined)
critical value of $\epsilon$
and by imposing the exactly known value
in two dimensions.
Curve (b) is based on an old result for the epsilon-expansion
which later turned out to
out to have the wrong 5th order term \protect{\cite{Kleinert}}.
Our own Borel-resummation (curve (a)) on the other hand
has been obtained using
the newest, presumably correct result for the fifth
order term \protect{\cite{Kleinert}}.
Partly by design, curve (b) agrees very well with the value of
${\protect{\nu}}$
in the equilibrium pure Ising model in $d-2$ dimensions.
Le Guillou and Zinn-Justin
quote an error due to truncation of the series, which is
increasingly larger
than $10 \%$ below $2.5$ dimensions.
Curve (a) agrees better with the numerical results
in our zero-temperature avalanche model,
indicated for the respective dimensions
by the black diamonds with error bars \protect{\cite{Perkovic}.}
It is believed that
the two resummation methods should
converge to the same results if taken to high enough order,
though this has never been proven.
%We note however that the same $\epsilon$-expansion
%is expected to have more than one underlying analytic function,
%due to its asymptotic character and nonperturbative corrections.
Also shown are values for $\nu$ for the {\it equilibrium} random
field Ising model in three dimensions (circles) from different
sources {\protect{\cite{equil-expont,Newman}}.}
\label{fig:two-Borel}}
\end{figure}

%XXXXXXX infinite-avalanche.app.tex XXXXX

\begin{figure}
 \caption
[Feynman diagram for the relevant correction to first order
in $\tilde \epsilon=8-d$
for systems with less than critical disorder $R<R_c$.]
{{\bf Feynman diagram.}
The correction to $O(\tilde{\epsilon})$ to the vertex w in
an expansion
about 8 dimensions, see eq.~(\protect\ref{w-loop2}) in the text.
 \label{fig:feynman-w}}
 \end{figure}

% XXXXXXXXXXXX avalanche-RG.app.tex XXXXXXXXX

\begin{figure}
\caption{
{\bf The function $P^{\it both}_{\it flip}$} defined in equation
({\protect{\ref{p-both-flip-def}}}), plotted as a function of $H_1$.
In the figure, $dH$ denotes the amplitude which is called $\Delta H$
in the text.
\label{fig:Pbothflip}}
 \end{figure}

%XXXXXXXXXXXXXXXXXXXXXXXXXXXXXXXXXXXXXXXXXXXXXXXXXX
%XXX  Table Captions and tables
%XXXXXXXXXXXXXXXXXXXXXXXXXXXXXXXXXXXXXXXXXXXXXXXXXX

\begin{table}[p]
\caption[Critical exponents of the hysteresis model
and the
equilibrium random field Ising model in three dimensions]
{{\bf Numerical results for the critical exponents
in three dimensions} for our
hysteresis model \protect{\cite{Perkovic,hysterIII}}
and for the equilibrium zero temperature random
field Ising model
\protect{\cite{Nattermann-review,equil-expont,Newman}}.
The breakdown of hyperscaling
exponent $\tilde\theta$ is calculated for the
the hysteresis model from the relation $\beta + \beta \delta =
(d-\tilde\theta) \nu$ (see section
\protect{\ref{Breakdown-of-hyperscaling}}
and references \cite{thesis,eps-big2}).
The values of the  critical exponents
of the two models remain within each other's errorbars (except for
$\nu$ and perhaps $\eta$, although Dayan, Schwartz, and Young
\protect{\cite{equil-nu1.4}} found that $\nu \simeq 1.4$ in the three
dimensional equilibrium random-field Ising model from real space
renormalization group calculations); the equality of the exponents was
conjectured by Maritan {\it et al} \protect{\cite{Banavar}}.
The numerical agreement may not be so surprising, if one remembers that the
$6-\epsilon$ expansion is the same for all exponents of
the two models. Nevertheless there is always room for nonperturbative
corrections, so that the exponents might still be different
in 3 dimensions (see section
\protect{\ref{sub:honest-dim-reduction}}, and
figure {\protect{\ref{fig:two-Borel}}}).
Physically, the agreement is rather unexpected, since the nature
of the two models is very different. While the hysteresis
model is far from equilibrium, occupying a history dependent,
metastable state, the equilibrium RFIM is always in the
lowest free energy state. One may speculate, however, about a
presumably universal crossover
from our hysteresis model to the equilibrium random field Ising
model as temperature fluctuations and a finite field-sweeping
frequency $\Omega$ are introduced (see appendix
\protect{\ref{ap:temperature}}).
\label{tab:equil-RFIM}}
\vspace{1.0cm}
\begin{tabular}{cr@{$\,\pm\,$}lp{2.1in}}
\hline
 &\multicolumn{2}{c}{Hysteresis loop \cite{Perkovic}}  &
\multicolumn{1}{c}{Equilibrium RFIM \cite{equil-expont,Newman}}\\
exponents  &\multicolumn{2}{c}{in 3 dimensions} &
\multicolumn{1}{c}{in 3 dimensions}  \\
& \multicolumn{2}{r}{(courtesy Olga Perkovi\'c)} &
\\ \hline
$\nu$ &  1.42 & 0.17 & 0.97, 1.30, $1.02 \pm 0.06$ \\
$\beta$&  0.0  & 0.43 & -0.1, 0.05, $0.06 \pm 0.07$ \\
$ \beta \delta$ & 1.81 & 0.36 & 1.6, $1.9\pm0.4$, $1.83 \pm 0.18$ \\
$\eta$ & 0.79 & 0.29 & 0.25, $0.5 \pm 0.5$, $0.14 \pm 0.067$ \\
$\tilde\theta$ & 1.5 & 0.5 & $1.45$, $1.5 \pm 0.45$, $1.851 \pm
0.067$ \\
$R_c$ (Gaussian) & 2.16 & 0.03 & $2.3 \pm 0.2$ \cite{Newman} \\
$ H_c(R_c)$ & 1.435 & 0.004 & 0 (by symmetry) \\
\hline
\end{tabular}
\end{table}


\begin{thebibliography}{100}

\bibitem{glasses} J.P. Sethna, J.D. Shore, M. Huang
{\it Phys. Rev. B} {\bf 44}, 4943 (1991), and references
therein; J.D. Shore,
Ph.D. thesis, Cornell University (1992), and
references therein;
T. Riste and D. Sherrington ``Phase Transitions and
Relaxation in Systems with Competing Energy Scales'',
{\it Proc. at NATO Adv. Study Inst.}, April 13-23 1993,
Geilo, Norway, and references therein;
M. M\'ezard, G. Parisi, M.A. Virasoro
``Spin Glass Theory And Beyond'', Word Scientific (1987),
and references therein;
K.H. Fischer and J.A. Hertz ``Spin Glasses'', Cambridge University
Press (1993), and references therein.

\bibitem{RFIM}
J.~Villain, {\it Phys. Rev. Lett.} 29, 6389, (1984);
J. Villain {\it Proc. at NATO Adv. Study Inst.}, April 8-19 1985,
Geilo, Norway, and references therein;
M. M\'ezard and A.P. Young, {\it Europhys. Lett.} {\bf 18}, 653
(1992), and references therein.

\bibitem{dynamics-RFIM}
G. Grinstein and J.F. Fernandez {\it Phys. Rev. B}
{\bf 29}, 6389 (1984);
J. Villain, {\it Phys. Rev. Lett.} {\bf 52}, 1543 (1984).

\bibitem{dyn-scal-RFIM}
A. J. Bray and M. A. Moore,
Journal of Physics C {\bf 18}, L927 (1985).
%also D.S. Fisher {\it Phys. Rev. Lett.} {\bf 56}, seperately
%listed under Fisher.

\bibitem{Fisher} D.~S. Fisher {\it Phys. Rev. Lett.} {\bf 56}, 416
(1986).

\bibitem{Barkhausen-review} J. C. McClure, Jr. and K. Schr\"oder, {
\it CRC Crit.
Rev. Solid State Sci.} {\bf 6}, 45 (1976).

\bibitem{Jiles} D. Jiles ``Introduction to
Magnetism and Magnetic Materials'', Chapman and Hall, 1991.

\bibitem{Martensites} ``Martensite'', G.~B.~Olson and
W.~S.~Owen, eds., ASM International; A.D. Bruce and R.A. Cowley,
``Structural Phase Transitions'', Taylor and Francis, London (1981);
P.F. Gobin and G. Guenin, ``Solid State Phase Transformations in
Metals and Alloys'', Aussois summer school, September 3-15, 1978,
Les \'Editions de Physique, Orsay, (1978).

\bibitem{non-SOC} D. Sornette, {\it J. Phys. I France} {\bf 4}, 209
(1994).

\bibitem{fiber-breaking}
references 16-25 in D. Sornette, {\it J. Phys. I France} {\bf 4}, 209
(1994); R.L. Smith, S.L. Phoenix, M.R. Greenfields, R.B. Hengstenburg,
R.E. Pitt,
{\it Proc. R. Soc. Lond A} {\bf 388}, 353 (1983) and references
therein;
P.C. Hemmer and A. Hansen {\it Theor. Phys. Seminar Trondheim}, {\bf
4} (1991).

\bibitem{Cote}
P. J. Cote and L. V. Meisel, {\it Phys. Rev. Lett.} 67, 1334, 1991;
L.V. Meisel and P.J. Cote,
{\it Phys. Rev. B} 46, 10822, 1992.

\bibitem{Stuart-Field}  S. Field, J. Witt,
F. Nori, and X. Ling, {\it Phys. Rev. Lett.} {\bf 74}, 1206 (1995).

\bibitem{Ortin-martensite-aval} E. Vives, J. Ort\'in, L. Ma\~nosa, I. R\`afols,
R. P\'erez-Magran\'e, and A. Planes {\it Phys. Rev. Lett.}
{\bf 72}, 1694 (1994).

\bibitem{Gutenberg-Richter} C.F. Richter {\it Ann. Geophys.} {\bf 9}
1 (1956).

\bibitem{Hallock} K.M. Godshalk and R.B. Hallock, {\it Phys. Rev. B}
{\bf 36}, 8294 (1987);
M.P. Lilly, P.T. Finley, and R.B. Hallock {\it Phys. Rev. Lett.}
{\bf 71}, 4186 (1993).

\bibitem{Coppersmith} S.N. Coppersmith {\it Phys. Rev. Lett.} {\bf
65}, 1044 (1990) has shown that on sufficiently large length scales
the elastic theory always breaks down.

\bibitem{FisherCDW} D.S. Fisher, {\it Phys. Rev. B} {\bf 31},
1396 (1985).

\bibitem{CDW-review} ``Charge Density Waves in Solids'', edited
by Gy. Hutiray and J. S\'olyom,
Lecture Notes in Physics Vol. 217 (Springer-Verlag, Berlin, 1984);
and ``Charge Density Waves in Solids'', edited by L.P. Gorkov and
G. Gr\"uner (Elsevier, Amsterdam, 1989).

\bibitem{NarayanI}
{O.~Narayan and D.~S. Fisher,
{\it Phys. Rev. Lett.} {\bf 68}, 3615 (1992) and {\it Phys. Rev. B}
{\bf 46}, 11520 (1992)}

\bibitem{NarayanII} O.~Narayan and A.A.~Middleton
{\it Phys. Rev. B} {\bf 49}, 244 (1994).

\bibitem{Myers}
C. Myers and J.P. Sethna {\it Phys. Rev. B} {\bf
47}, 11171 (1993), and {\it Phys. Rev. B} {\bf 47}, 11194 (1993)
(CDW).

\bibitem{Middleton} A.A.~Middleton and D.S.~Fisher {\it Phys. Rev.
Lett.} {\bf 66}, 92 (1991), and references therein;
{\it Phys. Rev. B} {\bf 47}, 3530 (1993) (CDW).

\bibitem{flux-lattice} A.I. Larkin and Yu.N. Ovchinikov, {\it J. Low
Temp. Phys.} {\bf 34},
409 (1979) (flux lattice).

\bibitem{flux-line} M.V. Feigel'man and V.M. Vinokur, {\it Phys.
Rev. B} {\bf 41}, 8986.

\bibitem{flux-line-theory} D.S. Fisher, and D.A. Huse, {\it Phys.
Rev. B} {\bf 43}, 130 (1991) and references therein.

\bibitem{Ertas2} D. Erta{\c s} and M. Kardar (preprint 1994).

\bibitem{fluid-invasion} M.A. Rubio, C. Edwards, A. Dougherty, and
J.P. Gollup, {\it Phys. Rev. Lett.} {\bf 63}, 1685 (1989); {\bf
65}, 1339 (1990);
V.K. Horvath, F. Family, and T. Vicsek, {\it Phys. Rev. Lett.}
{\bf 65}, 1388 (1990); {\it J. Phys. A} {\bf 24}, L25 (1991);
M.A. Rubio, A. Dougherty, and J.P. Gollup,
{\it ibid} {\bf 65}, 1389 (1990); V.K. Horvath, F. Family, and
T. Vicsek, {\it ibid} {\bf 67}, 3207 (1991); S. He, G.L.M.K.S.
Kahanda, and P-Z. Wong, {\it ibid} {\bf 69}, 3731 (1992).

\bibitem{Robbins}
M. Cieplak and M.O. Robbins,
{\it Phys. Rev. Lett.} {\bf 60}, 2042 (1988), and {\it Phys. Rev. B}
{\bf 41}, 11508 (1990); N. Martys, M. Cieplak, and M.O. Robbins,
{\it Phys. Rev. Lett.} {\bf 66}, 1058 (1991);
N. Martys, M.O. Robbins, and M. Cieplak,
{\it Phys. Rev. B} {\bf 44}, 12294 (1991);
B. Koiller, H. Ji, and M.O. Robbins, {\it ibid} {\bf 45}, 7762
(1992) (fluid invasion);
H. Ji and M.O. Robbins {\it ibid} {\bf 46},
14519 (1992); B. Koiller, H. Ji, M.O. Robbins,
{\it Phys. Rev. B} {\bf 46}, 5258 (1992)
(domain wall in random magnets).

\bibitem{NarayanIII} O.~Narayan and D.S.~Fisher {\it Phys. Rev. B}
{\bf 48}, 7030 (1993).

\bibitem{Nattermann} T. Nattermann, S. Stepanow, L.H. Tang and
H.Leschhorn
{\it J. Phys. II France} {\bf 2}, 1483 (1992)

\bibitem{Ertas}
D.~Erta{\c s} and M.~Kardar, {\it Phys. Rev.
E} {\bf 49}, R2532 (1994); J.F. Joanny and M.O. Robbins,
{\it J. Chem. Phys.} {\bf 92}, 3206 (1990) (contact lines).

\bibitem{flux-lattice-torn} A. Brass, H.J. Jensen, and A.J.
Berlinsky, {\it Phys. Rev. B} {\bf 39}, 102 (1989).

\bibitem{nonwetting} R. Lenormand and C. Zarcone, {\it Phys. Rev.
Lett.} {\bf 54}, 2226 (1985).

\bibitem{NarayanIV} O. Narayan and D.S. Fisher {\it Phys. Rev. B}
{\bf 49}, 9469 (1994), and references therein.

\bibitem{Chikazumi}
S. Chikazumi, ``Physics of Magnetism'', Wiley, New York (1964).

%\bibitem{Brandt-Dahmen} S. Brandt and H.D. Dahmen ``Physik'', Vol.2
%(Elektrodynamik), Springer Verlag (1986).

\bibitem{ferroelectrica} Aizu K., {\it Phys. Rev. B} {\bf 2}, 754 (1970),
{\it J. Phys. Soc. Japan} {\bf 44}, 334 and 683 (1978);
Martin ``Die Ferroelektrika'', Leipzig 1964.

\bibitem{Rudyak} V.M. Rudyak, {\it Bull. Acad. Sci. USSR
Phys. Ser.} {\bf 57}, 955 (1993) and references therein;
V.M. Rudyak {\it Bull. Acad. Sci. USSR Phys. Ser.} {\bf 45}, 1
(1981).

\bibitem{Nagel} G.L. Vasconcelos, M. de Sousa-Vieira, S.R. Nagel {\it Physica
A} {\bf 191}, 69 (1992).

\bibitem{earthquakes} R. Burridge and L. Knopoff {\it Bull. Seismol.
Soc. Am.} {\bf 57}, 341 (1967);
P.Bak and C. Tang, J. Geophys. Res.
{\bf 94} 15635 (1989);
Z. Olami, H.J.S. Feder, and K. Christensen,
{\it Phys. Rev. Lett.} {\bf 68}, 1244 (1992);
K. Christensen and
Z. Olami {\it Phys. Rev. A} {\bf 46}, 1829 (1992);
L. Pietronero, P. Tartaglia and Y.C. Zhang {\it Physica}
(Amsterdam) {\bf 137A}, 22 (1991); E.J. Ding, Y.N. Lu {\it Phys. Rev
Lett.} {\bf 70}, 3627 (1993).
%and Nagel,Gutenberg-Richter,slip-stick,Carlson

\bibitem{Carlson} J.M. Carlson and J.S. Langer {\it Phys. Rev.
Lett.} {\bf 62}, 2632 (1989).

\bibitem{hysterI} J.P.~Sethna, K.~Dahmen, S.~Kartha, J.A.~Krumhansl,
B.W.~Roberts, and J.D.~Shore, {\it Phys. Rev. Lett.} {\bf 70},
3347 (1993); K. Dahmen, S. Kartha, J.A. Krumhansl, B.W. Roberts,
J.P. Sethna, and J.D. Shore,
{\sl J.~Appl.\ Phys.} {\bf 75}, 5946 (1994).

\bibitem{hysterII}
K.~Dahmen and J.P.~Sethna, {\it Phys. Rev. Lett.} {\bf
71}, 3222 (1993).

\bibitem{Preisach} F.~Preisach, {\it Z.~Phys.} {\bf 94},277 (1935);
M.~Krasnoselskii
and A.~Pokrovskii, {\it Systems with Hysteresis}, (Nauka,
Moscow, 1983),
I.~D. Mayergoyz, {\it J.~Appl. Phys.} {\bf 57}, 3803 (1985);
{\it Mathematical Models of Hysteresis}, (Springer-Verlag,1991); P.C.
Clapp, {\it Materials Science and Engineering} {\bf A127}, 189-95
(1990).

\bibitem{OBrien-Weissman} K.P. O'Brien and M.B. Weissman,
{\it Phys. Rev. E} {\bf 50}, 3446 (1994) and references therein;
K.P. O'Brien and M.B. Weissman, {\it Phys. Rev. A} {\bf 46}, R4475
(1992).

\bibitem{Perkovic} O.~Perkovi\'c, K.A.~Dahmen, and J.P. Sethna,
(simulations: preprint in preparation).

\bibitem{Olson}  V. Raghavan in {\it Martensite}, G.~B.~Olson and
W.~S.~Owen, eds., (ASM International), 1992, p. 197.

\bibitem{Berger} A. Berger, unpublished.

\bibitem{Berger-setup} Details about the experimental setup used
by Berger are given in A.W. Pang, A. Berger, and H. Hopster,
{\it Phys. Reb B} {\bf 50}, 6457
(1994); A. Berger, A.W. Pang, H. Hopster,
{\it J. Magn. Magn. Mat.} {\bf 137}, L1 (1994);
A. Berger, A.W. Pang, and H. Hopster, {\it Phys. Rev. B} {\bf 52}
(1995, in print).

%\bibitem{Kachaturyan} A.G. Kachaturyan, ``Theory of Structural
%Transformations in Solids'' (Wiley, New York, 1983).


%\bibitem{Hallock-priv} R.B. Hallock and M.P. Lilly, private
%communication.

%\bibitem{Field-priv} S. Field and J. Witt, private communication.

\bibitem{Adams} W. Wu and P.W. Adams, {\it Phys. Rev. Lett.} {\bf
74}, 610 (1995).

\bibitem{Ortin-RBIM} E. Vives and A. Planes
{\it Phys. Rev. B} {\bf 50}, 3839 (1994) (simulations of the
two-dimensional random bond Ising model);
E. Vives, J. Goicoechea,
J. Ort\'in and A. Planes, preprint (1994) (simulations of the
RFIM and the RBIM in two and three dimensions, and of the
Blume-Emery-Griffiths model in two dimensions).

\bibitem{Chayes-private-com} L. Chayes, private communication.

\bibitem{RPME} Return point memory has been seen in ferromagnetism:
D.~C. Jiles, D.~L. Atherton, {\it J. Appl. Phys.} {\bf 55}, 2115
(1984);
J.A. Barker, D.E. Schreiber, B.G. Huth, and
D.H. Everett, {\it Proc. R. Soc. London A} {\bf 386}, 251-261
(1983).
Martensitic transformations (thermally and stress-induced) of
metallic alloys
J. Ort\'in {\it J. Appl. Phys.} {\bf 71}, 1454 (1992); {\it J.de
Phys. IV}, colloq. C4, {\bf 1}, C-65 (1991) and references therein.
In ammonium chloride: Smits {\it et al.} {\it Z. Phys. Chem. A}
{\bf 166}, 97 (1933); {\bf 175}, 359 (1936); {\it Z. Phys. Chem. B}
{\bf 41}, 215 (1938).
Spin transitions: E.W. M\"uller, H. Spiering, and P. G\"utlich,
{\it J. Chem. Phys.} {\bf 79}, 1439 (1983).
Adsorption of gases by porous solids: J. Katz, {\it J.
Phys. (Colloid) Chem.} {\bf 53}, 1166 (1949); Emmett and Cines, {\it
J. Phys.
(Colloid) Chem.}, {\bf 51}, 1248 (1947).  Charge--density waves:
Z.~Z.~Wang and N.~P.~Ong, {\it Phys. Rev. B} 34, 5967, 1986.
Cellular Automata: J.~Goicoechea and J.~Ort\'in, {\it Phys. Rev.
Lett} {\bf 72}, 2203 (1994), and references therein.

\bibitem{Goldenfeld} N.~Goldenfeld, ``Lectures on Phase transitions
and
the Renormalization Group'', Addison Wesley 1992, p.383.

\bibitem{Yeomans} J.M. Yeomans
``Statistical Mechanics of Phase Transformations''
Clarendon Press, Oxford (1992).

\bibitem{quantum} In this paper we consider a lattice of
classical spins. For an
illustrative quantum mechanical description of magnetic moments in
an external magnetic field see also S. Brandt and H.D. Dahmen
``The Picture Book of Quantum Mechanics'', second edition,
Springer Verlag, New York 1995.

\bibitem{vanKampen}
N.G. van Kampen ``Stochastic Processes in Physics and
Chemistry'', North Holland 1990.

\bibitem{Olami} S. Maslov and Z.Olami, preprint (1993).

\bibitem{Kartha} S. Kartha, Ph.D. thesis, Cornell University (1994).

\bibitem{Kleinert} H. Kleinert, J. Neu, V. Schulte-Frohlinde, and K.G.
Chetyrkin, {\it Phys. Lett. B} {\bf 272}, 39 (1991).

\bibitem{Krey} U. Krey, {\it J. Phys. C} {\bf 17}, L545-L549 (1984),
U. Krey, {\it J. Phys. C} {\bf 18}, 1455-1463 (1985), U. Krey and H.
Ostermeier, {\it Z. Phys. B} {\bf 66}, 219 (1987).

\bibitem{Stauffer} D. Stauffer and A. Aharony ``Introduction to
Percolation Theory'', Taylor and Francis (1992).

\bibitem{Nattermann-review} T. Nattermann and J. Villain,
{\it Phase Transitions} 1988, Vol. 11, pp 5-51, Gordon and Breach Science
publishers Inc. (1988), and references therein;
T. Nattermann and P. Rujan {\it Inter. J. Mod. Phys. B} {\bf 3}, 1597
(1989); D. P. Belanger and A. P. Young,
{\it J. Magn. Magn. Mat.} {\bf100}, 272 (1991);
E.B. Kolomeisky,
in {JETP Lett.} {\bf 52}, No.10, 538 (1990).

\bibitem{SchwartzSoffer} M. Schwartz, and A. Soffer, {\it Phys. Rev.
Lett.}
{\bf 55}, 2499 (1985).

\bibitem{Binney} J.J. Binney, N.J. Dowrick, A.J. Fisher, M.E.J.
Newman, ``The Theory of Critical Phenomena'',
Clarendon Press, Oxford (1992).

\bibitem{Ma} S.K. Ma ``Modern Theory of Critical Phenomena'',
Benjamin/Cummings Publishing Company Inc Reading, (1976).

\bibitem{Fisher-SA-notes} M. Fisher ``Critical Phenomena'',
Proc. Stellenbosch, South Africa 1982, Springer Verlag (1982).

\bibitem{Zinn-Justin} J. Zinn-Justin ``Quantum Field Theory and
Critical Phenomena'', 2nd edition, Clarendon Press, Oxford (1993).

\bibitem{Amit} D.J. Amit ``Field Theory, the Renormalization Group,
and Critical Phenomena'', World Scientific, 1984.

\bibitem{Halperin-review} P.C. Hohenberg, B.I. Halperin
``Theory of dynamic critical phenomena'' {\it Rev. Mod. Phys.} {\bf
49}, 435 (1977).

\bibitem{real-space} For a review of the real-space renormalization
group, which we do not employ here, see also
R. J. Creswick, H. A. Farach, and C. P. Poole, Jr.,
``Introduction to Renormalization--Group Methods in Physics'',
John Wiley and Sons, New York, (1992).

\bibitem{Lubensky} A. Brooks Harris and T.C. Lubensky, {\it Phys.
Rev. Lett.} {\bf 33}, 1540 (1974);
T.C. Lubensky, {\it Phys. Rev. B} {\bf 11}, 3573 (1975).

\bibitem{Middleton-thermal} A.A. Middleton {\it Phys. Rev. B},
{\bf 45}, 9465 (1992).

\bibitem{MSR} P.~C.~Martin, E.~Siggia and H.~Rose, {\it Phys. Rev. A}
{\bf 8}, 423 (1973); C.~De~ Dominicis, {\it Phys. Rev. B} {\bf 18},
4913
(1978); H.~Sompolinsky and A.~Zippelius, {\it Phys. Rev. B} {\bf 25},
6860 (1982); A.~Zippelius, {\it Phys. Rev. B} {\bf 29}, 2717 (1984).

\bibitem{BJW} R. Bausch, H.K. Janssen, and H. Wagner, {\it Z. Phys.
B} {\bf 24}, 113 (1976); U.C. T\"auber and F. Schwabl,
{\it Phys. Rev. B} {\bf 46}, 3337 (1992) and references therein.

\bibitem{dominicis} C. De Dominicis, {\it Phys. Rev. B} {\bf 18},
4913 (1978).

\bibitem{Zippelius} H.~Sompolinsky and A.~Zippelius, {\it Phys. Rev.
B} {\bf 25},
6860 (1982); A.~Zippelius, {\it Phys. Rev. B} {\bf 29}, 2717 (1984).

\bibitem{Ramond} P. Ramond ``Field Theory: A modern primer'',
Addison-Wesley Publishing Company, Inc., 1990.

\bibitem{Wilson}
K.~G.~Wilson and J.~Kogut, Physics Reports {\bf 12C}, 76 (1974).

\bibitem{Domb6} C. Domb, M.S. Green, ``Phase Transitions and
Critical Phenomena'', Vol 6, Academic Press (1976).

\bibitem{equ-of-state} E. Br\'ezin, D.J. Wallace, and K.G. Wilson,
{\it Phys. Rev. Lett.} {\bf 29}, 591 (1972), and {\it Phys. Rev. B}
{\bf 7}, 232 (1973); D. J. Wallace and R.P.K. Zia {\it J. Phys. C:
Solid Sstate Phys.} {\bf 7}, 3480 (1974).

\bibitem{mapping} A. Aharony, Y. Imry, S.K. Ma,
{\it Phys. Rev. Lett.} {\bf 37}, 1364 (1976);
A.P. Young, {\it J. Phys. A} {\bf 10}, L 257 (1977);
G. Parisi and N. Sourlas {\it Phys. Rev. Lett.} {\bf 43},
744 (1979).

\bibitem{Parisi} G.~Parisi, lectures given at the 1982 Les Houches
summer school XXXIX ``{\it Recent advances in field theory and
statistical mechanics}'' (North Holland), and references therein.

\bibitem{Tadic} B. Tadic {\it Z. Phys.} {\bf 41}, 13 (1981).

\bibitem{appendix-beta-delta}
{}From the RG description one deduces that
the exponents $\beta$ and $\delta$
are related to $\nu$, $\eta$, and $\bar\eta$
by the relations
$\beta = {\textstyle {\nu \over 2}}(d-4+\bar\eta)$,
$\delta = (d-2 \eta + \bar\eta)/(d-4+\bar\eta)$,
just as in the equilibrium
model \protect{\cite{Nattermann-review,Rieger-Young}}.
Using these relations one finds that
the inequality $\nu/\beta\delta \geq 2/d$ goes over into
the Schwartz-Soffer inequality $\bar\eta \leq 2 \eta$ that has been
derived for the corresponding equilibrium model
\protect{\cite{SchwartzSoffer}}.

\bibitem{Griffiths} R.B. Griffiths, {\it Phys. Rev.} {\bf 158}, 176
(1967).

\bibitem{Banavar}  A. Maritan, M. Cieplak, M.R. Swift and J. Banavar,
{\it Phys. Rev. Lett.} {\bf 72}, 946 (1994); J.P.~Sethna,
K. Dahmen, S. Kartha, J.A. Krumhansl, O. Perkovi\'c, B.W. Roberts,
and J.D. Shore, {\it Phys. Rev. Lett.} {\bf 72}, 947 (1994).

\bibitem{equil-expont} A.P. Young and M. Nauenberg
{\it Phys. Rev. Lett.} {\bf 54}, 2429, 1985;
A.T. Ogielski and D.A. Huse {\it Phys. Rev. Lett.} {\bf 56}, 1298,
1986.

\bibitem{Newman} M.E.J. Newman and G.T. Barkema,
Cornell Theory Center preprint \# CTC95TR218 (1995), submitted
to {\it Phys. Rev. E} for publication.

\bibitem{equil-nu1.4} I. Dayan, M. Schwartz, and A.P. Young,
{\it J. Phys. A} {\bf 26}, 3093 (1993).

\bibitem{Bender} C. Bender, S.A. Orszag, ``Advanced mathematical
methods for scientists and engineers'', Mc Graw Hill (1978).

%\bibitem{Hennecke} M. Hennecke {\it Phys. Rev. B} {\bf 48}, 6271
%(1993).

\bibitem{Stierstadt} K. Stierstadt and W. Boeckh,
{\it Z. Physik} {bf 186}, 154 (1965) (in German).
The results quoted here are extracted from replotting the data of
figure 4 in the form of histograms of the pulse areas
recorded withing a small interval of the external magnetic field
$H$.


\bibitem{Bertotti94} G. Bertotti, G. Durin, and A. Magni
{\it J. Appl. Phys.} {bf 75}, 5490 (1994).

\bibitem{Bittel} H. Bittel
{\it IEEE Trans. Magn.} {\bf 5}, 359 (1969).

\bibitem{Lieneweg} U. Lieneweg
{\it IEEE Trans. Magn.} {\bf 10}, 118 (1974).

\bibitem{Lieneweg-and-Grosse-Nobis} U. Lieneweg and W. Grosse-Nobis
{\it Intern. J. Magnetism} {\bf 3}, 11 (1972).

\bibitem{Bertotti90} G. Bertotti, F. Fiorillo, and A. Montorsi
{\it J. Appl. Phys.} {\bf 67}, 5574 (1990).

\bibitem{Urbach} J.S. Urbach, R.C. Madison, and J.T. Markert,
preprint 1994.

\bibitem{hysterIII} K.A.~Dahmen, O.~Perkovi\'c, and J.P.~Sethna,
preprint 1994, submitted for publication.

\bibitem{Montalenti} G. Montalenti
{\it Z. angew. Physik} {bf 28}, 295 (1970).

%\bibitem{Villain} J. Villain {\it J. Physique} {bf 46}, 1843 (1995);
%J. Villain in ``Scaling Phenomena in Disordered Systems'', ed. Roger
%Pynn and Arne Skjeltorp, Plenum Publishing Corporation, p.423.

%\bibitem{Birgeneau}
%M. Hagen, R. A. Cowley, S. K. Satija, H. Yoshizawa, G.
%Shirane, R.  J. Birgeneau, and H. J. Guggenheim,
%{it Phys. Rev. B} {\bf28}, 2602 (1983);
%R.J. Birgeneau, Y. Shapiro, G. Shirane,
%R.A. Cowley, and H. Yoshizawa, {\it Physica} {\bf 37B} and {\bf C}, 83
%(1986) and references therein; R.J. Birgeneau, R.A. Cowley,
%G. Shirane, and H. Yoshizawa {\it Phys. Rev. Lett.} {\bf 54}, 2147
%(1985).

\bibitem{BWR-thesis}
B.W. Roberts, Ph.D. thesis, Cornell University (1995), and
references therein.

\bibitem{BWR-Newman}
M. E. J. Newman, B. W. Roberts, G. T. Barkema, and J. P. Sethna,
{\it Phys. Rev. B} {\bf 48}, 16533 (1993), and references therein.

%\bibitem{Jaccarino}
%D. P. Belanger, A. R. King, V. Jaccarino, and J. L. Cardy,
%{\it Phys. Rev. B} {\bf28}, 2522 (1983).
%D.P. Belanger, A.R. King, and V. Jaccarino
%{\it Phys. Rev. B} {\bf 31}, 4538 (1985).

%\bibitem{Imry-Ma} Y. Imry and S.K. Ma {\it Phys. Rev. Lett.} {\bf
%35}, 1399 (1975).

%\bibitem{Grinstein-Ma} G. Grinstein and S.-K. Ma,
%{\it Phys. Rev. Lett.} {\bf49}, 685 (1982); G. Grinstein and S.-K.
%Ma, {\it Phys. Rev. B} {\bf28}, 2588 (1983).

\bibitem{Imbrie}
J.Z. Imbrie {\it Phys. Rev. Lett.} {\bf 53}, 1747 (1984);
{\it Commun. Math. Phys.} {\bf 98}, 145 (1985);
{\it Physica} {\bf 140 A}, 291 (1986).

\bibitem{Kupiainen} J. Bricmont and A. Kupiainen, {\it Phys. Rev.
Lett.} {\bf 59}, 1829 (1987); {\it Commun. Math. Phys.} {\bf 116},
539 (1988).

%\bibitem{Binder} K. Binder and D.W. Heermann, ``Monte Carlo
%Simulations in
%Statistical Physics'', Springer-Verlag Berlin Heidelberg 1992.

%\bibitem{Arcangelis} L. de Arcangelis, {\it J. Phys. A} {\bf 20},
%3057 (1987).

\bibitem{Chayes} J.T. Chayes, L. Chayes, D.S. Fisher, and T. Spencer.
{\it Phys Rev. Lett.} {\bf 57}, 2999, (1986).

%\bibitem{Brandt} S. Brandt, ``Statistical and Computational Methods
%in Data Analysis'', 2nd ed., North-Holland, Amsterdam 1976,
%p.168-171.

\bibitem{Ryder} L.H. Ryder, ``Quantum Field Theory'',
Cambridge University Press (1985).

\bibitem{Vladimirov}
A.~A.~Vladimirov, D.~I.~Kazakov,
and O.~V.~Tarasov, {\it Sov. Phys. JETP} {\bf 50} (3), 521 (1979)
and references therein.

\bibitem{Erdelyi} A. Erd\'elyi ``Asymptotic expansions'', Dover
Publications, Inc., New York.

\bibitem{Large-order-perturbation}
J.C. Le Guillou
and J. Zinn-Justin ``Large-Order Behaviour of Perturbation Theory'',
North Holland (1990).

\bibitem{Kazakov} D.I. Kazakov, O.V. Tarasov, and D.v. Shirkov,
{\it Teor. Mat. Fiz.} {\bf 38}, 15 (1979).

\bibitem{LeGuillou}
J.C. LeGuillou and J. Zinn-Justin {\it Phys. Rev. B} {\bf 21},
3976 (1980);
J.C. LeGuillou and J. Zinn-Justin {\it J. Physique Lett.} {\bf 46},
L137 (1985); J.C. LeGuillou and J.
Zinn-Justin, {\it J. Physique} {\bf 48}, 19 (1987).

\bibitem{Loeffel} J.J. Loeffel, Workshop on Pad\'e Approximants,
(eds. D. Bessis, J. Gilewicz, and P. Merry), CEA (1976)
and E. Br\'ezin, Review talk, European Particle Physics conference,
Budapest (1977).

%\bibitem{noise-stochastic} ``Noise and Stochastic Processes'',
%edited by N. Wax, Dover Publications, New York (1954).

\bibitem{porous-media-Robbins-ref-5-17}
J. Feder, ``Fractals'', Plenum, New York, 1988;
R. Lenormand and S. Bories, {\it C.R. Acad. Sci. Ser. B} {\bf 291},
279 (1980); R. Chandler, J. Koplik, K. Lerman, and J.F. Willemsen,
{\it J. Fluid Mech.} {\bf 119}, 249 (1982);
R. Lenormand, {\it J. Phys.:Cond. Mat.} {\bf 2}, SA79 (1990);
R. Lenormand, and C. Zarcone, {\it Phys. Rev. Lett.} {\bf 54}, 2226
(1985); J.P. Stokes, D.A. Weitz, J.P. Gollub, A. Dougherty, M.O.
Robbins, P.M. Chaikin, and H.M. Lindsay, {\it Phys. Rev. Lett} {\bf
57}, 1718 (1986); J.P. Stokes, A.P. Kushnik, and M.O. Robbins, {\it
Phys. Rev. Lett.} {\bf 60}, 1386 (1988).

\bibitem{Robbins-private} M. Robbins, private communication.

%\bibitem{slip-stick}
%A. Sornette and D. Sornette, Europhys.
%Lett. {\bf 9}, 197 (1989); H.J.S. Feder and J. Feder, {\it Phys. Rev.
%Lett.} {\bf 66}, 2669 (1991) (experiment); O. Pla and F. Nori,
%{\it Phys. Rev. Lett.} {\bf 67}, 919 (1991).

\bibitem{bootstrap-percolation}
J. Adler and A. Aharony, {\it J. Phys. A: Math. Gen} {\bf 21},
1387 (1988); J. Adler, {\it Physica A} {\bf 171}, 453 (1991)
and references therein; N.S. Branco, R.R. dos Santos, and S.L.A.
de Queiroz, {\it J. Phys. C} {\bf 21}, 2463 (1988).

\bibitem{Coram} C.M.~Coram, A.~Jacobs, N.~Heinig, and K.~B.~
Winterbon, {\it Phys. Rev. B} {\bf 40}, 6992 (1989).

\bibitem{Bertotti-SK-model} G. Bertotti and M. Pasquale
{\it J. Appl. Phys.} {\bf 69}, 5066 (1991).

\bibitem{SK} D. Sherrington and S. Kirkpatrick, {\it Phys. Rev.
Lett.} {\bf 35}, 1792 (1975).

\bibitem{BertottiI} G. Bertotti and M. Pasquale
{\it J. Appl. Phys.} {\bf 67}, 5255 (1990).

\bibitem{Bertotti-experiment}
B. Alessandro, G. Bertotti, A. Montorsi,{ \it J. Phys.
(Paris) Colloq.}
{\bf 49} C8, 1907 (1988).

\bibitem{dynamic-hysteresis}
D. Dhar and P.B. Thomas {\it J. Phys. A} {\bf 25},
4967 (1992); P.B. Thomas and D. Dhar {\it J. Phys. A:
Math. Gen.}, {\bf 26}, 3973 (1993); S. Gupta, preprint 1993;
J. Zemmouri, B. S\'egard,
W. Sergent, and B. Macke, {\it Phys. Rev. Lett.} {\bf 70}, 1135
(1993).

\bibitem{Rao-Krishnamurthy-Pandit}
M. Rao, H.R. Krishnamurthy, and R. Pandit
{\it Phys. Rev. B} {\bf 42}, 856 (1990) and references therein.

\bibitem{RFBEG} M. Blume, V.J. Emery, and R.B. Griffiths, {\it Phys.
Rev. A} {\bf 4}, 1071 (1971).

\bibitem{Ashcroft-Mermin} N.W. Ashcroft, N.D. Mermin,
``Solid State Physics'' W.B. Saunders Company (1976).

\bibitem{BWR} B.W. Roberts and J.P. Sethna (unpublished).

\bibitem{Rieger-Young} H. Rieger and A.P. Young, {\it J. Phys. A:
Math. Gen.}, {\bf 26}, 5279 (1993).

\bibitem{thesis} K.A. Dahmen,
Ph.D. thesis, Cornell University (1995).

\bibitem{ParisiCDW} G. Parisi and L. Pietronero,
{\it Europhys. Lett.} {\bf 16}, 321 (1991).

\bibitem{eps-big} K.A.~Dahmen, O. Perkovic, B.W. Roberts,
and J.P. Sethna, preprint in preparation.

\bibitem{eps-big2} K.A.~Dahmen, O. Perkovic, and
J.P. Sethna, preprint in preparation.


\bibitem{oopsI}
The $O(\epsilon^2)$ diagram which we constructed
in reference \cite{hysterII} and had argued to lead to
different results in $O(\epsilon^2)$ for the two models,
turned actually out to be irrelevant, as all other
diagrams which are different from the equilibrium model.

\bibitem{ToCome}
Further information is available via World Wide Web:\newline
http://www.lassp.cornell.edu/LASSP\underbar{\phantom{x}}Science.html.




\end{thebibliography}
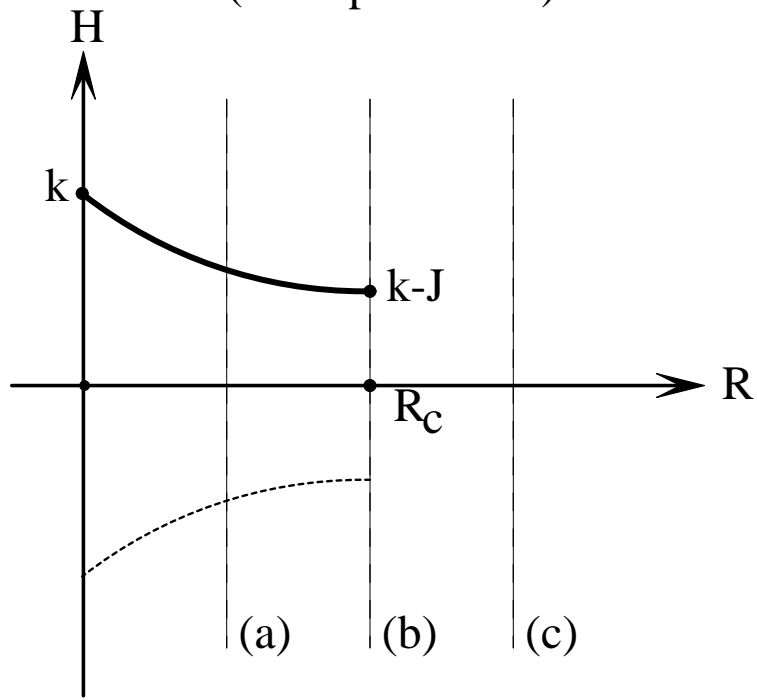
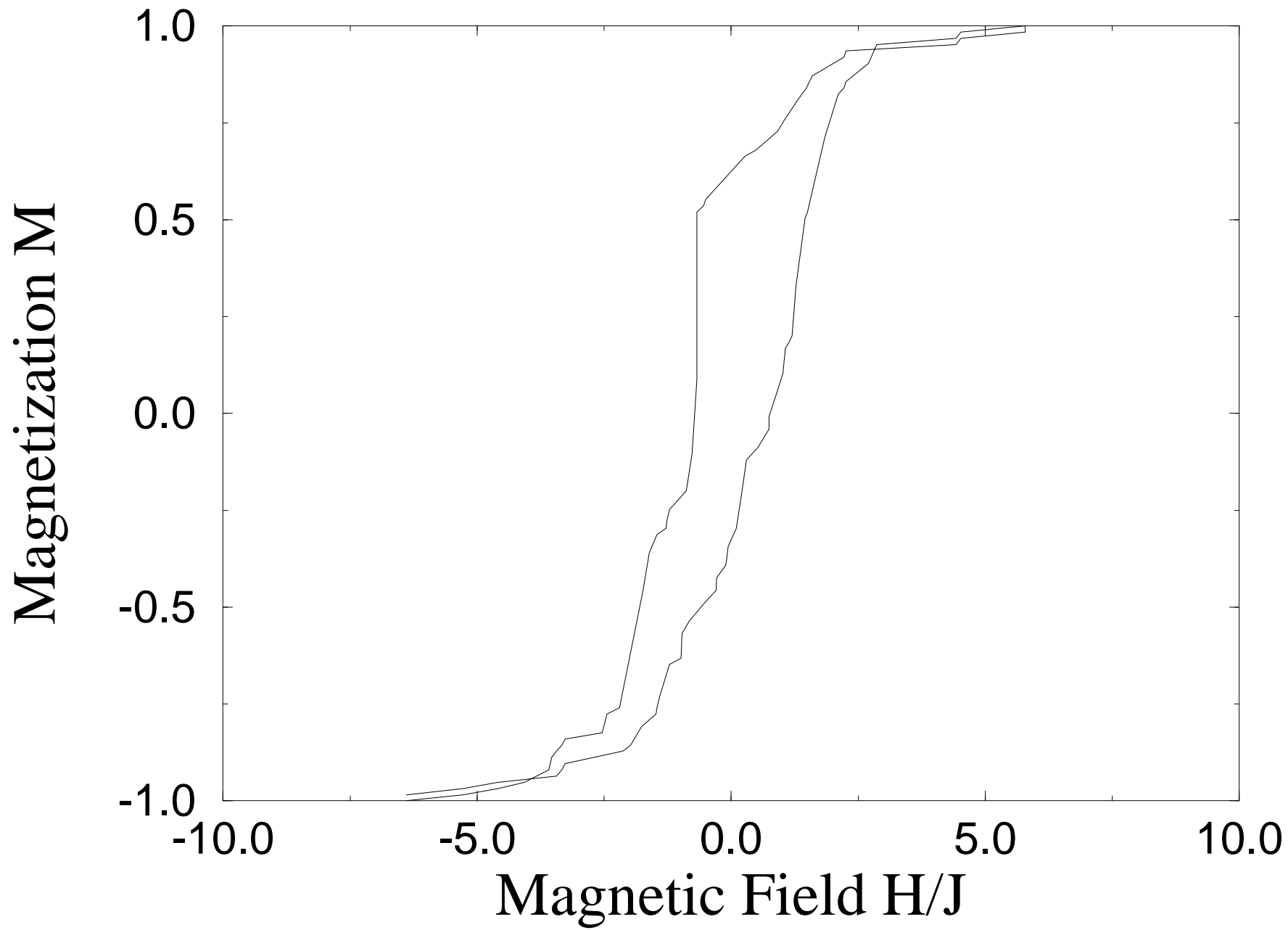
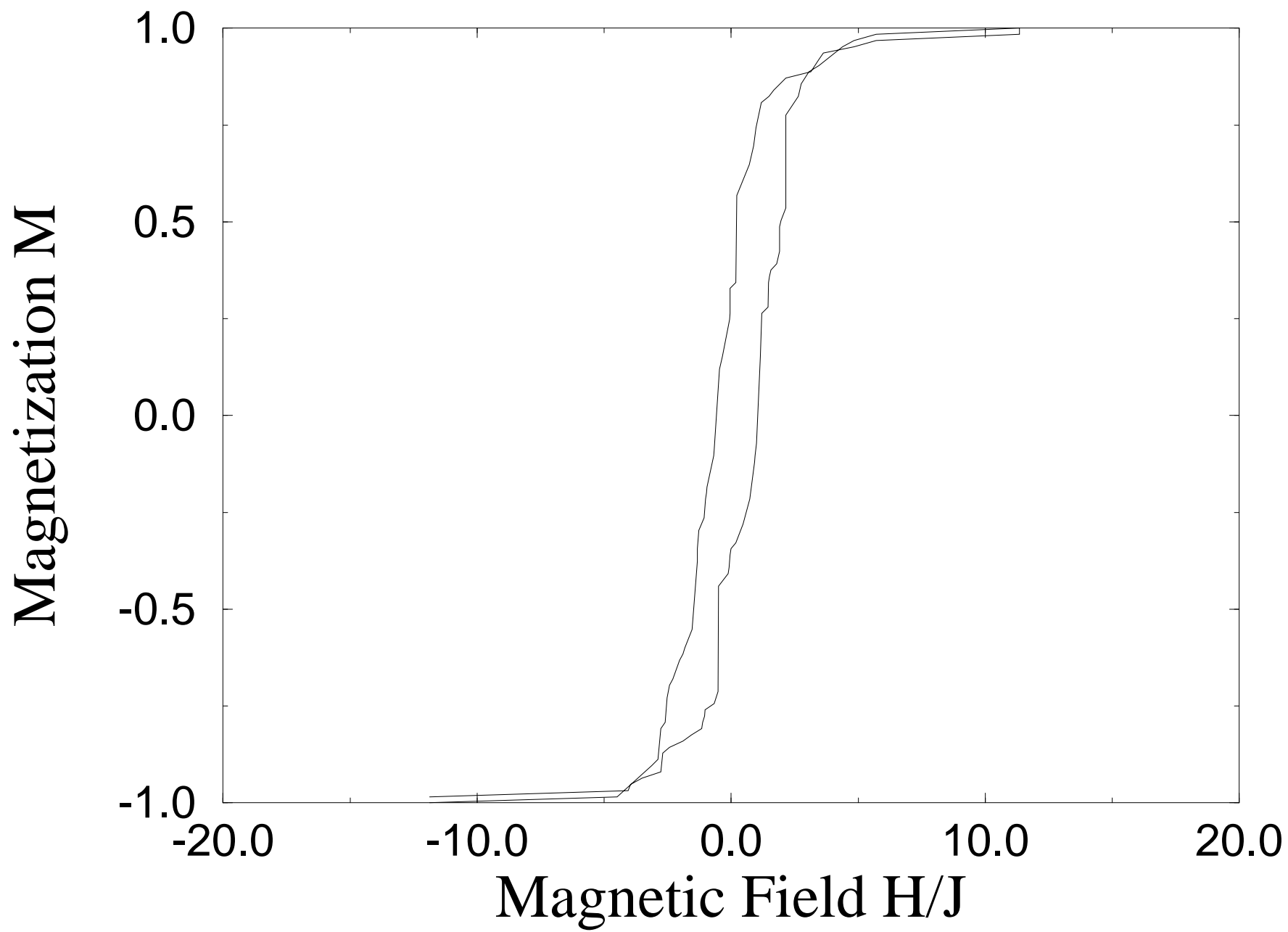
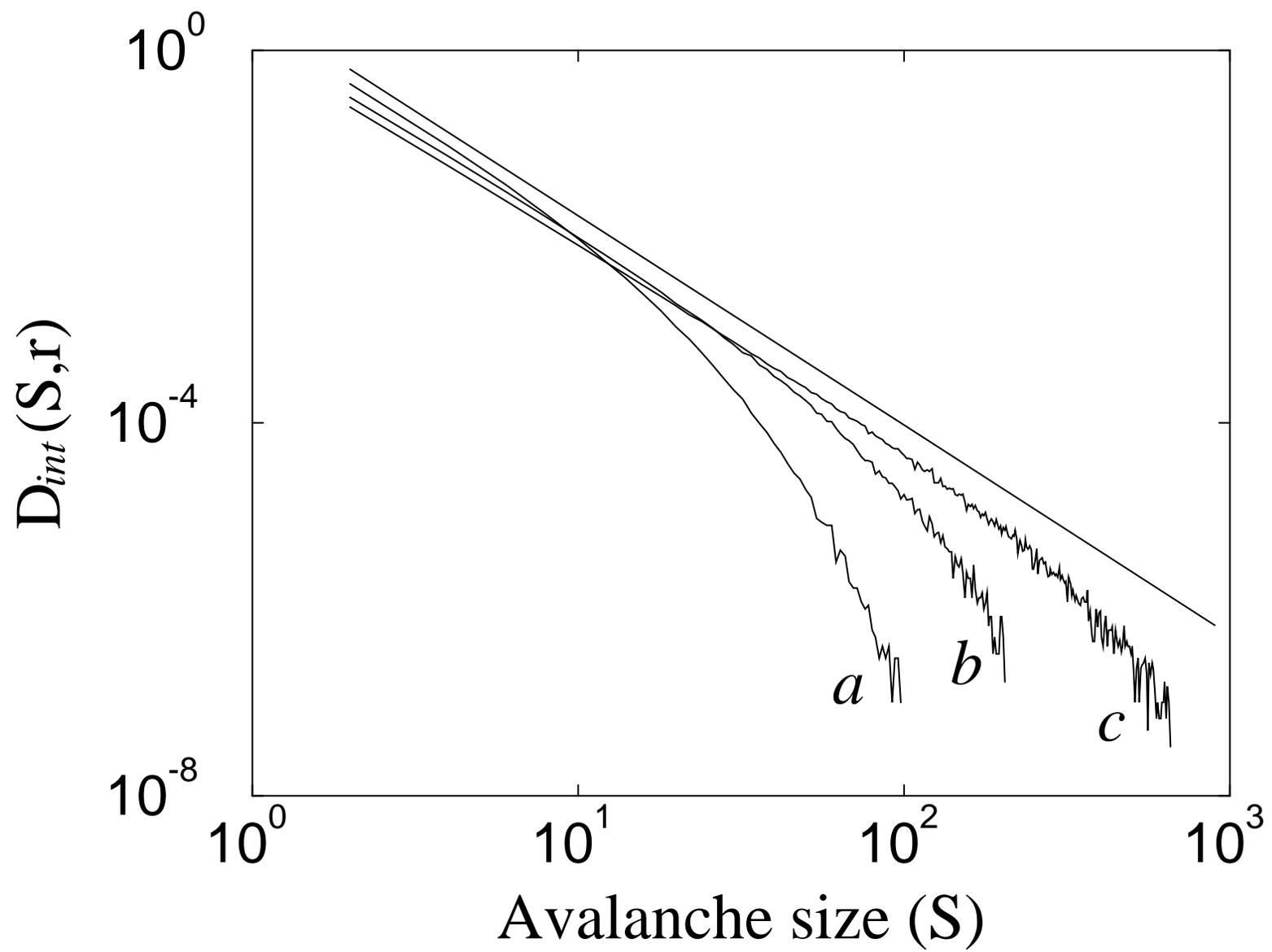
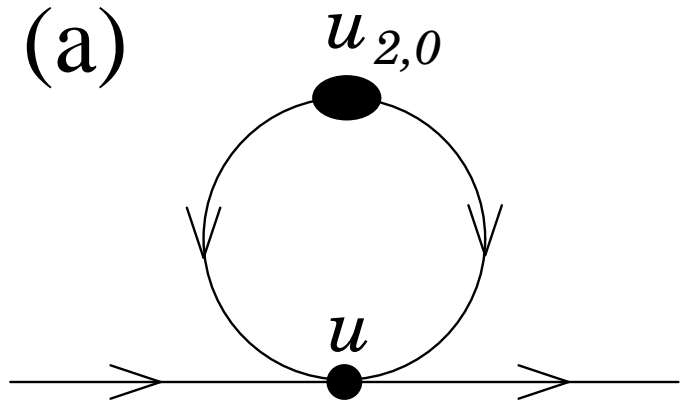
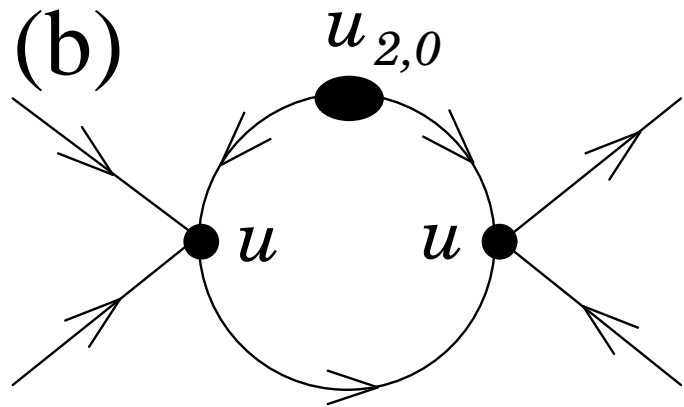
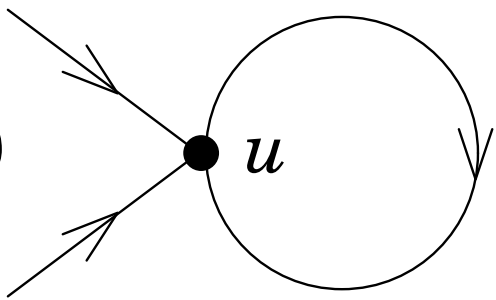
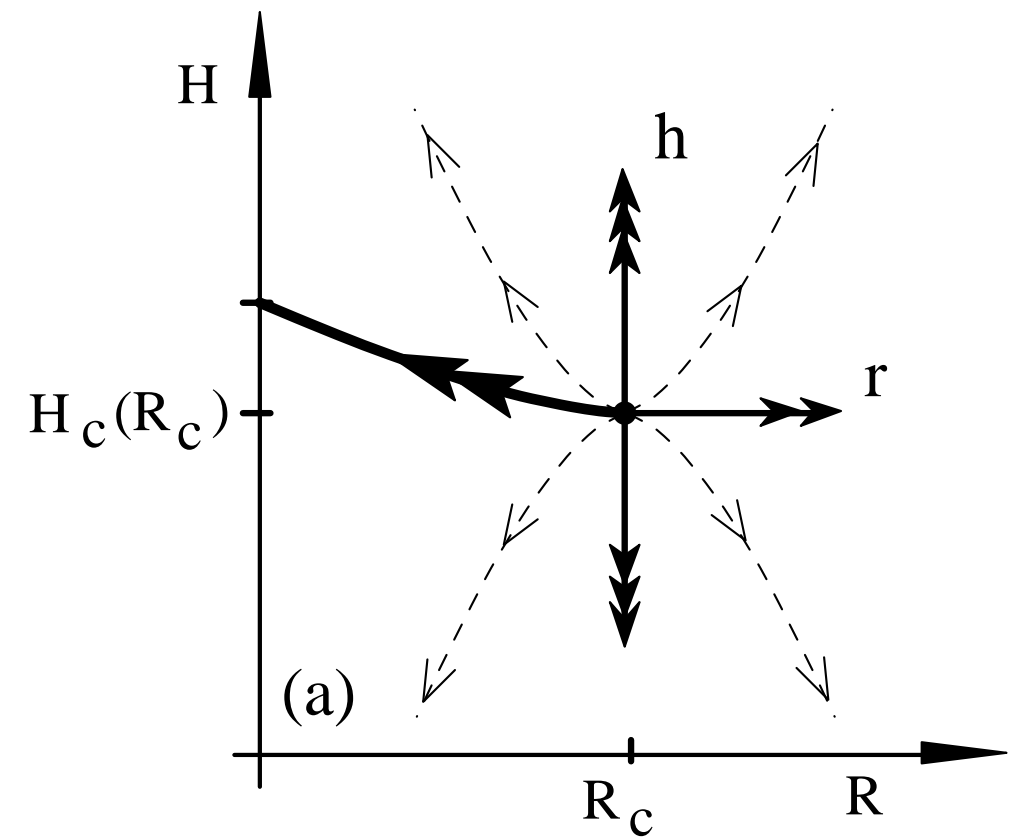
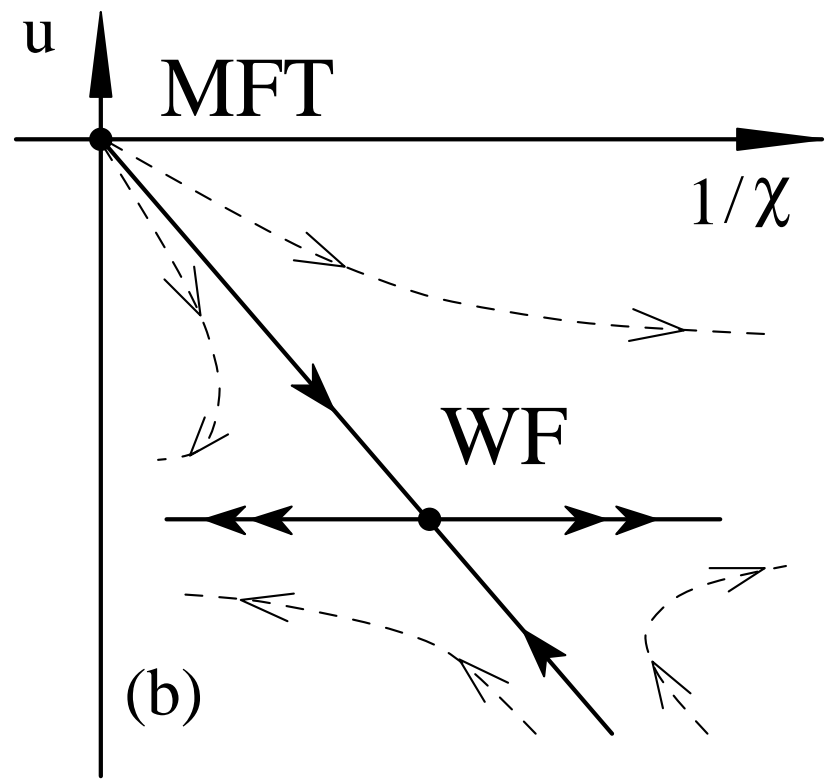
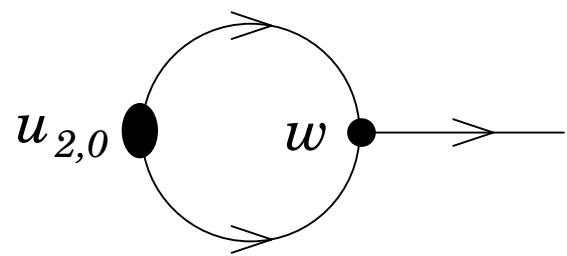
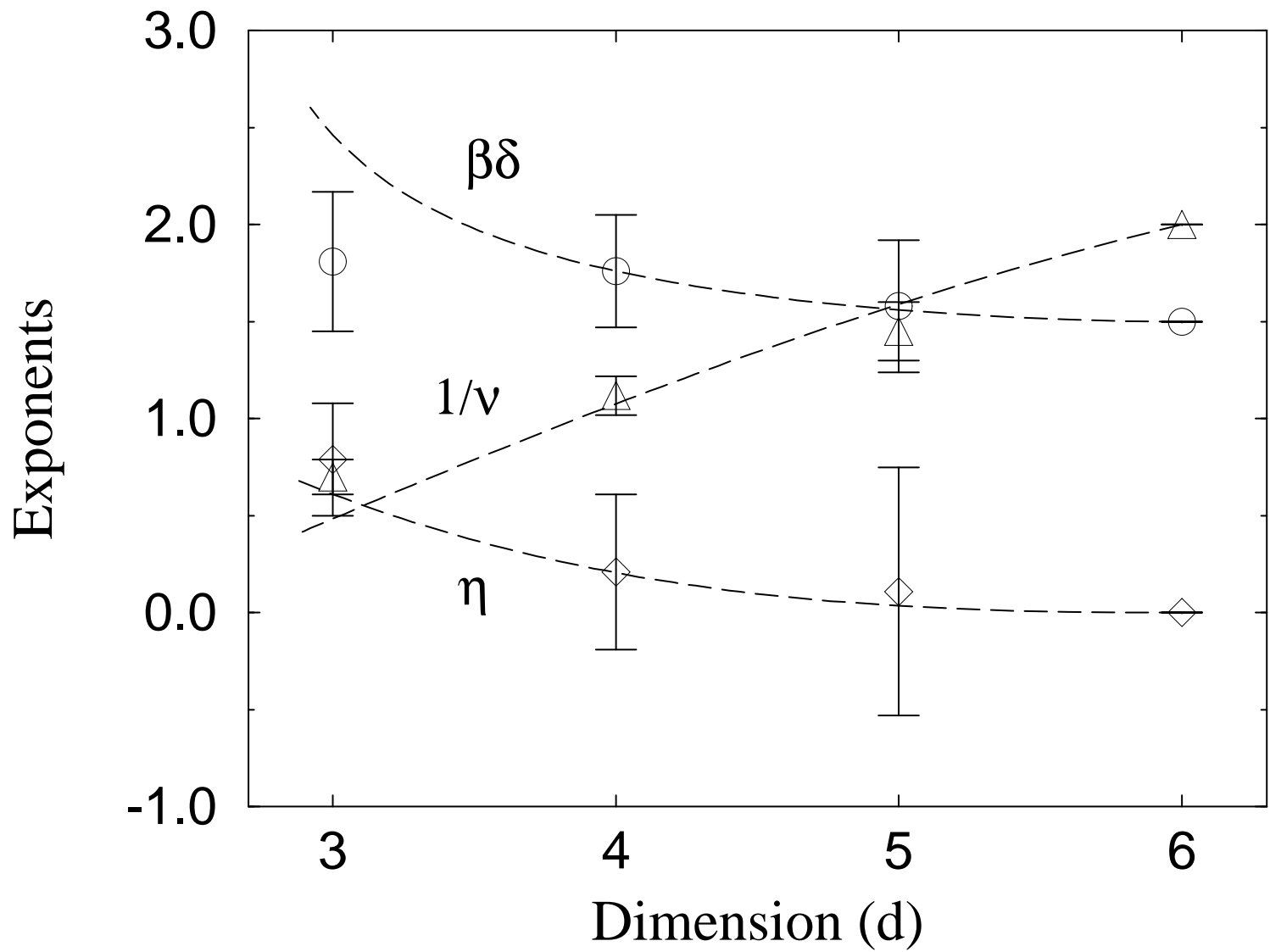
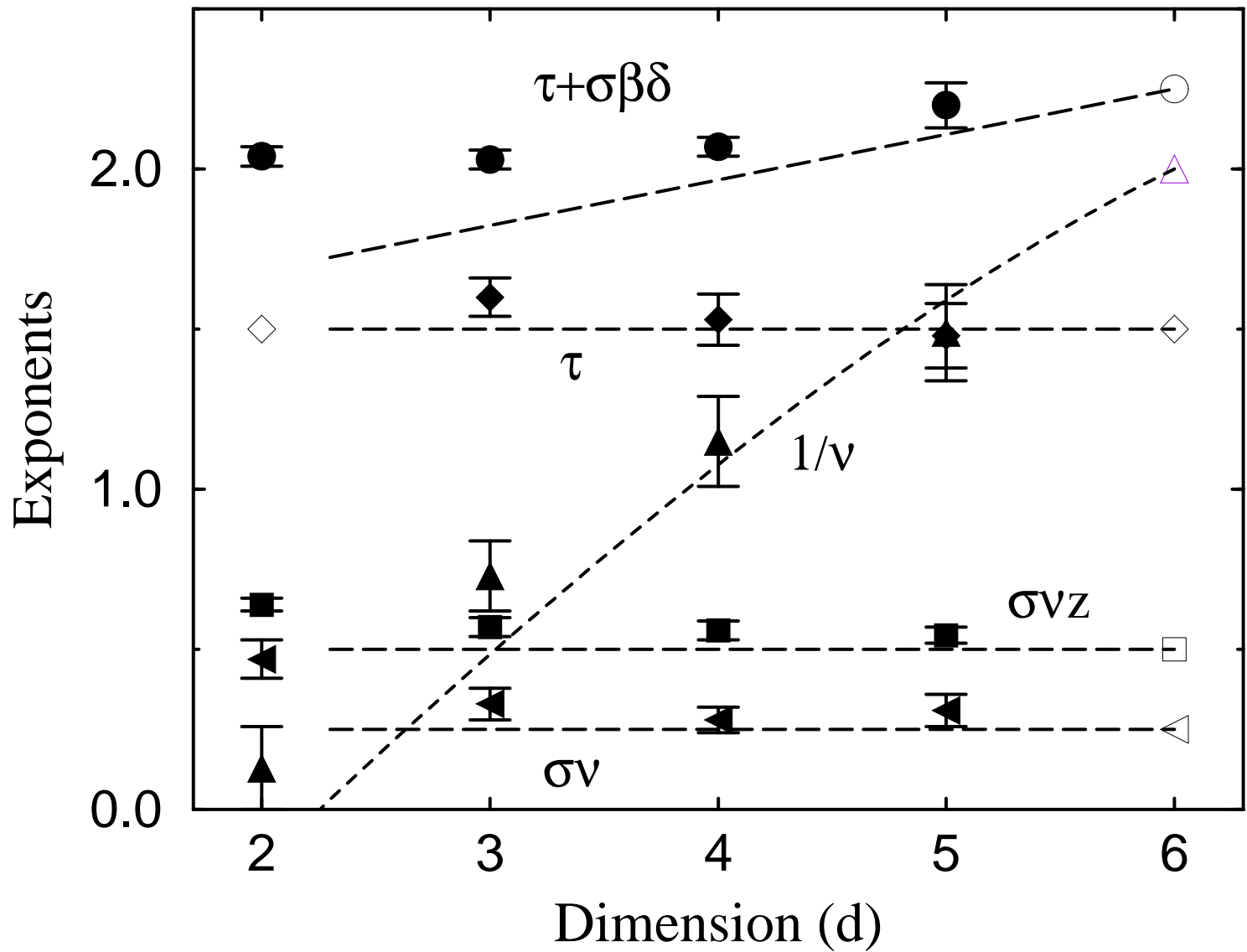
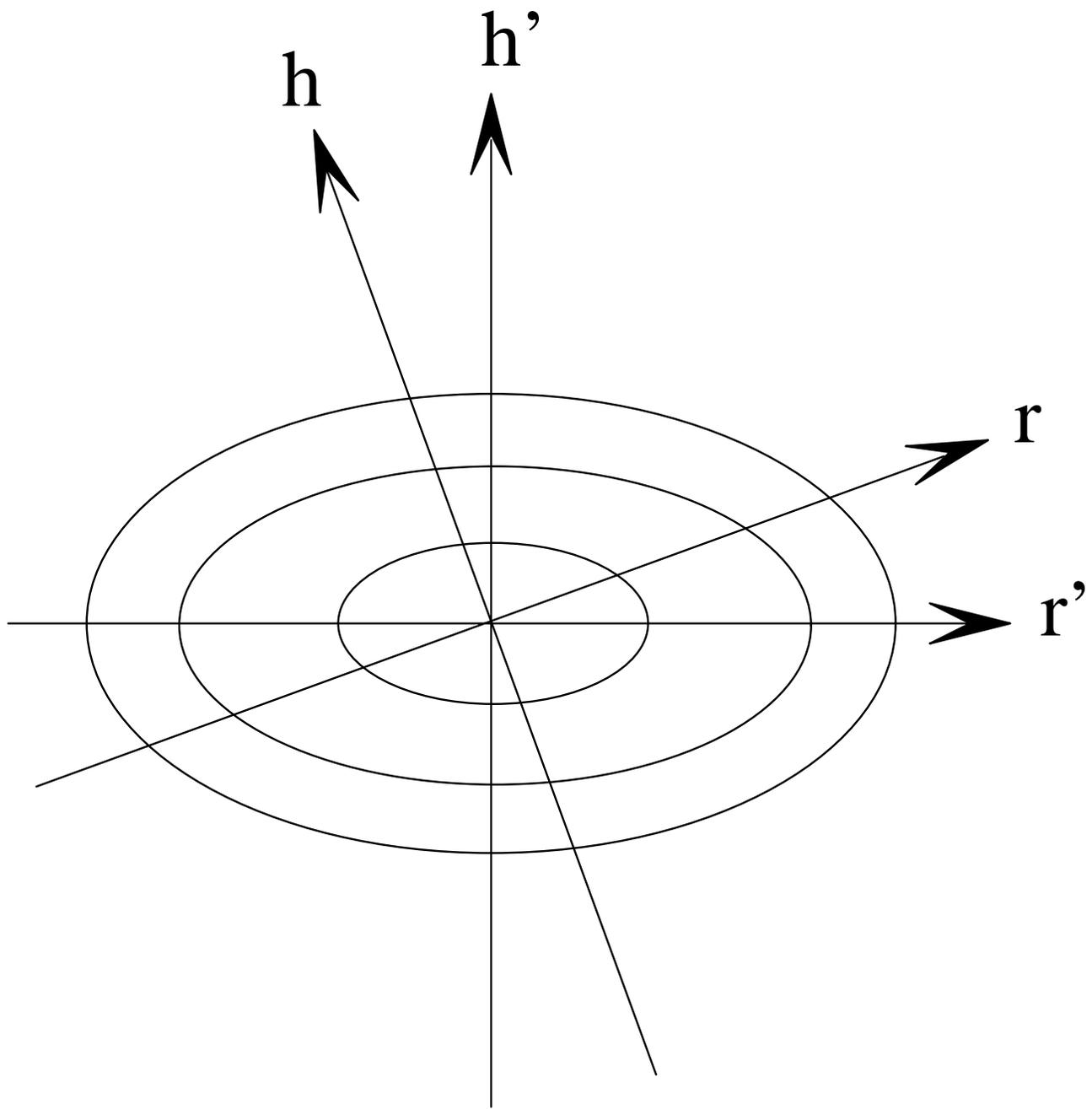
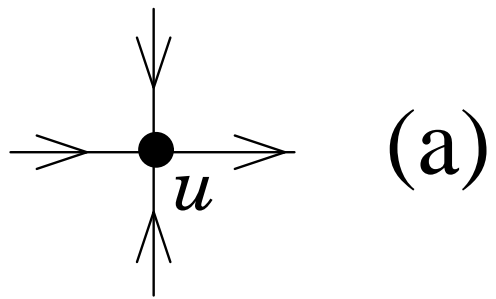
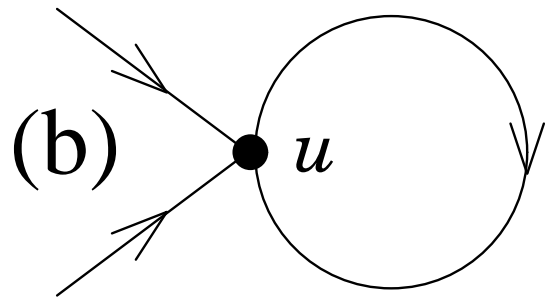
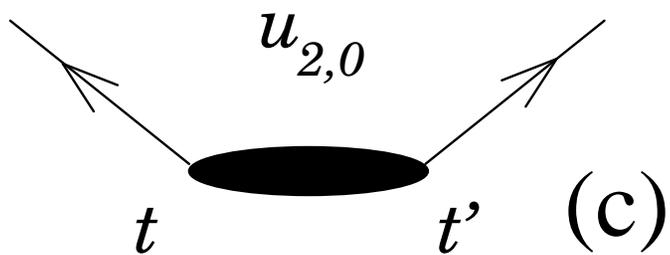
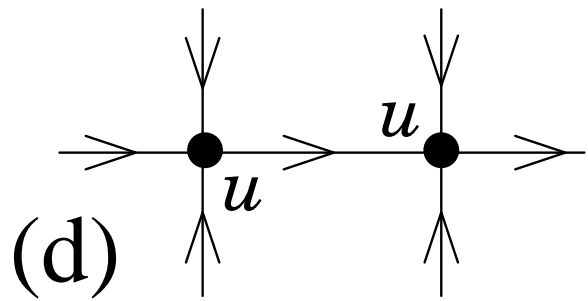
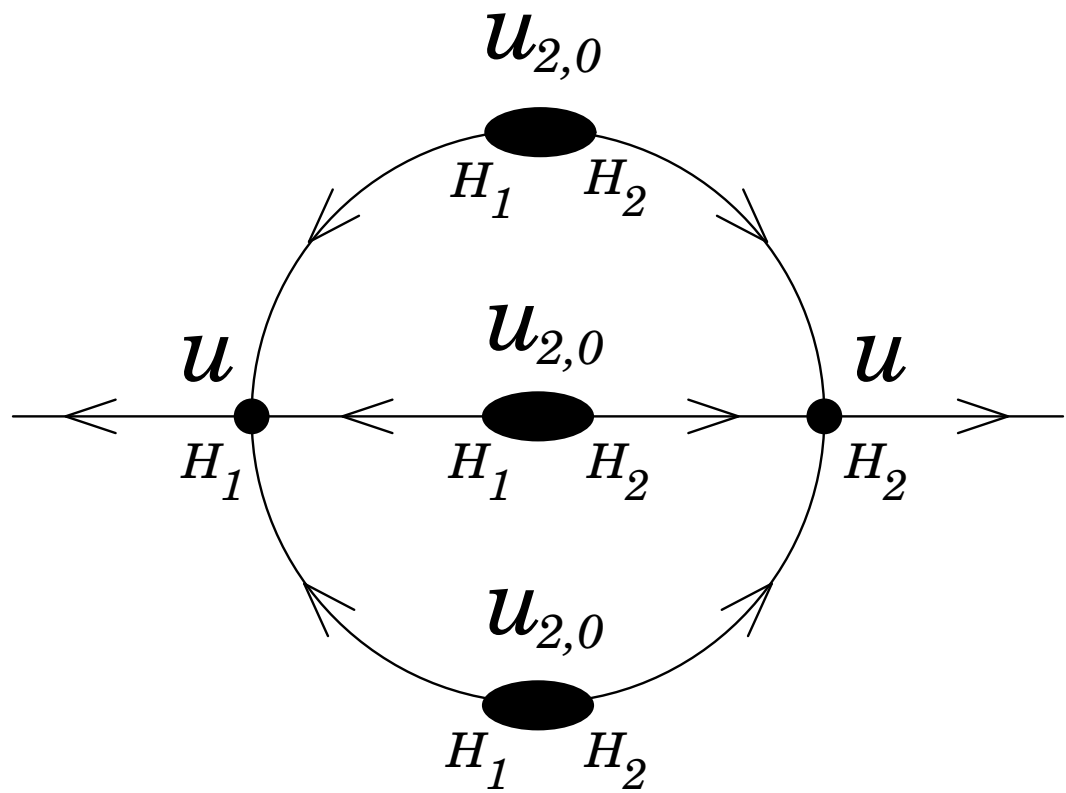
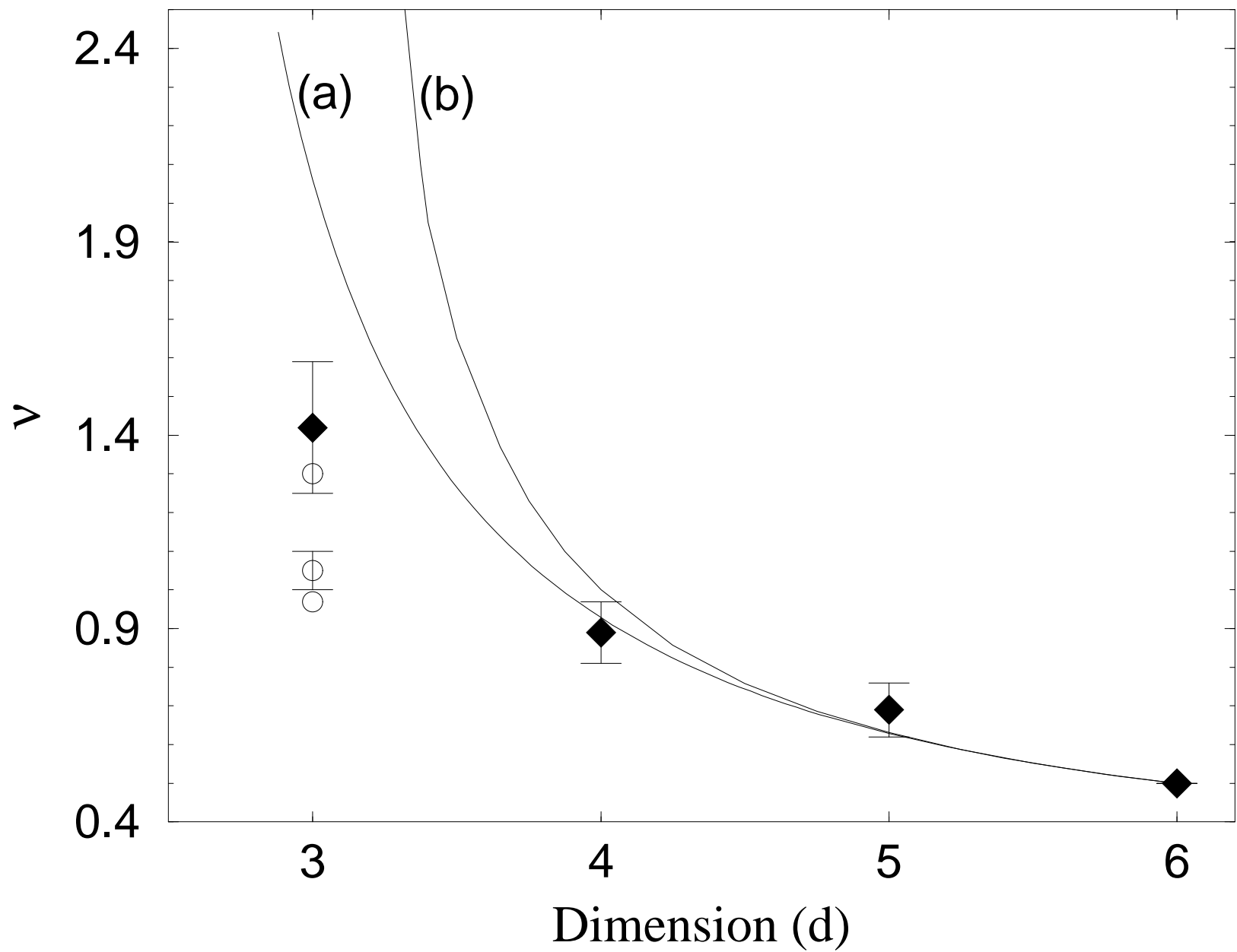
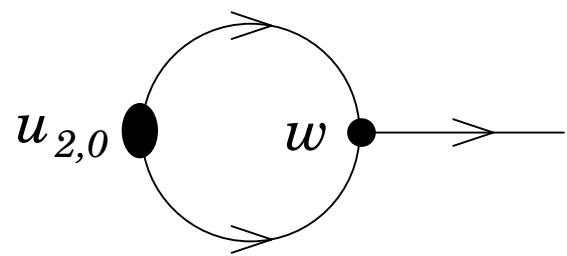
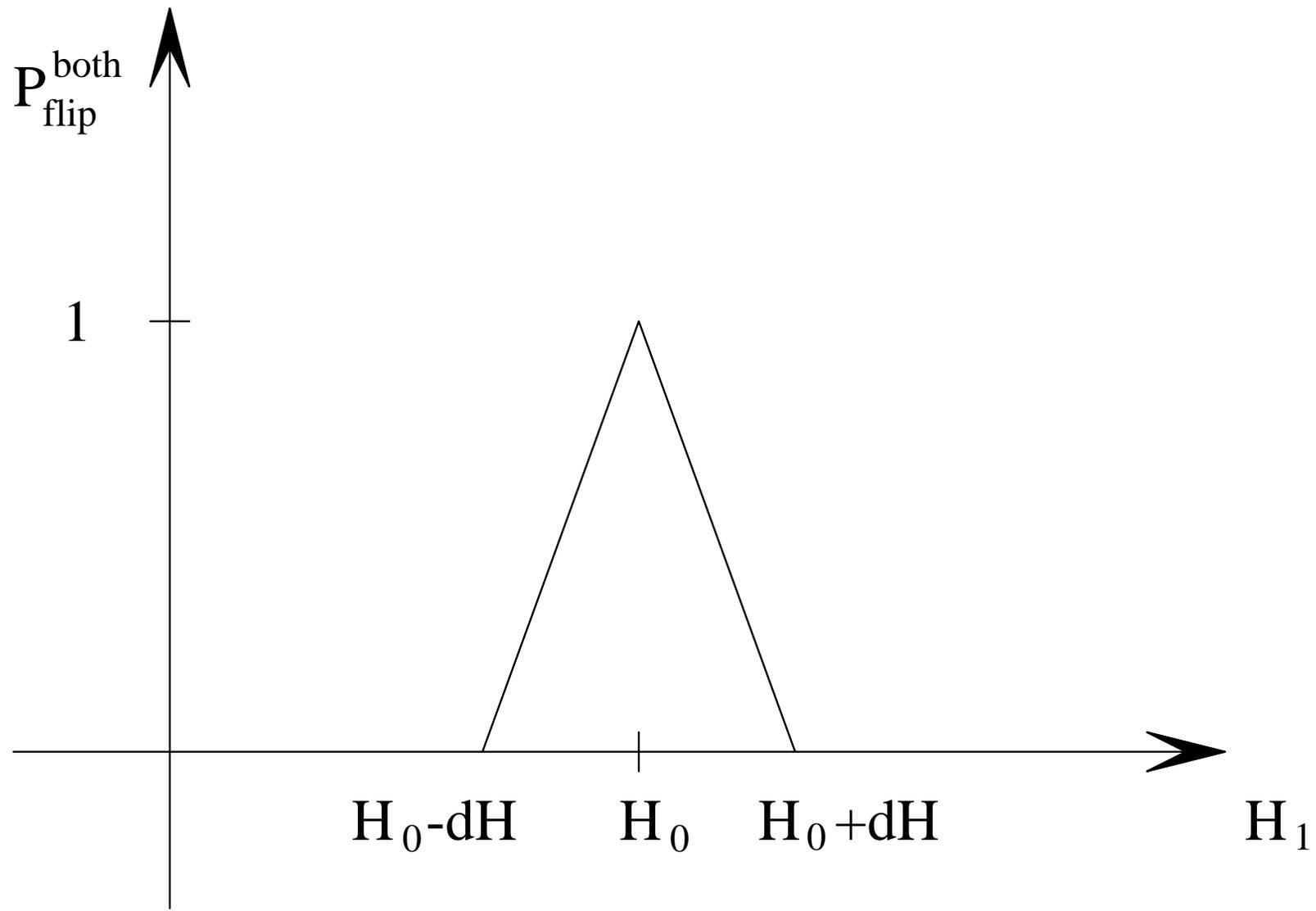
\end{document}